\documentclass{aastex631}
\usepackage{csquotes}

\shorttitle{Scylla II}
\shortauthors{Cohen et al.}

\begin{document}
\title{Scylla II. The Spatially Resolved Star Formation History of the Large Magellanic Cloud Reveals an Inverted Radial Age Gradient}

\correspondingauthor{Roger E. Cohen}
\email{rc1273@physics.rutgers.edu}

\author[0000-0002-2970-7435]{Roger E. Cohen}
\affiliation{Department of Physics and Astronomy, Rutgers the State University of New Jersey, 136 Frelinghuysen Rd., Piscataway, NJ, 08854, USA}

\author[0000-0001-5538-2614]{Kristen B. W. McQuinn}
\affiliation{Department of Physics and Astronomy, Rutgers the State University of New Jersey, 136 Frelinghuysen Rd., Piscataway, NJ, 08854, USA}
\affiliation{Space Telescope Science Institute, 3700 San Martin Drive, Baltimore, MD 21218, USA}

\author[0000-0002-7743-8129]{Claire E. Murray}
\affiliation{Space Telescope Science Institute, 3700 San Martin Drive, Baltimore, MD 21218, USA}
\affiliation{The William H. Miller III Department of Physics \& Astronomy, Bloomberg Center for Physics and Astronomy, Johns Hopkins University, 3400 N. Charles Street, Baltimore, MD 21218, USA}

\author{Benjamin F. Williams}
\affiliation{Department of Astronomy, University of Washington, Box 351580, U. W., Seattle, WA 98195-1580, USA}

\author{Yumi Choi}
\affiliation{NSF's National Optical-Infrared Astronomy Research Laboratory, 950 N. Cherry Avenue, Tucson, AZ 85719 USA}

\author[0000-0003-0588-7360]{Christina W. Lindberg}
\affiliation{The William H. Miller III Department of Physics \& Astronomy, Bloomberg Center for Physics and Astronomy, Johns Hopkins University, 3400 N. Charles Street, Baltimore, MD 21218, USA}
\affiliation{Space Telescope Science Institute, 3700 San Martin Drive, Baltimore, MD 21218, USA}

\author{Clare Burhenne}
\affiliation{Department of Physics and Astronomy, Rutgers the State University of New Jersey, 136 Frelinghuysen Rd., Piscataway, NJ, 08854, USA}

\author[0000-0001-5340-6774]{Karl D.\ Gordon}
\affil{Space Telescope Science Institute, 3700 San Martin Drive, Baltimore, MD 21218, USA}

\author{Petia Yanchulova Merica-Jones}
\affiliation{Space Telescope Science Institute, 3700 San Martin Drive, Baltimore, MD 21218, USA}

\author{Karoline M. Gilbert}
\affiliation{Space Telescope Science Institute, 3700 San Martin Drive, Baltimore, MD 21218, USA}

\author{Martha L. Boyer}
\affiliation{Space Telescope Science Institute, 3700 San Martin Drive, Baltimore, MD 21218, USA}

\author{Steven Goldman}
\affiliation{Space Telescope Science Institute, 3700 San Martin Drive, Baltimore, MD 21218, USA}

\author{Andrew E. Dolphin}
\affiliation{Raytheon, 1151 E. Hermans Road, Tucson, AZ 85756, USA}
\affiliation{University of Arizona, Steward Observatory, 933 North Cherry Avenue, Tucson, AZ 85721, USA}

\author{O. Grace Telford}
\affiliation{Department of Physics and Astronomy, Rutgers the State University of New Jersey, 136 Frelinghuysen Rd., Piscataway, NJ, 08854, USA}
\affiliation{Department of Astrophysical Sciences, Princeton University, 4 Ivy Lane, Princeton, NJ 08544, USA}
\affiliation{The Observatories of the Carnegie Institution for Science, 813 Santa Barbara Street, Pasadena, CA 91101, USA}
\altaffiliation{Carnegie-Princeton Fellow}

\begin{abstract}
    
The proximity of the Magellanic Clouds provides the opportunity to study interacting dwarf galaxies near a massive host, and spatial trends in their stellar population properties in particular, with a unique level of detail.  The Scylla pure parallel program has obtained deep (80\% complete to $>$1 mag below the ancient main sequence turnoff), homogeneous two-filter Hubble Space Telescope (\textit{HST}) imaging sampling the inner star-forming disk of the Large Magellanic Cloud (LMC), the perfect complement to shallower, contiguous ground-based surveys.  We harness this imaging together with extant archival data and fit lifetime star formation histories (SFHs) to resolved color-magnitude diagrams (CMDs) of 111 individual fields, using three different stellar evolutionary libraries.  We validate per-field recovered distances and extinctions as well as the combined global LMC age-metallicity relation and SFH against independent estimates.   
We find that the present-day radial age gradient reverses from an inside-out gradient in the inner disk to an outside-in gradient beyond $\sim$2 disk scalelengths, supported by ground-based measurements.  
The gradients become relatively flatter at earlier lookback times, while the location of the inversion remains constant over an order of magnitude in lookback time, from $\sim$1$-$10 Gyr.  This suggests at least one mechanism that predates the recent intense LMC-SMC interaction.  We compare observed radial age trends to other late-type galaxies at fixed stellar mass and discuss similarities and differences in the context of potential drivers, implying strong radial migration in the LMC.

\end{abstract}

\section{Introduction}

Dwarf galaxies represent ideal probes of the processes driving galaxy evolution due to their low masses (6$\lesssim$ Log$_{\rm 10}$ M$_{\star}$/M$_{\odot}$ $\lesssim$9) and correspondingly shallow potential wells.  Dwarfs are especially well suited for understanding the role of galaxy-galaxy interactions on their mass assembly over cosmic time.   
Compared to more massive galaxies, 
dwarfs are more common at all redshifts \citep[e.g.,][]{karachentsev13}, and they are preferentially star forming \citep[e.g.,][]{geha12}.  Simulations predict that dwarfs interact more frequently than massive galaxies within a given volume \citep{deason14}, and interacting dwarf galaxies are more likely to host star formation compared to isolated counterparts \citep{kadofong20,sun20}.  Furthermore, both simulations \citep{martin21} and observations of dwarf galaxy pairs throughout the local universe \citep{stierwalt15} reveal that dwarf-dwarf interactions cause enhancements in star formation even when they do not result in a merger.  However, among star-forming dwarf galaxies, the spatial distribution of star formation varies widely in terms of both central concentration and asymmetry \citep[e.g.,][]{mcquinn12,privon17,hunter18,annibali22}.  While several mechanisms have been proposed that can influence trends in stellar age within dwarf galaxies \citep[e.g.,][]{stinson09,radburnsmith12,mostoghiu18,graus}, the role of dwarf-dwarf interactions in setting intra-galaxy age trends is unclear.  
Therefore, any dwarf galaxy with independent constraints on its interaction history is a particularly critical target: If internal age gradients can be measured with sufficient resolution both spatially and temporally, we may constrain the impact of galaxy-galaxy interactions on its mass assembly over its entire lifetime.  

\subsection{The Large Magellanic Cloud as an Interacting Dwarf Galaxy}

The Magellanic Clouds are the archetypal example of interacting dwarf galaxies, and their proximity provides us with a uniquely detailed portrait of prolonged dwarf-dwarf interactions.  While the Clouds are likely on their first infall into the Milky Way, entering its virial radius $\sim$1$-$2 Gyr ago
\citep{nitya06a,besla07,mbk11,besla12,patel17,conroy21}, the LMC and SMC have been interacting with each other for several Gyr  
\citep{nitya06b,besla07,besla12,cullinane22a,cullinane22b}.    
This prolonged interaction includes a recent direct collision, occurring 140-160 Myr ago based on analytical orbit modeling \citep{zivick18}.   Adopting this timing constraint, a comparison of residual proper motions in the LMC disk with the \citet{besla12} models of the LMC-SMC interaction implies an impact parameter for the collision of $\sim$5 kpc \citep{choi22}.  The best-fitting \citet{besla12} models further predict that the LMC and SMC may have been interacting for as long as $\sim$6 Gyr (but see \citealt{vasiliev24}), in accord with coeval enhancements in their star formation rates beginning $\sim$3-4 Gyr ago \citep{hz09,weisz13,massana22}.

The prolonged LMC-SMC interaction has produced a plethora of signatures seen in the morphology and kinematics of stars and gas in the LMC.    
These include large-scale features such as the Leading Arm \citep{putman98}, Magellanic Stream 
\citep{mathewson74,stream1,stream2,chandra23} 
and Magellanic Bridge \citep[e.g.,][]{hindman63,harris07} as well as   
substructure near the LMC periphery \citep[e.g.,][]{besla16,mackey18,nidever19,gatto22a}.  In addition, while the gas disk is truncated towards the northeast, likely due to ram pressure stripping from the circumgalactic medium of the Milky Way \citep{salem15}, the stellar disk is truncated towards the southwest, likely resulting from multiple pericentric passages by the SMC \citep[e.g.,][]{belokuroverkal19}.  

Importantly, evidence of the extended LMC-SMC interaction is not restricted to the outer parts of the LMC, and numerous other interaction signatures are seen in the stellar disk, including its single spiral arm \citep{dev55}, warps and twists \citep{vdmcioni,olsen,subra10,oglelmcrr,yumimap,saroon22}, a ring-like overdensity \citep{yumiring}, and a tilted, off-centered bar \citep[e.g.][]{zhao00,zaritsky04,yumimap}.  In addition to morphological signatures, evidence of the LMC-SMC interaction is also borne out via stellar kinematics and chemistry \citep[e.g.][]{carrera17,cullinane20,grady21,choi22,cheng22,schmidt22,munoz23}.  

The LMC-SMC system is not strictly a benchmark for \textit{pairs} of interacting dwarf galaxies since the three-body LMC-SMC-Milky Way configuration is highly unusual in a cosmological context \citep{mbk11,liu11,robotham12,gonzalez13,besla18}.  However, the aforementioned features result mainly from prolonged interaction with the SMC rather than infall into the Milky Way based on both simulations \citep{murai80,besla12,besla16,yozin,williamson} and empirical evidence.  In particular, observations of LMC-SMC analogs find that interacting dwarfs retain substantial fractions of HI gas \citep{pearson16,fan23}, hosting LMC-like interaction signatures including gas bridges and one-armed spirals \citep{pearson18,luber22}.  These features have been detected not only for isolated dwarf pairs, but also for a possible LMC-SMC-Milky Way analog in at least one case \citep{paudel17}.

\subsection{Star Formation Histories from Resolved Color-Magnitude Diagrams}

In addition to clues gleaned from present-day morphology and kinematics, SFH fitting to resolved CMDs provides a powerful avenue to understand the evolution of dwarf galaxies.  
While ongoing or recent ($\lesssim$100 Myrs) star formation can be traced with H$\alpha$ and UV imaging \citep[e.g.,][]{kennicutt98,weisz12,cignoni19}, SFH fits to resolved CMDs extending faintward of the old ($\sim$13 Gyr) main sequence turnoff (oMSTO) have the potential to provide time-resolved \textit{lifetime} trends of both star formation rate and chemical enrichment (see \citealt{annibalitosi} for a review in the context of dwarf galaxies).  To date, observations reaching sufficient photometric depths to probe ancient lookback times have been restricted to the Local Group, and typically require spatial resolution achievable only with \textit{HST}, although there are some exceptions \citep[e.g.,][]{sand09,delpino13,santana16,bettinelli19}.
However, even just within the Local Group, lifetime SFHs have been obtained for low-mass galaxies over a range of environments, from Milky Way and M31 satellites \citep{skillman17,weisz19,savino23} to more isolated dwarfs \citep{cole07,cole14,monelli10a,monelli10b,hidalgo11,skillman14,albers19}.  This ensemble of lifetime SFHs has provided vast insight into galaxy evolution at low masses, elucidating the influence of environment on timescales for star formation and quenching \citep{gallart15,weisz15,aparicio16,sacchi21b,mcquinn23a,mcquinn23b}.

Ever since recognizing that the LMC and the Milky Way have fundamentally different SFHs \citep{hodge61}, SFH studies of the Magellanic Clouds have played a critical role in understanding the assembly history of the Local Group.  The simultaneous increase in star formation rate in the LMC and SMC beginning $\sim$3$-$4 Gyr ago and subsequent contemporaneous bursts \citep{hz09,weisz13,massana22} constrain the history of the LMC-SMC interaction.  Additionally, the reduced star formation rate during the $\sim$8 Gyr preceding the recent enhancement, particularly in the LMC, argues against multiple pericentric passages about the Milky Way, corroborating the first infall scenario.

\subsection{Spatially Resolved Star Formation Histories}

As a complement to galaxy-wide SFHs, spatially resolved lifetime SFHs within a galaxy can provide additional insight into the drivers of mass assembly via \textit{intra}-galaxy age gradients.  Such measurements can now be compared with simulations, which make predictions for internal age gradients and their relationship with other observables in dwarf galaxies \citep{graus}.  However, the role of dwarf-dwarf interactions in setting internal stellar population gradients is unclear, from either an observational or theoretical standpoint.  Models designed to assess the impact of perturbing companions are restricted to more massive disks, predicting that internal stellar population gradients should be measurably affected.  In particular, radial stellar migration \citep[e.g.,][]{sellwoodbinney}, 
a dominant process in the secular evolution of Milky Way-mass disks \citep{roskar12,bird13,frankel18,vincenzo20,patilbovy}, can be boosted by interactions with satellites \citep{quillen09,bird12,carr22} as well as additional factors including arm-bar resonances \citep{minchev10,minchev11,halle15} and high gas fractions \citep{bird12}.  The efficiency of such radial migration tends to increase with radius, preferentially populating the outer disk with old stars that migrated from more interior radii \citep[e.g.,][]{roskar08,minchev12,bird13}, as observed in the Milky Way \citep[e.g.,][]{lian22}.  
However, at LMC-like masses, the extent to which dwarf-dwarf interactions and other factors can boost radial migration remains unclear.  
Observationally, direct measurements of radial age gradients from SFH fits to resolved CMDs are only available for 12 Local Volume dwarfs 
\citep{williams09,cannon12,hidalgo13,delpino13,fusco14,santana16,bettinelli19,sacchi19,albers19,ruizlara21}, \textit{all} of which are either more isolated or less massive than the LMC.    
Therefore, the LMC serves as a crucial testbed to constrain the impact of dwarf-dwarf interactions on galaxy assembly by resolving its SFH both spatially and temporally over its entire lifetime.

Recognizing the value of characterizing spatial trends in the lifetime SFHs of the LMC, ground-based photometric surveys are producing contiguous SFH maps with ever-improving depth and fidelity.  These include the Magellanic Clouds Photometric Survey \citep{hz09}, 
the VISTA Survey of the Magellanic Clouds system \citep[VMC;][]{vmc,rubele12,mazzi21} and the Survey of the MAgellanic Stellar History \citep[SMASH;][]{ruizlara20,smashdr2,massana22}.  Dedicated imaging surveys are also mapping extensive portions of the LMC and SMC outskirts \citep{noel13,noel15,ripepi14,maglites,carrera17,cerny23,gatto24}, complemented by targeted spectroscopic observations \citep[e.g.,][]{lazygiants,cullinane20}.  
However, ground-based imaging is not always able to resolve individual stars faintward of the LMC oMSTO in crowded regions.  Space-based observations with \textit{HST} provide a crucial complement to ground-based imaging surveys, as its improved spatial resolution and well-characterized point spread function (PSF) can yield photometry reaching well down the unevolved main sequence at the distance of the LMC \citep{geha98,holtzman99,smeckerhane,weisz13}, 
although carefully designed, targeted ground-based observations are also capable of reaching sufficient photometric depths \citep{gallart08,meschin,nidever17,monteagudo}.

Thus far, deep imaging campaigns towards the LMC have yielded several insights on spatial variations in its SFH, constraining the longevity of the spiral arm to $\sim$2.5 Gyr \citep{ruizlara20}, while finding general consistency across SFHs in the bar and inner disk \citep{weisz13,monteagudo}.  Meanwhile, ground-based observations of three more distant fields (R$>$3 kpc) in the LMC's northern disk reveal an outside-in gradient correlated with HI column density.  Specifically, more distant fields with lower N(HI) are older in the mean, having already ceased star formation as much as $>$1 Gyr ago \citep{gallart08,meschin}.  This picture is qualitatively consistent with the age-radius relation from \citet{povick}, obtained by combining spectroscopically derived stellar parameters with multiband photometry and evolutionary models.  Insofar as the available spatial sampling could reveal, \citet{povick} also found hints that the southern disk was older, in the median, than the northern disk, at least out to R$\sim$7 kpc.  
However, a clear picture of spatially resolved star formation in the LMC probing back more than several Gyr to before the LMC-SMC interaction is still lacking.  Specifically, trends in stellar age as a function of distance from the center of the LMC can serve as a powerful diagnostic, both observationally and theoretically.  From an empirical standpoint, radial stellar age trends in other late-type LMC-mass galaxies show a \enquote{V-shaped} radial profile with an inversion in some cases \citep{williams09,sacchi19}, but the role of interactions and environment in setting such radial trends is unclear.  A similarly inverted \enquote{V-shaped} radial profile is suggested by the combination of existing ground-based results for the inner and outer LMC \citep{meschin,monteagudo}, but improved spatial coverage and a self-consistent analysis are needed to verify whether this trend holds over the full radial extent of the LMC's disk and enable comparisons to other galaxies.  Furthermore, available observations have yet to provide quantitative constraints on the significance of radial stellar age trends.  Previous studies of the LMC hint at azimuthal variations in the stellar age distribution within the LMC (i.e., at a fixed distance from its center; \citealt{ruizlara20,povick}), but lack the depth and/or spatial coverage to place such variations in the context of radial age gradients over the full extent of the disk.  At the same time, measurements of radial stellar age trends and their significance can be compared to cosmological simulations, which are now capable of making predictions for radial stellar age gradients as a function of global galaxy properties, including interaction history \citep[e.g.,][]{graus}. 

Here, we aim to leverage the SFH precision provided by deep, space based imaging together with the spatial resolution provided by numerous \textit{HST} imaging fields throughout the LMC disk that are now available.  By exploiting the high-fidelity CMDs yielded by homogeneous broadband \textit{HST} imaging, we can examine not only the spatial variations in SFH trends, but also the variation in these trends as a function of lookback time, probing back to before the LMC-SMC interaction.  While we focus on lifetime SFH trends across the LMC here, separate forthcoming studies include a companion paper on lifetime SFH trends within the SMC (Cohen et al.~2024) and the spatial and temporal distribution of starbursts in the LMC and SMC (C.~Burhenne et al., in prep.).

This paper is organized as follows: In Sect.~\ref{datasect}, we describe the new and archival imaging we employ and our PSF photometry and artificial star tests.  The methodology for fitting SFHs and the lifetime SFH metrics we use are presented in Sect.~\ref{sfhsect}.  In Sect.~\ref{obsgradsect} we present radial trends in lifetime SFH metrics, and discuss these results in the context of existing observations and simulations in Sect.~\ref{discusssect}.  We summarize our results and outstanding questions in Sect.~\ref{futuresect}, and provide an Appendix that includes tests of both internal and external consistency of our SFH results.

\section{Data \label{datasect}}

\subsection{Observations}

To gain insight into spatial trends in the lifetime star formation history of the LMC, we harness together an unprecedented number of \textit{HST} pointings, consisting of both new and archival imaging.  The on-sky spatial distribution of the fields we analyze here is shown in Fig.~\ref{map_fig}, with our target fields shown as filled circles color coded by their source.  The underlying stellar density map was built using the second data release of the SMASH survey \citep{smashdr2}, including only SMASH DR2 sources with \texttt{chi}$<$5, $|$\texttt{sharp}$|<$2, and 18.5$< g <$19.0 following \citet{yumiring}.  In the LMC, old stellar populations have a relatively smooth spatial distribution across the disk while younger populations show an increasingly fragmented smaller-scale structure \citep[e.g.][]{yumiring,ely19,gaia21,mazzi21}, so the SMASH magnitude cut is chosen to highlight the large- and intermediate-scale spatial structure in the inner disk of the LMC where many of our fields lie, including the bar and northern spiral arm.  To place our results in context, we also overplot fields from previous targeted observations focused on spatial trends of the LMC SFH, obtained from both the ground \citep{meschin, monteagudo} and with \textit{HST} \citep{weisz13}.  

\subsubsection{Scylla Fields: WFC3/UVIS F475W+F814W}

The new observations we use were obtained as part of the Scylla multi-cycle pure parallel imaging program (GO-15891, GO-16235, GO-16786; PI: C.~E.~Murray), accompanying the HST Ultraviolet Legacy library of Young Stars as Essential Standards (ULYSSES; \citealt{ulysses}) spectroscopic campaign.  Towards the LMC, ULYSSES obtained ultraviolet spectroscopy of massive O and B stars, and during these observations, Scylla obtained parallel imaging with Wide Field Camera 3 (WFC3).  When the ULYSSES primary observations 
required multiple orbits per target, Scylla obtained parallel imaging of a field in as many as seven broadband WFC3 filters from the ultraviolet (F225W) to the near-infrared (F160W).  However, to maximize scientific productivity, Scylla parallel exposures in F475W and F814W were prioritized in cases of shorter primary visits, yielding deep homogeneous 2-band photometry across all fields.  Additional details, including an overview of the Scylla pure parallel program and the science drivers motivating the observing strategy, as well as exposure-level information, are provided in Murray et al.~(2024).   Basic observational information for the fields we use is given in Table \ref{obstab}.  

\subsubsection{METAL Fields: WFC3/UVIS F475W+F814W}

We supplement the Scylla LMC parallel imaging fields with 
additional parallel imaging fields towards the LMC from the Metal Evolution, Transport and Abundance in the Large Magellanic Cloud (METAL) program \citep{metal}, obtained as part of GO-14675 (PI: Roman-Duval).  These fields were observed using the same instrument and filter combination as Scylla and a similar strategy (i.e., giving top priority to exposures in F475W and F814W), yielding compatible photometric depth (described in more detail below in Sect.~\ref{photsect}).  These fields are listed in Table \ref{obstab}, and for clarity their field numbers begin with the letter M.  

\subsubsection{Archival WFPC2 Fields: F555W+F814W}

We also exploit archival Wide Field Planerary Camera 2 (WFPC2) photometry from the Local Group Stellar Photometry Archive \citep[LGSPA;][]{lgspa}.  We retain only the LMC fields they denote as Priority 1 (a subset of which were analyzed by \citealt{weisz13}), meaning that they all have imaging in both the F555W and F814W filters extending faintward of the oMSTO.  We downloaded the raw photometry and artificial star catalogs provided on the LGSPA website, making cuts on per-filter photometric quality diagnostics identically as for the Scylla and METAL imaging (described below in Sect.~\ref{photsect}).  All of the LGSPA fields we use are listed in Table \ref{obstab} with field numbers beginning with the letter W.

\subsubsection{Outer Fields}

All of the imaging discussed thus far samples only the inner $\sim$4-5 kpc of the LMC disk.  To extend our analysis of radial trends within the LMC to larger distances and corroborate ground-based results from \citet{meschin}, we searched for fields imaged with two or more optical and/or near-infrared broadband filters to sufficient photometric depth for CMD-based SFH fitting (faintward of the oMSTO).  After discarding fields contaminated by known clusters, the search yielded four pointings: Three from GO-11700 (PI: Trenti), imaged in WFC3/UVIS F606W and WFC3/IR F160W, and one from GO-14688 (PI: Goudfrooij), imaged with ACS/WFC F435W and F814W.  These images were reduced and analyzed identically as the Scylla and METAL WFC3/UVIS pointings\footnote{Field LMC\_OF4 is the only field imaged with ACS/WFC, for which we found that more stringent photometric quality cuts optimized the rejection of spurious detections, so we required \texttt{crowd}$\leq$0.04 and $|$\texttt{sharp}$| \leq$ 0.1.} and they are listed in Table \ref{obstab} with field numbers beginning with OF. 

\vspace{0.5cm}

Basic per-field information relevant to our SFH fitting is listed for all fields in Table \ref{obstab}.  In addition to the field name in the first column, the second column lists the base filename for corresponding high level science products available in the MAST archive, or for the WFPC2 fields, the field name given in the LGSPA database.  After the positional information, we also list C80, the 80\% completeness limits in each filter used as the faint limit for SFH fitting (see Sect.~\ref{sfhsect}).  Lastly, we give the total number of stars passing our photometric quality cuts (after excising known clusters, see Sect.~\ref{clustersect}), and the number of these stars used for SFH fitting (brightward of C80 in both filters).  Our sample excludes five Scylla fields lacking usable imaging in F475W and/or F814W due to guide star acquisition failures, and 31 fields (22\%) with extreme differential extinction along their sightlines that hampers 
a reliable SFH fit.  We return to this point in the context of our SFH fitting technique in Sect.~\ref{matchsect}, noting that full star-by-star investigations of the dust properties towards our fields will be presented elsewhere (P.~Yanchulova Merica-Jones et al., in prep.). 

\startlongtable
\begin{deluxetable}{lcllccll}
\tabletypesize{\scriptsize}
\tablecaption{Observed Fields \label{obstab}}
\tablehead{
\colhead{Field} & \colhead{Archive Name} & \colhead{RA (J2000)} & \colhead{Dec (J2000)} & \colhead{C80 (blue)\tablenotemark{a}} & \colhead{C80 (F814W)} & \colhead{N(obs)} & \colhead{N(used)} \\ \colhead{} & \colhead{} & \colhead{$^{\circ}$} & \colhead{$^{\circ}$} & \colhead{mag} & \colhead{mag} & \colhead{} & \colhead{}}
\startdata
LMC\_2 & 15891\_LMC-9617ne-5147 & 82.707355 & -67.179013 & 27.22 & 25.03 & 8742 & 7263 \\ 
LMC\_3 & 15891\_LMC-9256ne-6744 & 82.881370 & -67.302077 & 27.49 & 25.18 & 7710 & 6466 \\ 
LMC\_4 & 15891\_LMC-3610se-7920 & 83.793703 & -69.801871 & 26.06 & 24.10 & 40954 & 29278 \\ 
LMC\_6 & 15891\_LMC-5082se-6540 & 83.068533 & -70.976740 & 27.09 & 24.94 & 27591 & 22784 \\ 
LMC\_9 & 15891\_LMC-12315nw-11221 & 73.972904 & -67.582014 & 26.61 & 24.56 & 10834 & 9177 \\ 
LMC\_13 & 15891\_LMC-4489se-10451 & 83.009413 & -70.790358 & 26.86 & 24.75 & 33716 & 26658 \\ 
LMC\_15 & 15891\_LMC-7454ne-11865 & 81.678511 & -67.707059 & 27.38 & 25.10 & 11233 & 9352 \\ 
LMC\_16 & 15891\_LMC-11456nw-12627 & 74.402606 & -67.738192 & 27.47 & 25.21 & 14831 & 12639 \\ 
LMC\_17 & 15891\_LMC-11384ne-12295 & 86.146871 & -67.327856 & 27.53 & 25.23 & 8890 & 7507 \\ 
LMC\_19 & 15891\_LMC-8680ne-12405 & 81.935129 & -67.378525 & 27.43 & 25.14 & 12768 & 10366 \\ 
LMC\_20 & 15891\_LMC-9679nw-13399 & 74.414712 & -68.498528 & 27.21 & 25.01 & 21904 & 17956 \\ 
LMC\_21 & 15891\_LMC-8532sw-13647 & 74.005938 & -70.048505 & 26.89 & 24.82 & 15890 & 13171 \\ 
LMC\_24 & 15891\_LMC-9690nw-13623 & 74.405282 & -68.498564 & 27.13 & 24.93 & 21514 & 17363 \\ 
LMC\_29 & 16235\_LMC-5812sw-7744 & 76.393030 & -70.306307 & 26.85 & 24.76 & 18468 & 13997 \\ 
LMC\_30 & 16235\_LMC-9740nw-7508 & 78.782341 & -67.177786 & 27.57 & 25.24 & 11917 & 9920 \\ 
LMC\_32 & 16235\_LMC-7623sw-22524 & 74.781195 & -70.186860 & 27.32 & 25.10 & 12788 & 10097 \\ 
LMC\_33 & 16235\_LMC-7234sw-22225 & 76.572560 & -71.203854 & 26.61 & 24.53 & 11302 & 8503 \\ 
LMC\_34 & 16235\_LMC-12057ne-22332 & 84.036835 & -66.647454 & 27.53 & 25.22 & 8790 & 7234 \\ 
LMC\_35 & 16235\_LMC-8599nw-23221 & 76.126318 & -68.157525 & 27.21 & 25.00 & 22213 & 17598 \\ 
LMC\_36 & 16235\_LMC-4763ne-26440 & 84.544744 & -69.433496 & 27.02 & 24.86 & 28132 & 21439 \\ 
LMC\_37 & 16235\_LMC-14421nw-26822 & 80.331012 & -65.756940 & 27.77 & 25.42 & 7369 & 5996 \\ 
LMC\_38 & 16235\_LMC-9173nw-28313 & 73.908891 & -69.186040 & 27.36 & 25.11 & 20342 & 16449 \\ 
LMC\_39 & 16786\_LMC-10028nw-33586 & 74.974443 & -68.065433 & 27.38 & 25.15 & 16493 & 13454 \\ 
LMC\_40 & 16235\_LMC-10728ne-8437 & 84.108754 & -67.052791 & 27.63 & 25.31 & 8581 & 7141 \\ 
LMC\_41 & 16235\_LMC-17892nw-9532 & 74.353245 & -65.586543 & 27.79 & 25.50 & 5217 & 4444 \\ 
LMC\_43 & 16786\_LMC-6222sw-15490 & 76.240886 & -70.514210 & 26.54 & 24.43 & 18504 & 13995 \\ 
LMC\_45 & 16786\_LMC-10253ne-6545 & 82.152110 & -66.950970 & 27.72 & 25.41 & 10497 & 8718 \\ 
LMC\_49 & 16786\_LMC-12311ne-5715 & 86.369148 & -67.083045 & 27.15 & 24.97 & 5963 & 4977 \\ 
LMC\_50 & 16786\_LMC-12141ne-5771 & 86.536879 & -67.190304 & 27.67 & 25.38 & 8237 & 6863 \\ 
LMC\_57 & 16786\_LMC-13556ne-15380 & 83.789680 & -66.176942 & 27.25 & 24.99 & 6312 & 5073 \\ 
LMC\_59 & 16786\_LMC-12269nw-24827 & 73.063050 & -68.024122 & 27.43 & 25.19 & 12406 & 10069 \\ 
LMC\_M1 & 14675\_LMC-13361nw-11112 & 72.675824 & -67.741186 & 25.08 & 23.36 & 4822 & 3594 \\ 
LMC\_M3 & 14675\_LMC-13557nw-35805 & 73.787232 & -67.191408 & 25.04 & 23.30 & 5629 & 4192 \\ 
LMC\_M7 & 14675\_LMC-15712nw-35293 & 74.108748 & -66.345979 & 26.63 & 24.56 & 5990 & 5033 \\ 
LMC\_M8 & 14675\_LMC-15978nw-34999 & 71.896837 & -67.041574 & 26.68 & 24.59 & 4237 & 3474 \\ 
LMC\_M9 & 14675\_LMC-16444nw-32449 & 75.379223 & -65.788319 & 25.05 & 23.31 & 3106 & 2312 \\ 
LMC\_M10 & 14675\_LMC-2723nw-30544 & 79.703327 & -69.128044 & 24.59 & 22.91 & 31709 & 19340 \\ 
LMC\_M11 & 14675\_LMC-2833ne-35147 & 81.936727 & -69.060475 & 26.22 & 24.15 & 31185 & 22940 \\ 
LMC\_M12 & 14675\_LMC-3773ne-33930 & 82.955801 & -69.012831 & 24.97 & 23.27 & 14266 & 10108 \\ 
LMC\_M13 & 14675\_LMC-4479ne-34301 & 84.298118 & -69.417817 & 24.91 & 23.19 & 15047 & 10645 \\ 
LMC\_M14 & 14675\_LMC-4853se-34056 & 82.742355 & -70.961725 & 26.44 & 24.34 & 22726 & 17870 \\ 
LMC\_M17 & 14675\_LMC-4999se-24927 & 82.564869 & -71.034105 & 25.01 & 23.23 & 13651 & 9336 \\ 
LMC\_M19 & 14675\_LMC-5441sw-1878 & 76.891800 & -70.450063 & 26.59 & 24.56 & 21929 & 17335 \\ 
LMC\_M23 & 14675\_LMC-5945ne-34377 & 85.555616 & -69.509344 & 26.43 & 24.35 & 18435 & 14770 \\ 
LMC\_M24 & 14675\_LMC-7231nw-24591 & 76.607795 & -68.505258 & 26.30 & 24.20 & 24294 & 18455 \\ 
LMC\_M25 & 14675\_LMC-7710se-32210 & 85.210976 & -71.395456 & 26.53 & 24.40 & 21294 & 16898 \\ 
LMC\_M26 & 14675\_LMC-7834ne-16086 & 81.385883 & -67.588229 & 25.03 & 23.21 & 5639 & 3800 \\ 
LMC\_M28 & 14675\_LMC-8576ne-10141 & 84.192328 & -67.728574 & 26.65 & 24.55 & 9258 & 7743 \\ 
LMC\_M29 & 14675\_LMC-9125ne-33311 & 83.461033 & -67.420937 & 26.58 & 24.47 & 8455 & 6954 \\ 
LMC\_M30 & 14675\_LMC-9166ne-31703 & 83.813224 & -67.469078 & 24.95 & 23.24 & 5358 & 3793 \\ 
LMC\_M31 & 14675\_LMC-9216nw-15936 & 75.288688 & -68.259820 & 26.55 & 24.43 & 17185 & 14014 \\ 
LMC\_M32 & 14675\_LMC-9441ne-25634 & 81.329185 & -67.137412 & 24.99 & 23.27 & 4444 & 3153 \\ 
LMC\_W1 & lmc\_u65006 & 81.025001 & -68.813331 & 25.02 & 23.82 & 17021 & 11397 \\ 
LMC\_W2 & lmc\_u2o901 & 81.070831 & -69.780555 & 22.35 & 21.51 & 20355 & 6157 \\ 
LMC\_W3 & lmc\_u4b107 & 78.258331 & -71.278892 & 25.34 & 23.93 & 7993 & 5353 \\ 
LMC\_W4 & lmc\_u65005 & 76.583335 & -70.972503 & 25.48 & 24.15 & 8507 & 6151 \\ 
LMC\_W5 & lmc\_u65003 & 86.345832 & -71.145835 & 25.23 & 23.92 & 13179 & 9168 \\ 
LMC\_W6 & lmc\_u63s01 & 75.483329 & -68.622222 & 25.24 & 23.92 & 13120 & 9219 \\ 
LMC\_W7 & lmc\_u65007 & 73.595832 & -70.032775 & 25.43 & 24.02 & 8493 & 5959 \\ 
LMC\_W8 & lmc\_u2o902 & 89.574996 & -68.347503 & 26.32 & 24.59 & 2807 & 1980 \\ 
LMC\_W10 & lmc\_u2hw04 & 83.841667 & -69.271667 & 24.48 & 23.25 & 10477 & 6146 \\ 
LMC\_W11 & lmc\_u3y501 & 83.666664 & -70.411666 & 24.63 & 23.58 & 23206 & 14186 \\ 
LMC\_W13 & lmc\_u2pu02 & 76.066665 & -66.575836 & 24.40 & 23.01 & 4870 & 2207 \\ 
LMC\_W15 & lmc\_u4zn07 & 75.295829 & -69.121109 & 24.44 & 23.31 & 15014 & 8841 \\ 
LMC\_W16 & lmc\_u4zn04 & 75.991668 & -69.551109 & 24.53 & 23.42 & 15994 & 9661 \\ 
LMC\_W17 & lmc\_u4zn02 & 79.154167 & -70.481109 & 24.92 & 23.72 & 15393 & 9971 \\ 
LMC\_W18 & lmc\_u4zn01 & 79.329170 & -70.778892 & 24.92 & 23.71 & 12885 & 8030 \\ 
LMC\_W19 & lmc\_u4zn06 & 80.062500 & -69.249443 & 23.44 & 22.57 & 26431 & 12319 \\ 
LMC\_W20 & lmc\_u4zn05 & 81.262496 & -69.798889 & 23.12 & 22.30 & 29832 & 13390 \\ 
LMC\_W21 & lmc\_u4zn03 & 81.579170 & -70.350555 & 23.88 & 22.97 & 19567 & 10397 \\ 
LMC\_W22 & lmc\_u4zn08 & 83.487503 & -70.853057 & 24.72 & 23.59 & 14876 & 9245 \\ 
LMC\_W23 & lmc\_u4b112 & 80.737503 & -69.781387 & 23.40 & 22.54 & 31970 & 14980 \\ 
LMC\_W24 & lmc\_u4b113 & 80.987503 & -69.823890 & 23.31 & 22.45 & 27548 & 13272 \\ 
LMC\_W25 & lmc\_u4b114 & 81.012496 & -69.854446 & 23.34 & 22.45 & 29143 & 14113 \\ 
LMC\_W26 & lmc\_u4b115 & 80.729164 & -69.747497 & 23.23 & 22.33 & 24128 & 11067 \\ 
LMC\_W27\tablenotemark{b} & lmc\_u4b101 & 77.675003 & -71.215835 & 25.37 & 23.92 & 7507 & 4995 \\ 
LMC\_W28\tablenotemark{b} & lmc\_u4b102 & 77.608329 & -71.192779 & 25.31 & 23.85 & 7813 & 5195 \\ 
LMC\_W29\tablenotemark{b} & lmc\_u4b103 & 77.541664 & -71.169998 & 25.27 & 23.86 & 7724 & 5060 \\ 
LMC\_W30\tablenotemark{b} & lmc\_u4b104 & 77.741668 & -71.238609 & 25.28 & 23.82 & 7571 & 4903 \\ 
LMC\_W31\tablenotemark{b} & lmc\_u4b105 & 77.808334 & -71.261390 & 25.45 & 24.02 & 7993 & 5505 \\ 
LMC\_W32\tablenotemark{b} & lmc\_u4b106 & 77.879165 & -71.284164 & 25.44 & 23.99 & 7723 & 5212 \\ 
LMC\_W33\tablenotemark{b} & lmc\_u4b108 & 78.054168 & -71.210556 & 25.40 & 23.92 & 8148 & 5473 \\ 
LMC\_W34\tablenotemark{b} & lmc\_u4b109 & 77.987503 & -71.187774 & 25.36 & 23.94 & 7820 & 5241 \\ 
LMC\_W35\tablenotemark{b} & lmc\_u4b110 & 77.916664 & -71.163887 & 25.47 & 24.04 & 8437 & 5843 \\ 
LMC\_W36 & lmc\_u63s03 & 75.579170 & -68.570274 & 25.23 & 23.90 & 12577 & 8784 \\ 
LMC\_W37\tablenotemark{b} & lmc\_u63s04 & 75.433334 & -68.648056 & 25.19 & 23.84 & 13063 & 9054 \\ 
LMC\_W38\tablenotemark{b} & lmc\_u63s05 & 75.383331 & -68.673889 & 25.21 & 23.93 & 13698 & 9776 \\ 
LMC\_W39\tablenotemark{b} & lmc\_u63s06 & 75.333335 & -68.699722 & 25.16 & 23.83 & 8725 & 5916 \\ 
LMC\_W40 & lmc\_u65004 & 74.324996 & -69.457221 & 25.16 & 23.83 & 11202 & 7650 \\ 
LMC\_W41 & lmc\_u65008 & 76.658332 & -69.340553 & 24.62 & 23.59 & 14573 & 8921 \\ 
LMC\_W42 & lmc\_u65002 & 76.454170 & -69.723609 & 24.97 & 23.80 & 18617 & 12431 \\ 
LMC\_W44 & lmc\_u6e806 & 84.004165 & -69.291664 & 24.42 & 23.06 & 11044 & 6227 \\ 
LMC\_W45 & lmc\_u6m706 & 84.004165 & -69.291114 & 24.51 & 23.10 & 11479 & 6522 \\ 
LMC\_W46 & lmc\_u6mi30 & 83.887496 & -69.355834 & 24.06 & 22.80 & 11476 & 6033 \\ 
LMC\_W48 & lmc\_u4ax15 & 77.879165 & -65.479446 & 26.58 & 25.04 & 5577 & 3844 \\ 
LMC\_W49 & lmc\_u4ax20 & 77.579170 & -65.446662 & 26.22 & 24.53 & 4701 & 3398 \\ 
LMC\_W51 & lmc\_u4ax27 & 88.966667 & -73.916946 & 26.39 & 24.60 & 2318 & 1591 \\ 
LMC\_W52 & lmc\_u4ax84 & 93.933334 & -69.806388 & 25.59 & 23.97 & 2089 & 1364 \\ 
LMC\_W53 & lmc\_u4ax86 & 93.779167 & -69.815277 & 26.76 & 25.12 & 2874 & 1974 \\ 
LMC\_W54 & lmc\_u4ax93 & 93.220832 & -69.816665 & 26.25 & 24.44 & 2425 & 1658 \\ 
LMC\_W55 & lmc\_u4ax73 & 82.212501 & -73.507499 & 25.61 & 24.14 & 2975 & 2022 \\ 
LMC\_W57 & lmc\_u4ax82 & 81.829170 & -73.547500 & 26.26 & 24.50 & 3524 & 2450 \\ 
LMC\_W60 & lmc\_u4ax10 & 75.420829 & -66.002777 & 26.07 & 24.37 & 4535 & 3183 \\ 
LMC\_W61 & lmc\_u4ax29 & 75.849998 & -66.342498 & 25.38 & 24.05 & 4844 & 3437 \\ 
LMC\_W62 & lmc\_u4ax31 & 76.258331 & -66.421112 & 26.40 & 24.91 & 6924 & 4883 \\ 
LMC\_W63 & lmc\_u4ax38 & 75.758331 & -66.401664 & 25.94 & 24.40 & 5486 & 3870 \\ 
LMC\_W64 & lmc\_u4ax95 & 92.633331 & -69.033332 & 25.50 & 23.98 & 2366 & 1560 \\ 
LMC\_OF1 & 11700\_E121-20-25164se-17241& 95.946162 & -64.664004 & 26.93 & 24.60\tablenotemark{c} & 944 & 632 \\
LMC\_OF2 & 11700\_LMC-23472ne-16872 & 88.271113 & -64.089937 & 26.91 & 24.54\tablenotemark{c} & 3122 & 1942 \\ 
LMC\_OF3 & 11700\_LMC-21118ne-18087 & 84.87939 & -64.153484 & 26.62 & 24.28\tablenotemark{c} & 3997 & 2375 \\
LMC\_OF4 & 14688\_LMC-18355nw-16051 & 76.447066 & -65.016922 & 25.94 & 23.95 & 4799 & 3413 \\
\enddata
\tablenotetext{a}{For all Scylla and METAL fields (LMC\_2-LMC\_M32), the blue filter is WFC3/UVIS F475W, while for all WFPC2 LGSPA fields (LMC\_W1-LMC\_W64), the blue filter is WFPC2 F555W.  For outer fields LMC\_OF1-LMC\_OF3, the blue filter is WFC3/UVIS F606W, and for LMC\_OF4, the blue filter is ACS/WFC F435W.}
\tablenotetext{b}{For SFH analysis, SFH results from contiguous fields LMC\_W27-W32 were combined and designated LMC\_WC1, SFH results from LMC\_W33-W35 were combined into LMC\_WC2, and LMC\_W37-W39 were combined into LMC\_WC3.}
\tablenotetext{c}{For imaging of fields LMC\_OF1-LMC\_OF3, the redder of the two filters is WFC3/IR F160W.}
\end{deluxetable}

\begin{figure}
\gridline{\fig{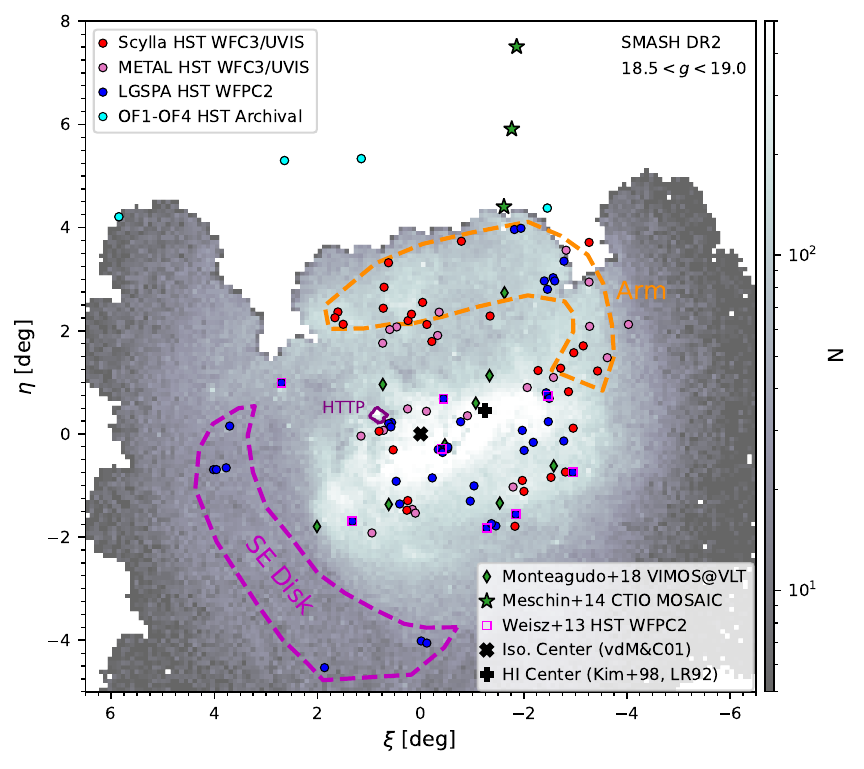}{0.99\textwidth}{}}
\caption{Locations of our fields overlaid on a stellar density map from SMASH DR2 (see text for details), in tangent plane coordinates assuming an LMC center of (RA,Dec) = (82.25$^{\circ}$,$-$69.5$^{\circ}$) based on near-infrared stellar isodensity contours in the outer disk \citep{vdmcioni}.  The dynamical center of the HI disk \citep{luks92,kim98} used for our analysis of radial age gradients in Sect.~\ref{obsgradsect} is shown as a cross.  North is up, east is to the left, and the SMC is located to the lower right (well outside the plot).  The fields we analyze are shown as filled circles, color-coded by their source as indicated in the upper left.  Ground-based imaging fields analyzed by \citet{monteagudo} and \citet{meschin} are shown as green diamonds and stars, and the subset of archival \citet{lgspa} WFPC2 LGSPA fields analyzed by \citet{weisz13} are overplotted as pink squares.  Dashed lines indicate the subsets of fields in the southeast disk (magenta) and the spiral arm (orange) discussed in Sect.~\ref{obsgradsect}, and the Hubble Tarantula Treasury Project (HTTP) footprint \citep{httpcat} is shown in purple and labeled.}
\label{map_fig}
\end{figure}

\subsection{Photometry \label{photsect}}

For Scylla imaging, preprocessing and PSF photometry is described in detail in Murray at al.~(2024), so here we briefly summarize the salient details, bearing in mind that we also include archival imaging from additional instrument/filter combinations not included in the Scylla program.  
For each target field, we downloaded individual \texttt{flc} images, which have undergone standard 
pipeline processing, including corrections for charge transfer inefficiency.  We used the \texttt{astrodrizzle} task within \texttt{drizzlepac} \citep{drizzlepac} to construct a deep, distortion-corrected, drizzled reference image for each filter.  We performed additional preprocessing and point-spread function fitting (PSF) photometry using \texttt{Dolphot} \citep{dolphot,dolphot2}, which uses model PSFs customized to each filter of each instrument onboard \textit{HST}.  Briefly, \texttt{Dolphot} preprocessing for each image includes applying a pixel area map, masking bad pixels, splitting into individual chips, and calculating a sky frame.  Details of how \texttt{Dolphot} performs PSF photometry can be controlled using various input parameters, and we use the parameters recommended by \citet{williams_phat,phatter}, which are based on testing over an extensive range of crowding conditions.

The photometric catalogs output by \texttt{Dolphot} are calibrated to the Vegamag photometric system, and have several diagnostic parameters which are useful to eliminate poorly measured, non-stellar and/or spurious detections.  These parameters are provided for each star as global values, per-filter values, and per-image values.  We impose the following cuts on our photometric catalogs using the per-filter diagnostic parameters to retain only well-measured stellar sources:

\begin{enumerate}
    \item Signal-to-noise ratio SNR $\geq$5 in each filter.
    \item Photometric quality flag $=$ 0 or 2 in each filter.  This is a bitwise flag that we use to eliminate stars that have too many bad or saturated pixels to obtain a meaningful PSF fit.
    \item The \texttt{sharp} parameter measures whether a source is more or less centrally concentrated than the model PSF, with positive values representing more concentrated sources (i.e., cosmic rays) and negative values representing more extended sources (i.e., blends or background galaxies).  We retain sources with $|$\texttt{sharp}$| \leq$ 0.25.
    \item The \texttt{crowd} parameter measures how much brighter (in magnitudes) a star would have been if its neighbors had not been fit with a PSF and subtracted, and is particularly useful for eliminating artifacts from diffraction spikes.  We require \texttt{crowd} $\leq$ 0.25.
\end{enumerate}

The procedure of calculating the best-fit SFH for any particular observed CMD relies on our ability to apply the noise properties of our data to synthetic photometry.  Therefore, we perform artificial star tests to quantify incompleteness, photometric error and bias (offset) as a function of color, magnitude, and spatial location.  For each field, we insert $>$10$^5$ artificial stars, noting that artificial stars are photometered individually, so there is no practical restriction on their proximity to each other.  The artificial stars are assigned a flat spatial distribution, and input magnitudes generated using the Bayesian Extinction and Stellar Tool\footnote{\url{https://github.com/BEAST-Fitting/beast}} \citep{beast} assuming PARSEC evolutionary models \citep{parsec} and a \citet{kroupa} initial mass function.  To ensure adequate CMD coverage, we use 
broad, flat input distributions in heliocentric distance 39 $\leq$ (D$_{\sun}$/kpc) $\leq$ 69, 7.0 $\leq$ (Log Age/yr) $\leq$ 10.13, metallicity $-$2.1 $\leq$ [M/H] $\leq$ $-$0.3 and extinction 0$\leq$$A_{V}$$\leq$1.  For each field, an additional subset of the artificial stars are drawn from another flat $A_{V}$ distribution with a higher maximum $A_{V}$, but given the modest ($A_{V}$$<$1) extinction towards the fields we analyze (see Fig.~\ref{cmd_fig} and the discussion in Sect.~\ref{distcompsect} below), this subset comprises $<$20\% of the input artificial stars in all cases.  Artificial stars are considered recovered if they pass all of the aforementioned quality cuts applied to the observations.

All stars passing our quality cuts are shown in the CMDs in the left (for Scylla and METAL imaging) and right (for WFPC2 LGSPA imaging) panels of Fig.~\ref{cmd_fig}, and the CMDs of the four outer fields (with different instrument and filter combinations) are shown in Fig.~\ref{cmd_of_fig}.  We have overplotted PARSEC isochrones \citep{parsec} to indicate the ranges of stellar mass and age accessible in our photometry in various stellar evolutionary phases.  Our Scylla imaging extends faintward of the oMSTO down to $\sim$0.5M$_{\odot}$, as deep or deeper than previous HST campaigns designed to measure lifetime SFHs in the LMC and SMC \citep[e.g.][]{weisz13,cignoni}.  However, due to observational constraints, the saturation limit of our imaging also typically occurs faintward of recent ground-based surveys used to map SFH trends in the LMC \citep[e.g.][]{smashdr2,mazzi21}.  Shallower ground-based imaging may therefore be optimal for resolving the recent ($\lesssim$500 Myr; see Fig.~\ref{cmd_fig}) star formation history of the LMC, although the use of logarithmic time bins in our SFH fitting  mitigates the impact of our saturation limits on \textit{cumulative lifetime} SFH metrics that we use (see Sect.~\ref{sfhsect}).

\begin{figure}
\gridline{\fig{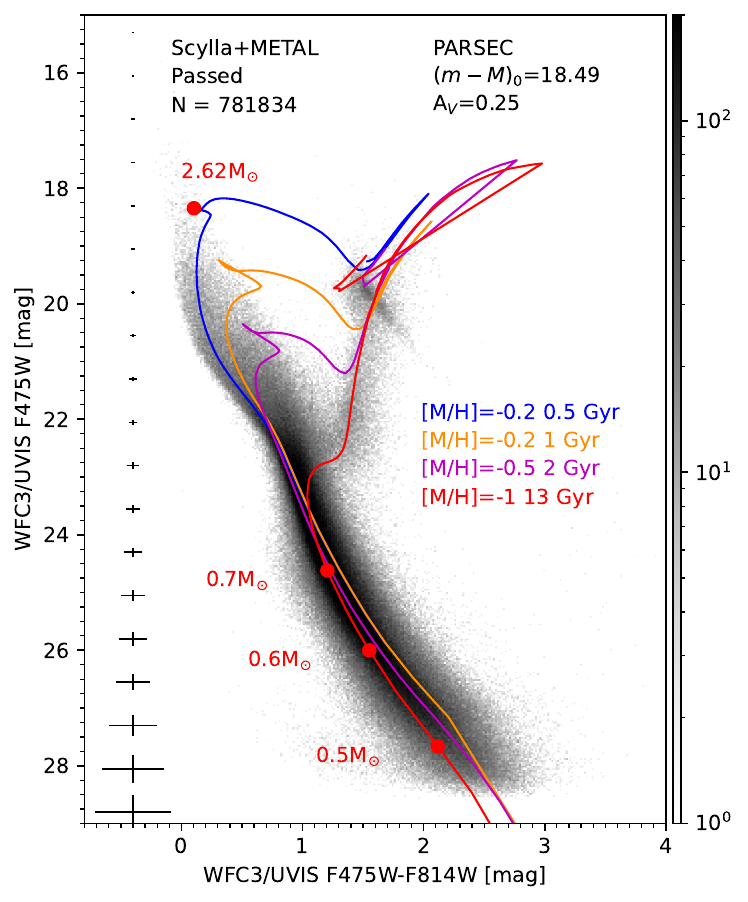}{0.5\textwidth}{}
          \fig{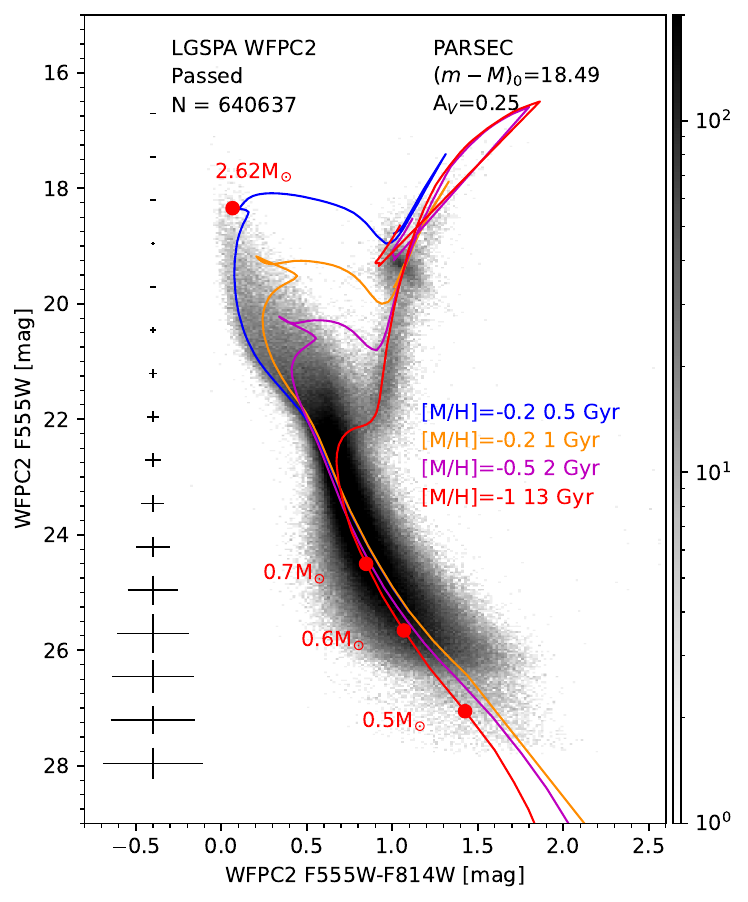}{0.5\textwidth}{}}
\caption{Stacked CMD of all stars passing our photometric quality cuts, from Scylla fields and archival METAL fields, both of which were observed with WFC3/UVIS F475W and F814W filters (left) and archival LGSPA photometry observed with WFPC2 F555W and F814W (right).  No correction for field-to-field differences in extinction or distance has been applied.  Median photometric errors in magnitude bins, assessed using artificial star tests, are shown along the left-hand side of each CMD.  PARSEC isochrones have been overplotted, indicating the main sequence stellar mass range typical of our photometric catalogs.} 
\label{cmd_fig}
\end{figure}

\begin{figure}
\gridline{\fig{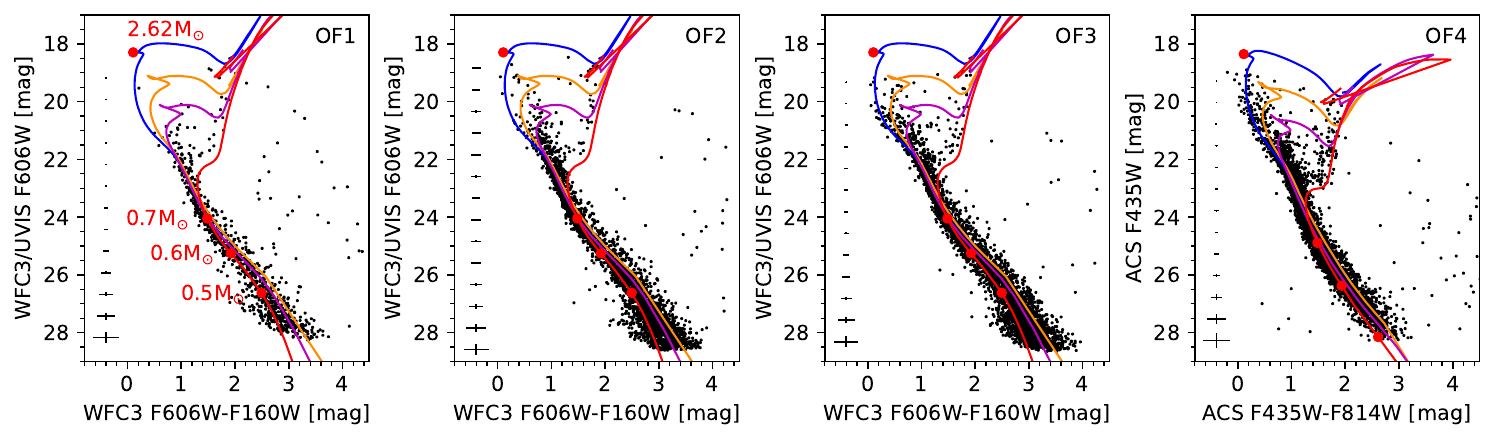}{0.99\textwidth}{}}
\caption{CMDs of the four outer fields LMC\_OF1-LMC\_OF4 used to extend our radial coverage of the LMC, with isochrones overplotted as in Fig.~\ref{cmd_fig}.}
\label{cmd_of_fig}
\end{figure}

\begin{figure}
\gridline{\fig{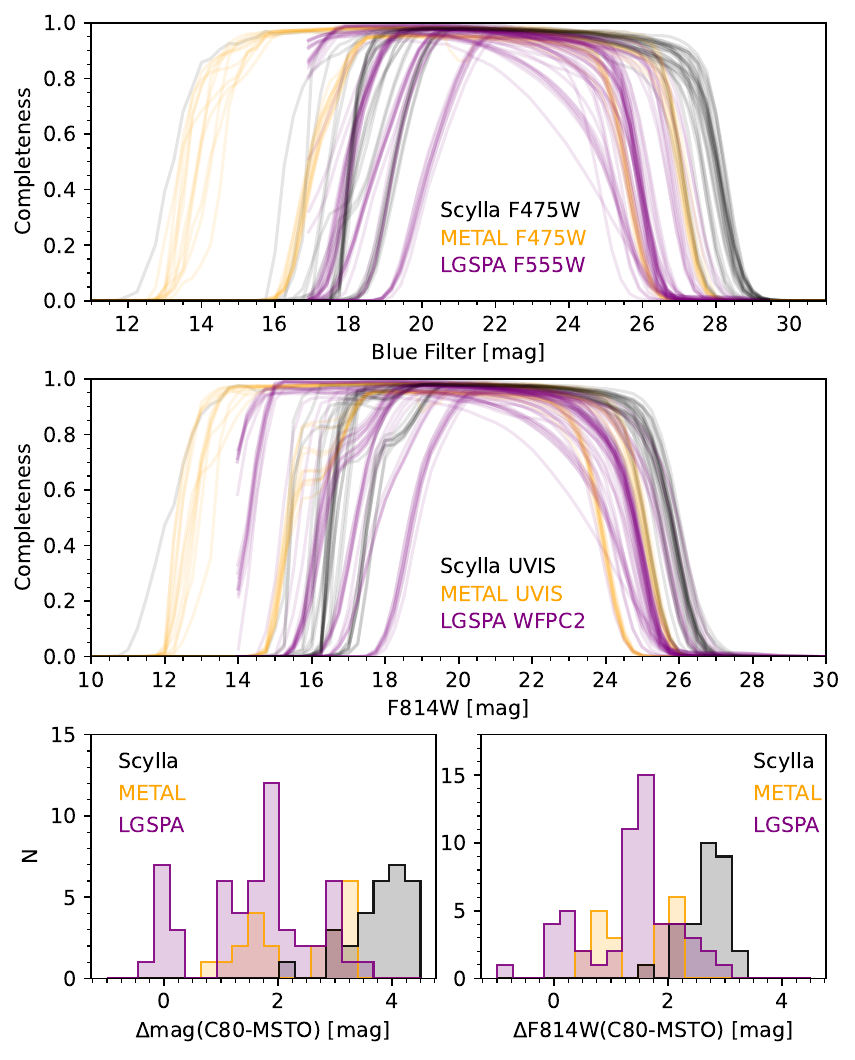}{0.75\textwidth}{}}
\caption{Photometric completeness versus magnitude in the bluer of the two filters (top) and F814W (middle).  Completeness curves are shown for each individual field, color-coded to indicate whether that field was observed as part of our Scylla program, or incorporated from archival METAL or LGSPA imaging.  The panels in the bottom row illustrate the distribution of the faint magnitude limits imposed for SFH fitting (C80, corresponding to the 80\% faint completeness limit) compared to the metal-poor oMSTO (13 Gyr, [M/H]=$-$1) from a PARSEC isochrone, calculated using per-field distance and extinction values.}
\label{ast_fig}
\end{figure}

\section{Star Formation History Fitting \label{sfhsect}}

For each target field, we determine the best-fit SFH using the well-established technique of synthesizing the observed CMD using linear combinations of single stellar populations (see \citealt{tolstoy09} for a review).  In the present case, due to the proximity of the LMC, there are several issues affecting both the selection of the input stellar sample and our SFH fitting methodology that warrant further detail.  

\subsection{Input Catalogs \label{clustersect}}

Before performing SFH fits, 
we make three modifications to our input photometric catalogs after applying the photometric quality cuts described above in Sect.~\ref{photsect}.  First, known star clusters \citep[][and references therein]{bica08}\footnote{We also verified that there were no recently-discovered clusters present in our fields from LMC star cluster catalogs updating the \citet{bica08} compilation \citep[e.g.][]{glatt10,nayak,sitek16,palma16,narloch22}}
were excised to avoid biasing our SFH results for the LMC field population.  Specifically, for each cluster, we removed all stars within a circular spatial region centered on the cluster, with a radius equal to the catalog semi-major axis.  The excised area due to overlap with known clusters was $<$40\% of the full field of view in all cases, $<$20\% in all but three cases, and affected our total sampled area in the LMC by $<$4\%.  The locations of these clusters are listed with the radius of the excluded area and the affected fields in Appendix \ref{clusterapp}.

The second restriction applied to our input catalogs is that for SFH fitting, we retain only stars brightward of the 80\% faint completeness limit in both filters, denoted as C80 (assessed using the artificial star tests).  
The reason for this choice of faint magnitude limit is shown in the upper two panels of Fig.~\ref{ast_fig}, where we plot individual completeness curves for all of our fields, illustrating that photometric completeness drops off quite steeply with magnitude close to the detection limit in each field.  In the bottom row of Fig.~\ref{ast_fig}, we plot histograms of the difference between C80 and the magnitude of the oMSTO, assessed using a 13 Gyr PARSEC isochrone with [M/H]=$-$1 and distances and extinctions from \citet{yumimap} (in excellent agreement with our independently derived values, see Sect.~\ref{distcompsect}), illustrating that the vast majority ($>$85\%) of our fields are at least 80\% complete to $>$1 mag faintward of the oMSTO in both filters\footnote{When recovering input SFHs and chemical enrichment histories from synthetic CMDs, \citet{piatti85comp} found that input SFHs and age-metallicity relations are recovered to within their uncertainties down to a completeness limit of $\sim$85\%, nearly identical to the value we employ, albeit using a different SFH code.}.  We verified that removal of the shallowest fields (i.e., those with C80 $\lesssim$0.5 mag faintward of the oMSTO in either filter in the bottom row of Fig.~\ref{ast_fig}) would not impact our results since these are all archival WFPC2 fields in the inner bar showing SFH results consistent with other inner bar fields, including field-to-field scatter (discussed in Sect.~\ref{obsgradsect}).  Because completeness is, in a strict sense, dependent on the complex interplay of CMD location, crowding and exposure time, we also calculated the completeness at various CMD locations for each of our fields, finding completeness fractions at the red clump of $>$90\% in all cases and completeness fractions at the oMSTO of $>$90\% for $>$84\% of our fields (with the remainder consisting of the aforementioned WFPC2 fields).

Lastly, in cases of three or more contiguous fields, SFH results were statistically combined to mitigate biased spatial sampling of the LMC in our data, although this only occurred in three cases\footnote{Specifically, LMC\_W27-W32 were combined, LMC\_W33-W35 were combined, and LMC\_W37-W39 were combined.} and does not affect our results beyond their uncertainties.  Similarly, we excluded the shallower of a pair of fields with near-total spatial overlap (LMC\_20 and LMC\_24; we use this pair as an internal consistency check in Appendix~\ref{amrsect}).   

\subsection{Methodology \label{matchsect}}

We calculate SFHs by fitting to the observed CMDs of each field using the software code \texttt{MATCH} \citep{match}.  For SFH fitting, the CMD is divided into bins with a width of 0.05 mag in color and 0.1 mag in magnitudes, and \texttt{MATCH} convolves model-predicted synthetic photometry with observational effects ascertained from the artificial star tests to calculate the best-fit star formation rate and mean metallicity in each time bin, seeking the best fit using a Poisson maximum likelihood statistic.  Random (statistical) uncertainties are calculated using a Hybrid Monte Carlo approach detailed in \citet{dolphin_randerr}, and systematic uncertainties are calculated as in \citet{dolphin_syserr}, performing 50 Monte Carlo iterations where the CMD is re-fit after perturbing $T_{\rm eff}$ and $M_{\rm bol}$ by randomly drawn Gaussian deviates.  We place no prior restrictions on the age-metallicity relations (AMRs) in our SFH fits, and compare the AMR recovered by \texttt{MATCH} to independent metallicity measurements in Appendix \ref{amrsect} as a validation of our SFH fitting procedure.

For SFH fitting, we assume the following values: A \citet{kroupa} initial mass function, a binary fraction of 0.35 with a flat mass-ratio distribution, a time resolution of $\Delta$Log Age/yr = 0.05 dex from Log Age/yr = 7.2 (the youngest age available across all three sets of evolutionary models we employ) to Log Age/yr = 10.15, and a metallicity spread in each time bin of $\Delta$[M/H] = 0.15 dex over an allowed metallicity range of $-$2.0$\leq$[M/H]$\leq$+0.2.  These values were chosen to be consistent with the vast database of Local Volume dwarf galaxy SFHs calculated using \texttt{MATCH} \citep[e.g.,][]{weisz11,weisz14,geha15,mcquinn18,savino23} as well as our companion study of the SMC (Cohen et al.~2024, submitted).  However, the shape of the SFH (which we use as our quantitative metric for measuring stellar age gradients; see Sect.~\ref{csfhsect}) is insensitive to variations in the chosen input values for these quantities \citep{monelli10b,hidalgo11,cignoni12,cole14}.  For each field, we perform three SFH fits to the observed photometric catalog, each assuming a different set of stellar evolutionary models, allowing us to directly assess the impact of model choice on our SFH results.  Specifically, we perform fits assuming isochrones from PARSEC \citep{parsec}, MIST \citep{mist1,mist2}, and the updated version of the BaSTI isochrones \citep{basti}, which we denote BaSTI18 to avoid confusion with earlier BaSTI models \citep{basti_old}.  

\subsubsection{Distance and Extinction \label{distcompsect}}

For each stellar model set, we let \texttt{MATCH} find the best-fit SFH solution within a grid of distance modulus $(m-M)_{0}$, foreground extinction $A_{V}$, and internal differential extinction $\delta A_{V}$, with a grid spacing of 0.05 mag in all three of these quantities\footnote{This is essentially an expansion of the grid-search technique applied to M31 by \citet{lewis15} and M33 by \citet{lazzarini} to include $(m-M)_{0}$ in addition to foreground extinction $A_{V}$ and internal extinction $\delta A_{V}$.}.  Towards our LMC fields, independent constraints on these values are available in many cases \citep[e.g.,][]{inno,yumimap,cusano21,skowron,bell22}.  However, with an eye towards future self-consistent analyses of other targets (including the SMC) lacking such high-fidelity information, we choose to let per-field values of distance, foreground extinction and internal differential extinction float in our SFH solutions, instead exploiting the detailed information available for our LMC sightlines to assess our ability to recover these values.

\begin{figure}
\gridline{\fig{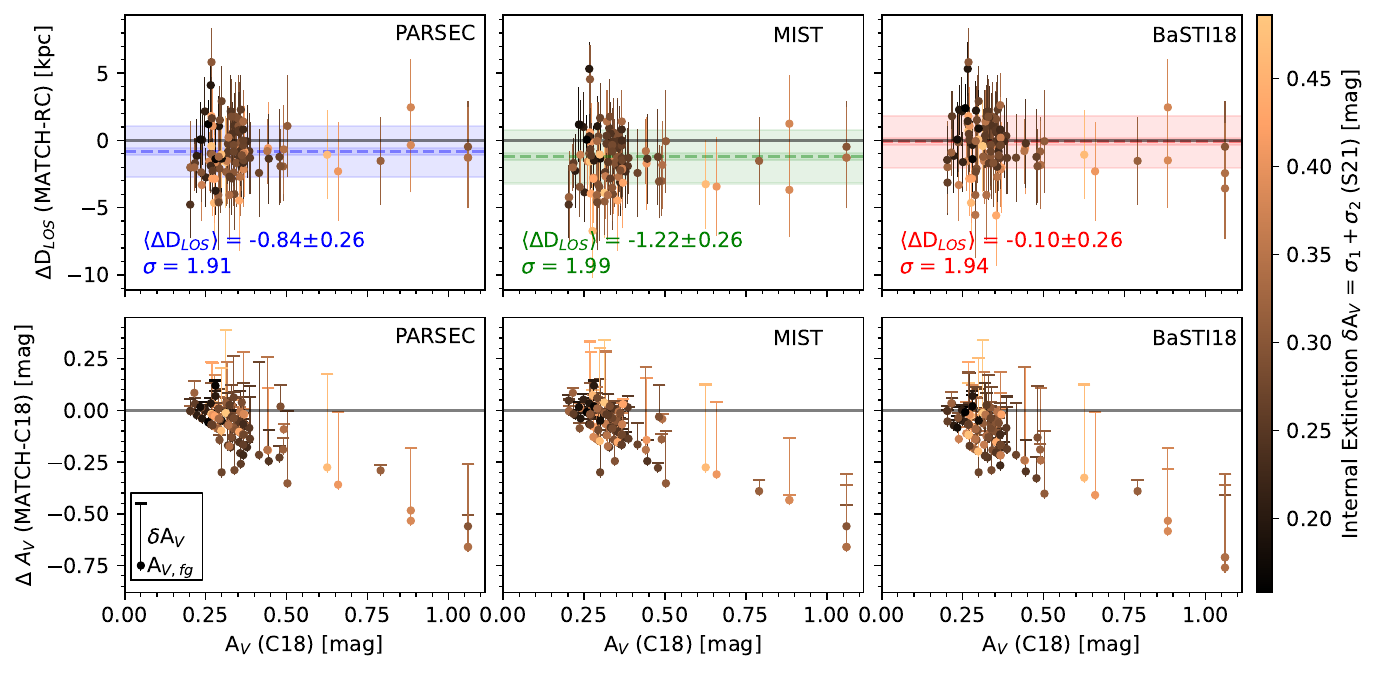}{0.99\textwidth}{}}
\caption{\textbf{Top Row:} Comparison between the best-fitting distance from our \texttt{MATCH} SFH fits versus the distance (and uncertainty) measured independently from the red clump in SMASH photometry as in \citet{yumimap}.  Differences are plotted as a function of median line-of-sight extinction to the LMC red clump in each field from \citet{yumimap}, and points are color-coded by the line-of-sight extinction internal to the LMC = $(\sigma_{\rm 1}+\sigma_{\rm 2})$ from \citet{skowron}.  Distance offsets, shown for three different evolutionary models from left to right, are smaller than the red clump distance uncertainties, and uncorrelated with total or differential extinction.  The dashed line and dark colored shading in each panel represents the (inverse square-weighted) mean and its standard deviation, and the lighter shading indicates the standard deviation of the individual offsets.  \textbf{Bottom Row:} Difference between the best-fit extinction from \texttt{MATCH} and the median extinction measured using LMC red clump stars as in \citet{yumimap}.  Because \texttt{MATCH} fits for both the foreground Galactic extinction $A_{V,fg}$ and additional internal differential extinction $\delta$A$_{V}$, the vertical location of the filled circles corresponds to the best-fit foreground Galactic extinction $A_{V,fg}$ and the vertical lines represent the best-fitting additional differential extinction $\delta$$A_{V}$ internal to the LMC for each field from our SFH fits, illustrated in the box in the lower left panel.  When considering both foreground and internal extinction resulting from our SFH fits with \texttt{MATCH}, our results are in good agreement with the independent red clump-based measurements of the \textit{median} extinction, with outliers at the $\lesssim$0.3 mag level (see text for discussion).}
\label{comp_dist_av_fig}
\end{figure}

In the top row of Fig.~\ref{comp_dist_av_fig}, we compare the best-fit distance returned by \texttt{MATCH} for each of our fields against independent values from \citet{yumimap}, which were calculated using the red clump in SMASH photometry as a tracer.  For direct comparison, these values were recalculated by one of us (Y.~Choi) at a finer spatial resolution of $\sim$40 pc, approximately the size of a target (i.e., WFC3/UVIS or WFPC2) field of view.  

We find remarkable agreement between our \texttt{MATCH} SFH-based distances and the independent red clump-based distances, 
with mean offsets $<$1.5 kpc ($<$3\%) for all three sets of assumed stellar evolutionary models (shown as dashed horizontal lines in the top row of Fig.~\ref{comp_dist_av_fig}).  The scatter of the individual field-to-field distance offsets, represented by the (weighted) standard deviation (light shading in the top row of Fig.~\ref{comp_dist_av_fig}), is $\sigma$$<$2 kpc in all cases, similar to the uncertainties of the red clump distances themselves, which have a median of 2.6 kpc across our fields.  The model-to-model differences in the distance offsets are quite modest, with a difference in the mean offsets of $<$1.2 kpc ($<$3\%) across the three model sets we include, smaller than the standard deviation of field-to-field distance offsets when assuming any one particular evolutionary model.  Accordingly, the best-fit distances agree well across different sets of evolutionary models, with model-to-model differences of $\leq$0.05 mag (one grid step in our fitting) in distance for 81\% of our fields (and $\leq$0.05 in foreground extinction $A_{V,fg}$ for 73\% of our fields; see below).  We also verified that the distance offsets are not correlated with distance, faint magnitude limits (i.e., C80), or foreground or differential extinction.  As an additional check, if we assume per-field distances based on a fit of an inclined disk to the distribution of red clump stars \citep{yumimap} rather than letting per-field distances float as a free parameter, our results are unaffected to within their uncertainties.

Regarding extinction, \texttt{MATCH} implements a fixed per-field value of the Galactic foreground extinction $A_{V}$ and, optionally, a per-field value of additional internal differential extinction $\delta$$A_{V}$ with a line-of-sight distribution applied using straightforward analytical functions.  When differential extinction within a field becomes sufficiently dramatic, 
this analytical treatment of foreground and differential extinction breaks down, indicated by statistically poor SFH fits and, often, unphysical trends in chemical evolution history.   
We have excluded 31 fields (22\%) from our sample (also excluded from Figs.~\ref{map_fig}-\ref{comp_dist_av_fig} and Tables \ref{obstab}-\ref{tautab}) with SFH fits that are drastically affected by these issues, noting that they correlate tightly with independent measurements of dust properties: All of our rejected fields have either $(\sigma_{\rm 1}+\sigma_{\rm 2})$$\geq$0.5\footnote{\citet{skowron} quantify differential extinction internal to the LMC by fitting half-Gaussians to the distribution of the red clump along the reddening vector, with standard deviations $\sigma_{\rm 1}$ and $\sigma_{\rm 2}$ on the near and far side of the LMC respectively, and we have translated these values to the $V$ band using coefficients from \citet{sf11}.} from \citet{skowron} and/or $\Sigma_{\rm dust}$$\geq$0.13 M$_{\odot}$/pc$^{2}$ from the maps of \citet{dusttemp}.  Unsurprisingly, these highly extincted fields are clustered almost exclusively (84\% within $<$30$\arcmin$) around the two largest HII regions in the LMC, the Tarantula Nebula and N11 \citep{henize56}, so their exclusion comes at a negligible cost to our spatial coverage.  Excluding these highly extincted fields, \texttt{MATCH} is reasonably adept at disentangling foreground Galactic extinction $A_{V,fg}$ from internal differential extinction $\delta$$A_{V}$.  This is illustrated in the bottom row of Fig.~\ref{comp_dist_av_fig}, where we compare the \textit{median} extinction calculated as in \citet{yumimap} (transforming their values to the $V$ band using coefficients from \citealt{sf11}) against the foreground and additional differential extinction output by \texttt{MATCH}.  
The location of each point on the vertical axis corresponds to the value that would be obtained considering only the foreground extinction $A_{V,fg}$ from \texttt{MATCH}, and the vertical line represents the best-fit additional differential extinction $\delta$$A_{V}$ output by \texttt{MATCH}.  While there are a few cases of discrepancies at the $<$0.3 mag level, this is an expected consequence of assumptions on star-dust geometry and radial trends in intrinsic LMC red clump properties (e.g., fig.~12 of \citealt{yumimap}), preferentially affecting the inner part of the LMC disk where the majority of our fields lie.  Furthermore, we recover median values of the foreground Galactic extinction 0.2$\leq$$A_{V,fg}$$\leq$0.3 
for all three evolutionary model sets, 
in good agreement with values inferred from comparisons with dust emission maps in the outer regions of the LMC where they remain reliable \citep[e.g.][]{sf11}.    

\subsubsection{Foreground and Background Contamination}

When finding the combination of ages and metallicities that produces the best SFH fit, \texttt{MATCH} also allows the inclusion of additional foreground/background populations with a fixed color-magnitude distribution but an amplitude that is allowed to vary as part of the SFH solution.  We produced the predicted foreground Milky Way color-magnitude distribution for each field by querying the TRILEGAL galaxy model \citep{trilegal}, using input coordinates corresponding to the center of the field, the \citet{skowron} foreground extinction value (i.e., their median$-$$\sigma_{1}$) and an area of 1 deg$^2$ to produce sufficient number statistics.  
The best-fit foreground Milky Way projected stellar densities returned by \texttt{MATCH} $\Sigma_{\rm MATCH}$ were in good agreement with the TRILEGAL-predicted values $\Sigma_{\rm TRILEGAL}$ given their uncertainties: Across our target fields, the ratio $\Sigma_{\rm MATCH} / \Sigma_{\rm TRILEGAL}$ has a median and standard deviation of (1.1, 0.9, 1.2) $\pm$ (3.0, 2.9, 3.0) corresponding to assumed isochrone libraries of (PARSEC, MIST, BaSTI18).  We found that the impact of foreground contaminants on the  
SFH fits was ultimately negligible, 
and the SFH results were unaffected beyond their uncertainties by the inclusion or exclusion of any foreground Milky Way component at all, in accord with the above values and uncertainties of $\Sigma_{\rm MATCH} / \Sigma_{\rm TRILEGAL}$.  Bearing in mind that Galaxy models may not be excellent predictors of the potentially complex stellar populations along sightlines towards the Magellanic clouds \citep[e.g.][]{nidever19}, 
we made the conservative choice to include the model-predicted foreground component with a floating amplitude in the SFH fit primarily to capture the contribution of its uncertainty to the formal errors on the SFH solutions, although this was ultimately negligible.

Our stellar sample can also be contaminated by unresolved background galaxies that appear sufficiently compact to be well-fit by a stellar PSF.   
 To measure contamination by background galaxies, we searched the HST archive for sparse, high Galactic latitude fields observed with the same instrument+filter combination as our Scylla and METAL fields (WFC3/UVIS F475W+F814W) and similar exposure times.  We located four fields with 35.59$\leq$$|b|$$\leq$53.44 (GO-12233, PI: Courbin), performing PSF photometry and quality cuts identically as for our target fields\footnote{Conservatively, we excluded sources with 5$\arcsec$ of the target lensing QSO in each field \citep{courbin10,courbin12}, although this choice did not impact the results.  In any case, any included lensed sources will bias our result pessimistically.}.  We assumed the TRILEGAL Galaxy model is representative of the foreground Galactic population, and compared the color-magnitude distribution of background galaxies (i.e., all sources passing our photometric quality cuts that are unaccounted for by the TRILEGAL model) to the observed color-magnitude distribution in each target field, correcting for field area and incompleteness (for details of this procedure, see \citealt{cohen_amiga,cohen_sombrero}).  We found that for all of our WFC3/UVIS fields, the predicted number of background galaxies passing our photometric quality cuts and inadvertently included in our SFH analysis is statistically equivalent to zero, and the predicted fractional contribution of background galaxies to the stellar sample used for SFH fitting is $<$0.2\%.  

\subsection{Cumulative Star Formation History Metrics \label{csfhsect}}

\begin{figure}
\gridline{\fig{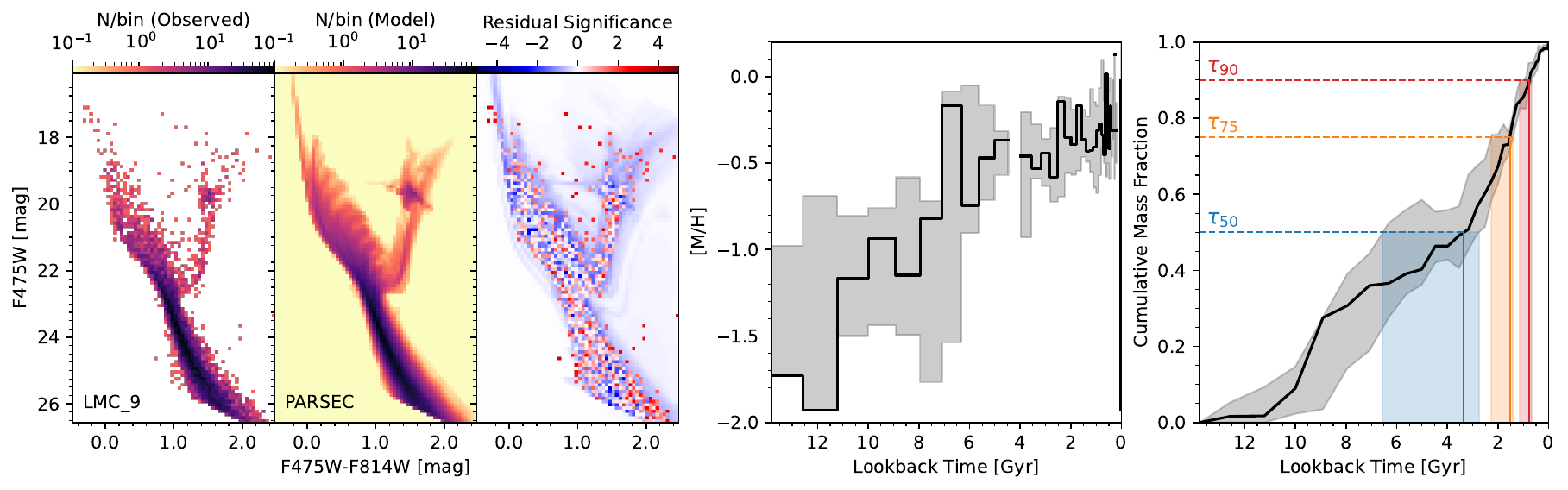}{0.99\textwidth}{}}
\vspace{-1cm}
\gridline{\fig{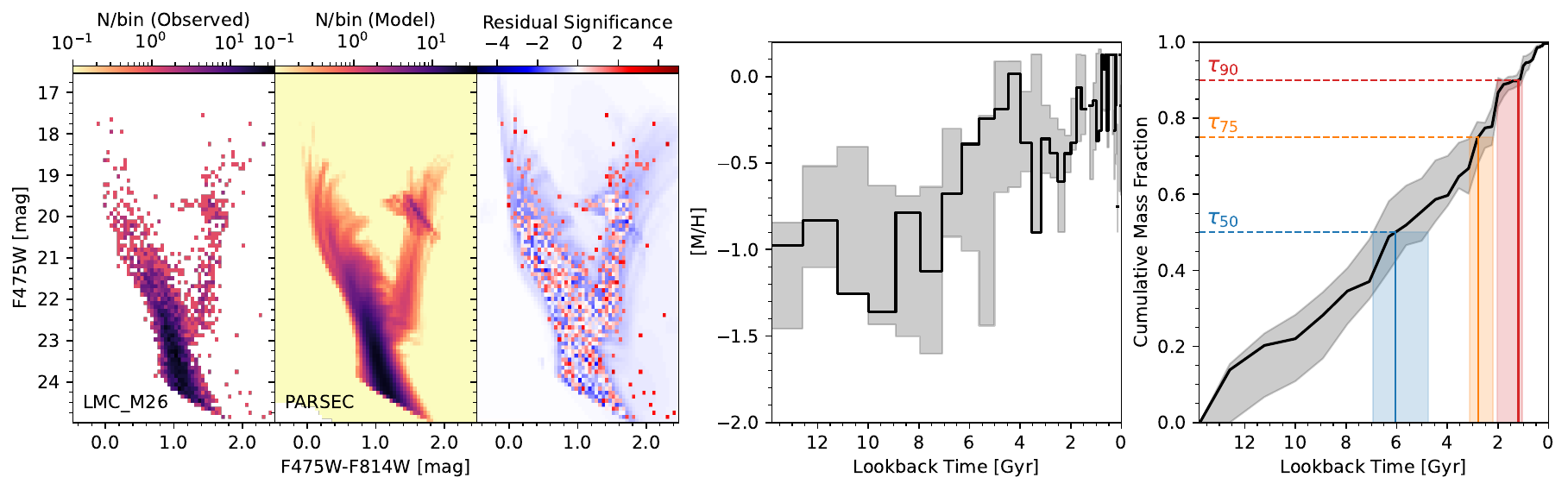}{0.99\textwidth}{}}
\caption{Results of SFH fits for two example fields with data quality typical of our sample, LMC\_9 (top row) and LMC\_M26 (bottom row).  These fits assume PARSEC stellar evolutionary models, although we also perform SFH fits for all fields assuming two additional sets of evolutionary models.  Each row shows, from left to right: Hess diagrams (i.e., two-dimensional binned color-magnitude histograms) of the observed sample used for SFH fitting (including only stars brightward of the 80\% completeness limit, ascertained from artificial star tests), the best-fit model, and the statistical significance of the fit residuals \textbf{in terms of Poisson standard deviations}, in the sense (observed-model).  The middle panel in each row shows the best-fit age-metallicity relation 
and the right-hand panel shows the best-fit cumulative star formation history, with 1-$\sigma$ uncertainties enclosed by grey shading, including both random and systematic uncertainty contributions.  In the right-hand panel, the CSFH metrics we use are illustrated, using a different color for each: The vertical solid lines indicate the lookback times 
$\tau_{50}$, $\tau_{75}$ and $\tau_{90}$, by which 
50\%, 75\% and 90\% of the total stellar mass ever formed had already been formed.  The shading around each vertical line indicates the uncertainties on these CSFH metrics, inferred directly from the uncertainties on the CSFH.} 
\label{sfhexamplefig}
\end{figure}

Example SFH fits assuming PARSEC models are shown in Fig.~\ref{sfhexamplefig} for two typical fields to illustrate the quality of our photometry and SFH fits on a per-field basis.  A useful diagnostic of SFH fit quality is a color-magnitude map of the SFH fitting residuals in terms of Poisson standard deviations (third plot from left in each row of Fig.~\ref{sfhexamplefig}).  In particular, poor fits can result from large-amplitude residuals in particular CMD regions containing features that are not well reproduced by the best-fit SFH.  Similarly, low-amplitude residuals that are uncorrelated with CMD location and produce a \enquote{checkerboard} pattern indicate reliable SFH fits that reproduce the observed stellar sequences.

To quantify the mass assembly history of our LMC fields, we use metrics calculated from cumulative star formation histories (CSFHs).  Specifically, we examine the lookback times by which (50\%, 75\%, 90\%) of the cumulative stellar mass had been formed, denoted ($\tau_{50}$, $\tau_{75}$, $\tau_{90}$).  This metric is illustrated in the right-hand panels of Fig.~\ref{sfhexamplefig}, and the per-field values for ($\tau_{50}$, $\tau_{75}$, $\tau_{90}$) are accessible via Table \ref{tautab}.  Comparing these values across the three sets of stellar evolutionary models that we use for SFH fitting, we find excellent agreement, with different assumed evolutionary models yielding results in agreement to within their 1-$\sigma$ uncertainties in the vast majority ($>$92\%) of cases.

We also combine the SFH results for all of our individual fields together (details of statistically combining SFHs are discussed in \citealt{weisz11,dolphin_syserr,weisz13}) to produce a global CSFH for the LMC, provided in Table \ref{csfh_global_tab}, in order to perform several consistency checks against recent literature results.  These consistency checks are detailed in Appendix \ref{amrsect}, but we briefly point out that globally, our results are consistent with LMC SFHs from other recent surveys, as well as $\tau_{50}$ and $\tau_{80}$ values for star-forming dwarf galaxies at 0.01$<$$z$$<$0.15 (most with 8.5$\lesssim$Log$_{10}$ M$_{\star}$/M$_{\odot}$$\lesssim$9 and about half within 1 Mpc of a massive host; see Fig.~16 of \citealt{mandala}) and CSFHs from IllustrisTNG simulations of LMC-mass satellites with a recent ($<$2 Gyr) infall into a Milky Way-mass host \citep{engler22}, especially in light of substantial galaxy-to-galaxy differences even across similar environments also seen in other simulations \citep{garrisonkimmel}.

\begin{deluxetable}{lccccccccc}
\tablecaption{Lookback Times to Form 50\%, 75\% and 90\% of Cumulative Stellar Mass for Three Evolutionary Models\label{tautab}}
\tablehead{
\colhead{Model:} & \multicolumn{3}{c}{PARSEC} & \multicolumn{3}{c}{MIST} & \multicolumn{3}{c}{BaSTI18} \\
\colhead{Field} & \colhead{$\tau_{50}$} & \colhead{$\tau_{75}$} & \colhead{$\tau_{90}$} & \colhead{$\tau_{50}$} & \colhead{$\tau_{75}$} & \colhead{$\tau_{90}$} & \colhead{$\tau_{50}$} & \colhead{$\tau_{75}$} & \colhead{$\tau_{90}$} \\ \colhead{} & \colhead{Gyr} & \colhead{Gyr} & \colhead{Gyr} & \colhead{Gyr} & \colhead{Gyr} & \colhead{Gyr} & \colhead{Gyr} & \colhead{Gyr} & \colhead{Gyr} }
\startdata
LMC\_2 & 3.16$^{+2.07}_{-0.29}$ & 1.48$^{+0.40}_{-0.10}$ & 0.30$^{+0.27}_{-0.27}$  & 4.98$^{+3.20}_{-1.03}$ & 1.48$^{+0.34}_{-0.06}$ & 0.39$^{+0.45}_{-0.19}$  & 4.61$^{+2.10}_{-1.10}$ & 1.54$^{+0.26}_{-0.11}$ & 0.39$^{+0.52}_{-0.14}$ \\
LMC\_3 & 4.65$^{+0.98}_{-1.15}$ & 1.65$^{+0.31}_{-0.10}$ & 0.49$^{+0.61}_{-0.33}$  & 5.75$^{+1.44}_{-1.97}$ & 1.85$^{+0.22}_{-0.47}$ & 0.54$^{+0.37}_{-0.28}$  & 3.73$^{+3.10}_{-0.46}$ & 1.59$^{+0.75}_{-0.10}$ & 0.47$^{+0.62}_{-0.18}$ \\
LMC\_4 & 5.54$^{+1.76}_{-0.63}$ & 2.66$^{+0.79}_{-0.08}$ & 1.31$^{+0.17}_{-0.10}$  & 6.07$^{+2.29}_{-0.50}$ & 2.81$^{+1.07}_{-0.39}$ & 1.09$^{+0.16}_{-0.03}$  & 6.54$^{+1.48}_{-0.82}$ & 2.94$^{+1.49}_{-0.20}$ & 1.43$^{+0.16}_{-0.21}$ \\
LMC\_6 & 5.80$^{+4.34}_{-0.57}$ & 2.96$^{+1.02}_{-0.18}$ & 1.51$^{+0.80}_{-0.09}$  & 7.47$^{+2.44}_{-0.99}$ & 2.98$^{+1.97}_{-0.48}$ & 1.61$^{+0.36}_{-0.15}$  & 7.02$^{+2.63}_{-1.91}$ & 2.75$^{+2.26}_{-0.12}$ & 1.64$^{+0.60}_{-0.22}$ \\
LMC\_9 & 3.34$^{+3.20}_{-0.59}$ & 1.53$^{+0.74}_{-0.13}$ & 0.76$^{+0.37}_{-0.09}$  & 4.66$^{+2.69}_{-1.61}$ & 1.60$^{+0.30}_{-0.23}$ & 0.80$^{+0.09}_{-0.12}$  & 4.76$^{+2.53}_{-1.87}$ & 1.64$^{+0.23}_{-0.25}$ & 0.71$^{+0.24}_{-0.06}$ \\
LMC\_13 & 4.94$^{+2.01}_{-0.36}$ & 3.05$^{+0.42}_{-0.67}$ & 1.50$^{+0.06}_{-0.08}$  & 7.01$^{+2.22}_{-0.69}$ & 3.34$^{+1.88}_{-0.38}$ & 1.44$^{+0.40}_{-0.19}$  & 6.49$^{+1.28}_{-1.28}$ & 3.23$^{+1.12}_{-0.31}$ & 1.55$^{+0.37}_{-0.08}$ \\
LMC\_15 & 5.72$^{+1.74}_{-0.87}$ & 2.70$^{+1.66}_{-0.35}$ & 1.31$^{+0.17}_{-0.14}$  & 7.83$^{+1.93}_{-2.56}$ & 2.84$^{+1.80}_{-0.44}$ & 1.47$^{+0.31}_{-0.41}$  & 6.22$^{+2.15}_{-1.30}$ & 3.12$^{+1.73}_{-0.30}$ & 1.45$^{+0.36}_{-0.29}$ \\
LMC\_16 & 4.49$^{+1.48}_{-0.79}$ & 1.65$^{+0.68}_{-0.10}$ & 0.90$^{+0.36}_{-0.15}$  & 4.62$^{+1.45}_{-1.04}$ & 1.81$^{+0.55}_{-0.21}$ & 0.92$^{+0.09}_{-0.24}$  & 4.56$^{+1.13}_{-1.38}$ & 1.82$^{+0.34}_{-0.14}$ & 0.79$^{+0.38}_{-0.12}$ \\
LMC\_17 & 4.45$^{+1.91}_{-1.04}$ & 1.61$^{+0.40}_{-0.29}$ & 0.57$^{+0.17}_{-0.10}$  & 5.00$^{+1.70}_{-1.21}$ & 1.56$^{+0.56}_{-0.10}$ & 0.42$^{+0.19}_{-0.05}$  & 4.50$^{+1.06}_{-1.09}$ & 1.61$^{+0.14}_{-0.39}$ & 0.49$^{+0.19}_{-0.04}$ \\
LMC\_19 & 5.44$^{+1.43}_{-0.85}$ & 1.85$^{+0.50}_{-0.20}$ & 0.50$^{+0.65}_{-0.01}$  & 5.94$^{+1.66}_{-1.26}$ & 1.86$^{+0.50}_{-0.31}$ & 0.73$^{+0.22}_{-0.29}$  & 5.40$^{+2.31}_{-0.25}$ & 2.04$^{+0.57}_{-0.33}$ & 1.01$^{+0.06}_{-0.47}$ \\
\enddata
\tablecomments{This Table is published in its entirety in machine-readable format, a portion is shown here for guidance regarding its form and content.}
\end{deluxetable}

\begin{deluxetable}{cccc}
\tablecaption{Global Combined LMC Star Formation History: Lookback Times Versus Cumulative Mass Fraction \label{csfh_global_tab}}
\tablehead{
\colhead{Model:} & \colhead{PARSEC} & \colhead{MIST} & \colhead{BaSTI18} \\ \colhead{Mass Fraction} &  \colhead{Gyr} & \colhead{Gyr} & \colhead{Gyr}}
\startdata
0.05 & 13.17$^{+0.20}_{-0.21}$ & 13.36$^{+0.09}_{-0.15}$ & 13.27$^{+0.17}_{-0.06}$ \\
0.10 & 12.58$^{+0.40}_{-1.02}$ & 12.96$^{+0.18}_{-0.30}$ & 12.77$^{+0.35}_{-0.13}$ \\
0.15 & 10.30$^{+2.29}_{-0.34}$ & 12.50$^{+0.33}_{-0.93}$ & 11.89$^{+0.91}_{-0.37}$ \\
0.20 & 9.15$^{+1.56}_{-0.21}$ & 10.97$^{+1.34}_{-0.52}$ & 10.78$^{+1.31}_{-0.47}$ \\
0.25 & 8.27$^{+1.10}_{-0.15}$ & 9.81$^{+1.15}_{-0.33}$ & 9.56$^{+0.99}_{-0.46}$ \\
0.30 & 7.57$^{+0.81}_{-0.11}$ & 8.84$^{+0.97}_{-0.17}$ & 8.41$^{+0.73}_{-0.26}$ \\
0.35 & 6.96$^{+0.60}_{-0.10}$ & 8.21$^{+0.60}_{-0.22}$ & 7.57$^{+0.49}_{-0.21}$ \\
0.40 & 6.39$^{+0.44}_{-0.10}$ & 7.58$^{+0.36}_{-0.20}$ & 6.76$^{+0.44}_{-0.16}$ \\
0.45 & 5.77$^{+0.41}_{-0.08}$ & 6.94$^{+0.26}_{-0.20}$ & 5.84$^{+0.63}_{-0.04}$ \\
0.50 & 5.02$^{+0.54}_{-0.01}$ & 6.20$^{+0.27}_{-0.14}$ & 5.07$^{+0.46}_{-0.04}$ \\
0.55 & 4.59$^{+0.31}_{-0.02}$ & 5.53$^{+0.15}_{-0.20}$ & 4.19$^{+0.50}_{-0.02}$ \\
0.60 & 3.83$^{+0.51}_{-0.01}$ & 4.87$^{+0.04}_{-0.30}$ & 3.58$^{+0.28}_{-0.01}$ \\
0.65 & 3.19$^{+0.35}_{-0.01}$ & 3.81$^{+0.24}_{-0.04}$ & 3.11$^{+0.29}_{-0.01}$ \\
0.70 & 2.84$^{+0.28}_{-0.01}$ & 2.94$^{+0.31}_{-0.01}$ & 2.70$^{+0.29}_{-0.01}$ \\
0.75 & 2.54$^{+0.14}_{-0.03}$ & 2.38$^{+0.26}_{-0.01}$ & 2.33$^{+0.21}_{-0.01}$ \\
0.80 & 2.06$^{+0.17}_{-0.05}$ & 1.96$^{+0.17}_{-0.01}$ & 1.91$^{+0.14}_{-0.02}$ \\
0.85 & 1.55$^{+0.08}_{-0.02}$ & 1.47$^{+0.13}_{-0.01}$ & 1.57$^{+0.05}_{-0.02}$ \\
0.90 & 1.18$^{+0.02}_{-0.02}$ & 1.06$^{+0.02}_{-0.01}$ & 1.09$^{+0.04}_{-0.02}$ \\
0.95 & 0.65$^{+0.05}_{-0.01}$ & 0.66$^{+0.03}_{-0.01}$ & 0.58$^{+0.07}_{-0.01}$ \\
\enddata
\end{deluxetable}

\section{Spatial Trends in CSFH Metrics \label{obsgradsect}}

\begin{figure}
\gridline{\fig{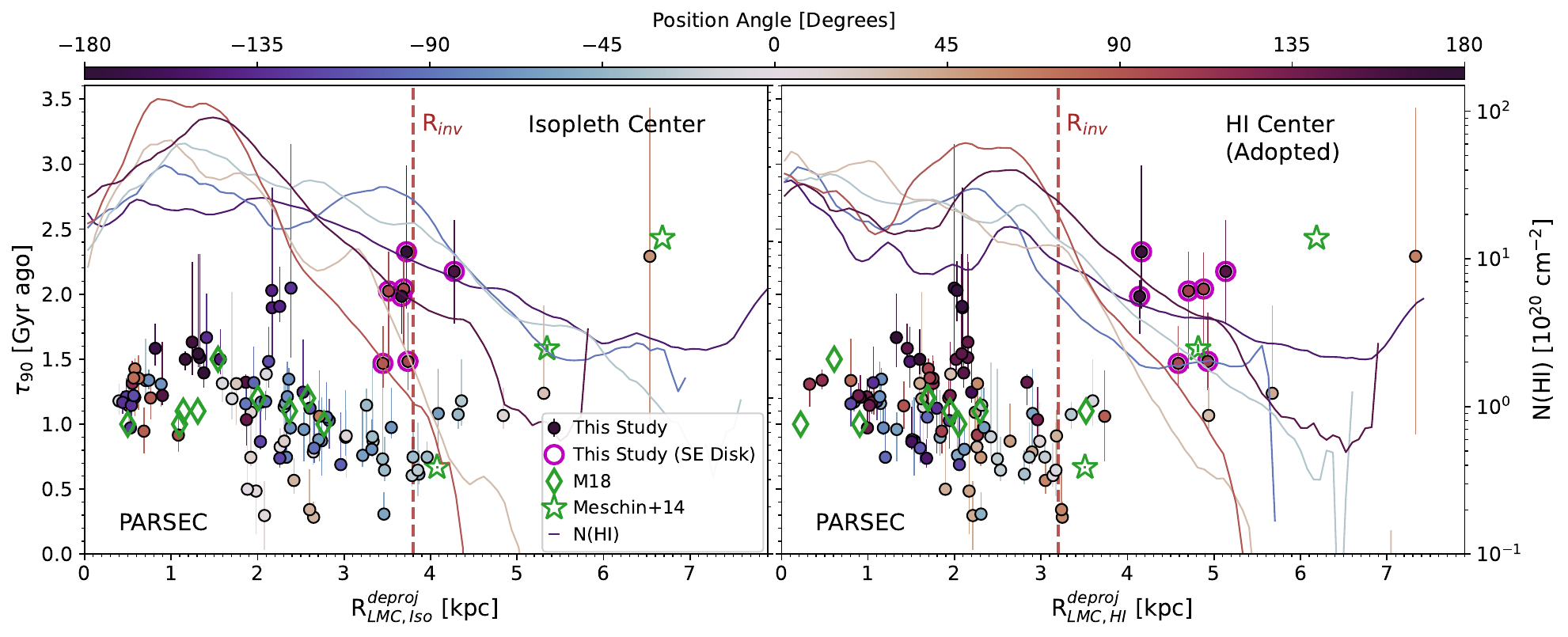}{0.99\textwidth}{}}
\caption{The lookback time to form 90\% of each field's cumulative stellar mass (denoted $\tau_{90}$) as a function of in-plane distance from the center of the LMC.  We compare trends assuming the \citet{vdmcioni} isopleth-based center location (left panel) versus the HI kinematic center we adopt (right panel).  Individual fields from this study are shown as filled circles, color-coded by their position angle east of north relative to the adopted center.  The subset of our fields located in the southeast disk (corresponding to the magenta dashed lines in Fig.~\ref{map_fig}) are circled in magenta.  Literature results from the ground-based fields of \citet{monteagudo} and \citet{meschin} are overplotted as green diamonds and stars respectively.  The location of the inversion in radial age trends R$_{inv}$ is shown as a vertical dashed brown line.  Radial profiles of HI column density \citep{kim03} in azimuthal slices are shown as lines, also color-coded by position angle, corresponding to the right-hand vertical axis.  Radial trends for both the per-field $\tau_{90}$ values from our SFH fitting as well as the HI column density profiles show less scatter when adopting the HI kinematic center, consistent with models predicting that LMC-SMC-like interactions can produce an offset photometric distribution.}
\label{csfh_hi_fig}
\end{figure}

We analyze radial trends in mass assembly over the lifetime of LMC by examining the per-field characteristic times $\tau_{50}$, $\tau_{75}$, and $\tau_{90}$ 
as a function of in-plane deprojected radius from the center of the LMC R$_{LMC}^{deproj}$.  To calculate R$_{LMC}^{deproj}$, we assume viewing angles from \citet{yumimap}, including a line-of-sight distance of 49.9 kpc \citep{lmcdist} to the center location from \citet{vdmcioni}. An example of the radial trends we find is illustrated in Fig.~\ref{csfh_hi_fig}, where we plot $\tau_{90}$ versus R$_{LMC}^{deproj}$.  The most prominent feature, detailed below, is an inversion from an inside-out gradient in the inner disk to an outside-in gradient in the outer disk, occurring at a value of R$_{LMC}^{deproj}$ denoted R$_{inv}$.

Various center locations have been measured for the LMC depending on the adopted tracer population and methodology (e.g., Fig.~9 of \citealt{niederhofer22}).  The two panels of Fig.~\ref{csfh_hi_fig} compare radial trends in $\tau_{90}$ versus R$_{LMC}^{deproj}$ assuming two different locations for the LMC center: The HI kinematic center, which is an average of the values from \citet{kim98} and \citet{luks92}, and the isopleth-based center from \citet{vdmcioni}, chosen because it effectively samples the opposite extremum of literature values, differing from the kinematic center by 1.16 kpc at the distance of the LMC.  For our analysis, we adopt the HI kinematic center ($\alpha_{0}$, $\delta_{0}$) = (78.77$^{\circ}$, $-$69.01$^{\circ}$), which is consistent with the \textit{stellar} dynamical center found by \citet{vdmk14}, and stellar proper motions from the VMC survey \citep{niederhofer22} also yield a center consistent with the \citet{vdmk14} location.  More generally, our use of the kinematic center is supported by simulations, finding that a photometric center offset from the kinematic center is a common feature of LMC-SMC-like interactions \citep{bekki09,besla12,yozin,pardy16}.

One advantage of using the kinematic center to analyze radial gradients in stellar population properties is that they can be directly compared to radial trends in gas properties (we return to this point in Sect.~\ref{discusssect}).  In Fig.~\ref{csfh_hi_fig}, we overplot radial profiles of HI column density \citep{kim03}, calculated in six azimuthal slices, for comparison with the radial trend in $\tau_{90}$ assuming the two different LMC center locations.  
When adopting the HI kinematic center, not only does the azimuthal symmetry of the HI column density profiles improve, but we also find that the southeast disk fields (circled in magenta in Fig.~\ref{csfh_hi_fig}) are in better agreement with radial trends seen at other position angles.  While some azimuthal scatter remains in the outer disk, there are at least two potential causes related to the placement of available fields: First, the outermost \citet{meschin} ground-based field samples an overdensity on the north side of the LMC disk at 7$^{\circ}$$<$R$<$8$^{\circ}$ detected by \citet[][see their fig.~4]{yumiring}, which also appears preferentially metal-rich in the residual metallicity maps of \citet[][see their fig.~12]{grady21}.  Second, the outermost of our four outer fields (R$_{LMC}^{deproj}$$>$6 kpc in both panels of Fig.~\ref{csfh_hi_fig}) samples the warp in the northeastern stellar disk detected by \citet{saroon22}. 

In the left column of Fig.~\ref{csfh_rad_fig}, we show radial trends for all three of our CSFH metrics ($\tau_{90}$, $\tau_{75}$, $\tau_{50}$) adopting the HI kinematic center.  The location of the inversion, at R$_{inv}$$\approx$3.2 kpc, remains fixed over all of the lookback times we sample, and the radial gradients appear relatively flatter at earlier lookback times.  This result is independent of the choice of stellar evolutionary model or adopted center location, so throughout our analysis we present results assuming PARSEC models and the HI kinematic center. In Appendix \ref{altmodelsect}, we present results recalculated assuming alternate evolutionary models (MIST and BaSTI18) and/or an alternate (isopleth-based) center location, finding that the location of R$_{inv}$ changes slightly to $\approx$3.8 kpc adopting the isopleth-based center but our results are qualitatively unaffected (we found that adopting intermediate center locations simply resulted in trends and R$_{inv}$ values that were intermediate between those resulting from the two center locations we explore).  With this in mind, we now address the inside-out trend in the inner disk and the inversion in the outer disk in more detail.

\subsection{The Inner Disk \label{innerdisksect}}

\begin{figure}
\gridline{\fig{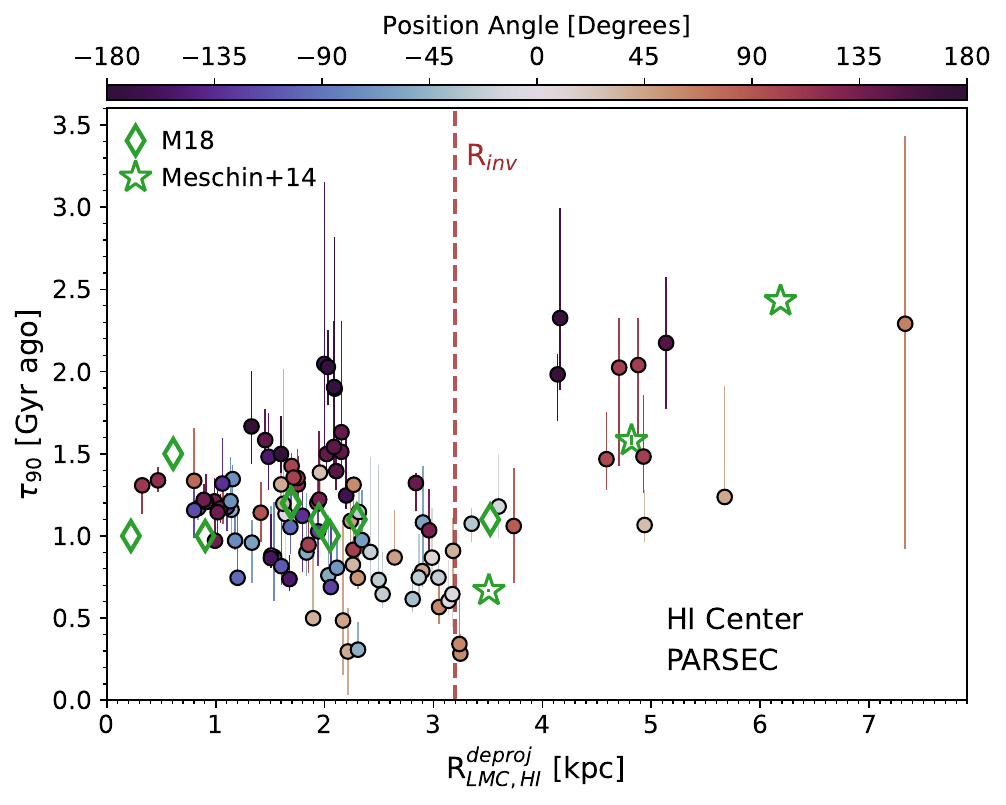}{0.5\textwidth}{}
          \fig{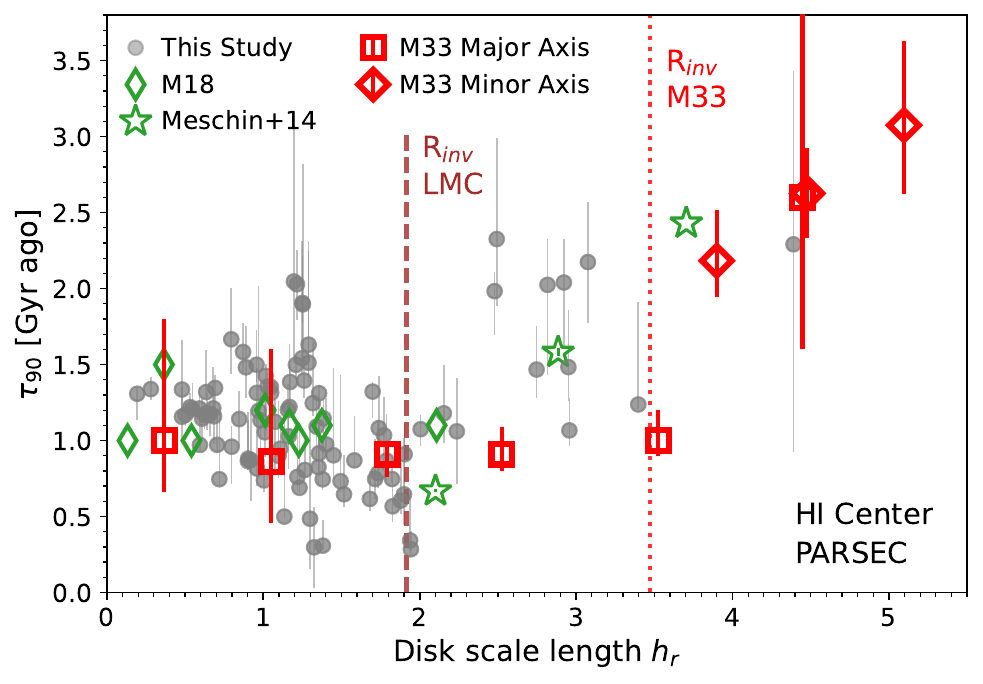}{0.5\textwidth}{}}
\vspace{-1.1cm}
\gridline{\fig{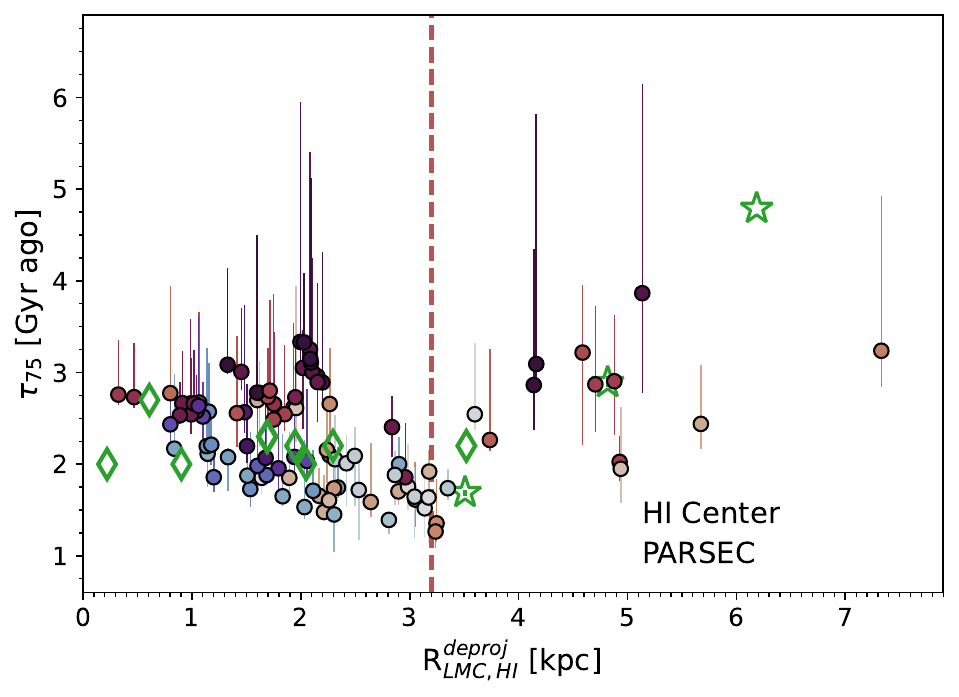}{0.49\textwidth}{}
          \fig{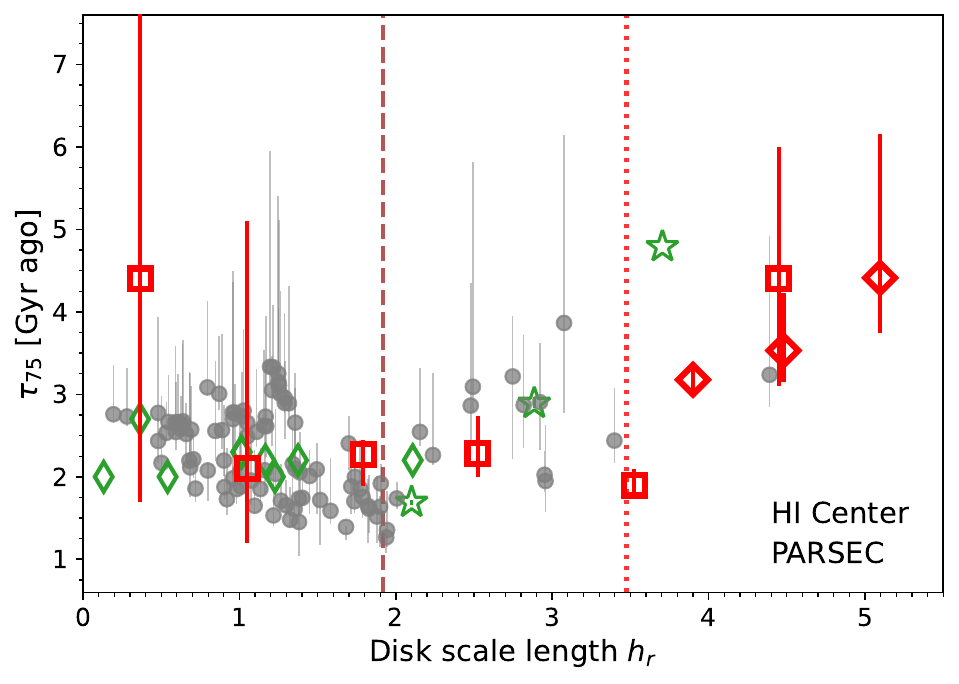}{0.49\textwidth}{}}
\vspace{-1.1cm}
\gridline{\fig{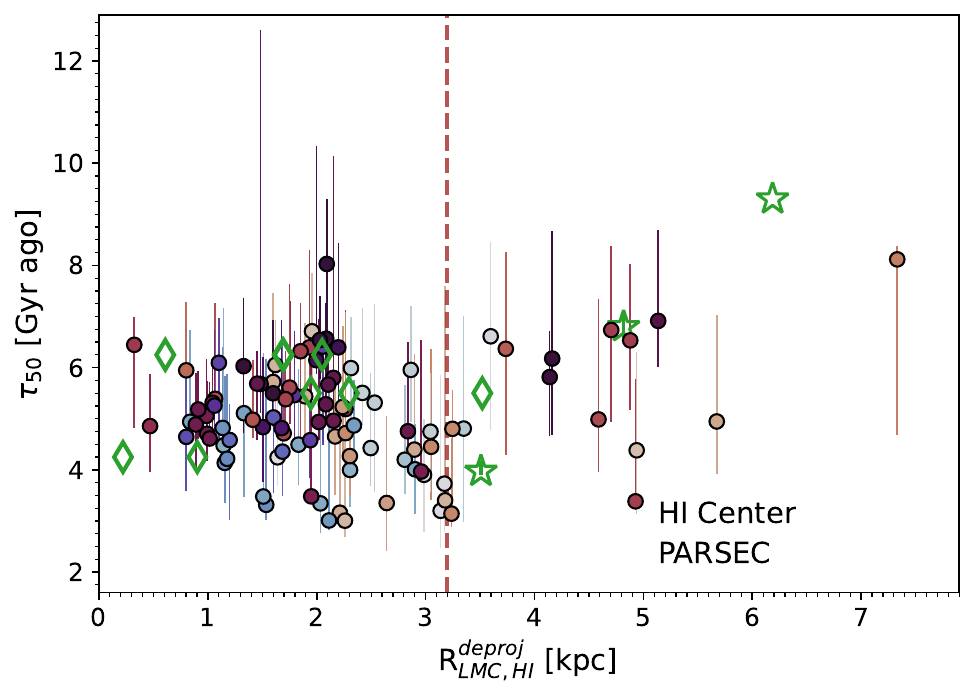}{0.49\textwidth}{}
          \fig{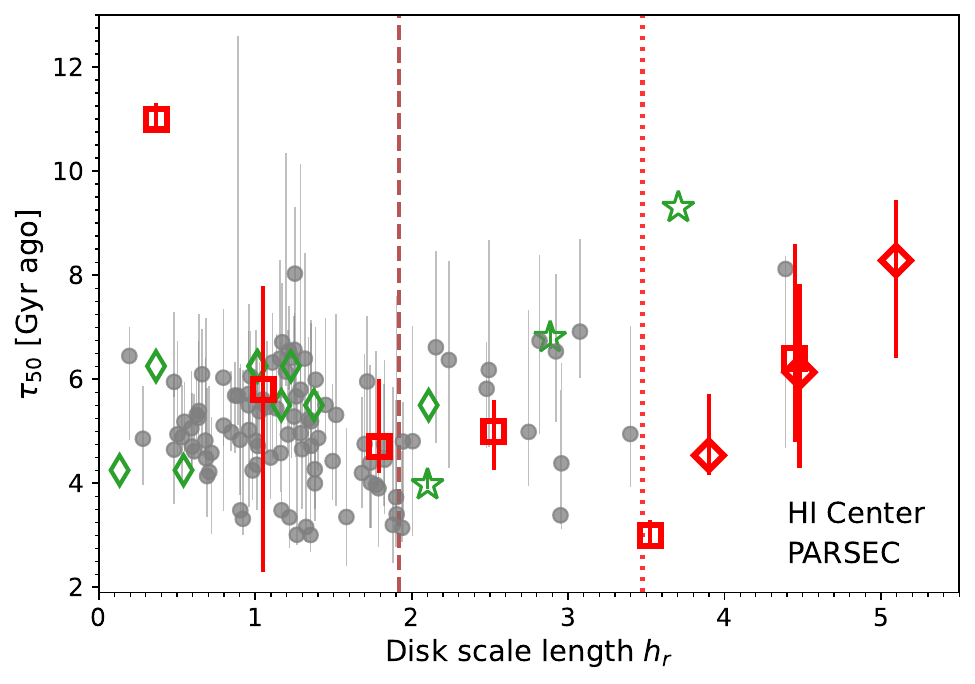}{0.49\textwidth}{}}
\vspace{-1.0cm}
\caption{Lifetime star formation history metrics versus in-plane distance from the center of the LMC, assuming viewing angles from \citet{yumimap} and the LMC center location based on HI kinematics.  From top to bottom, the plots show the trend in lookback time by which 90\%, 75\% and 50\% of the cumulative stellar mass had formed.  In the left-hand column, individual fields are color-coded by their position angle east of north, with symbols as in Fig.~\ref{csfh_hi_fig}.  In the right-hand column, our fields are shown in grey and values for M33 \citep{barker07,williams09,barker11} are overplotted in red, with deprojected radial distances normalized to the disk scalelength $h_{r}$ for each galaxy.  These results assume PARSEC evolutionary models, and in Appendix \ref{altmodelsect} we demonstrate that the inversion in radial age trends, and its constancy over the range of lookback times shown here, is robust to choice of stellar evolutionary models and adopted LMC center location.}
\label{csfh_rad_fig}
\end{figure}

The trends in the inner LMC disk (R$\leq$R$_{inv}$) in Fig.~\ref{csfh_rad_fig} are generally consistent with an inside-out gradient such that the innermost fields assembled a larger fraction of their stellar mass earlier \citep[e.g.,][]{white91,mo98,bird13}, extending out to R$_{inv}$$\approx$3.2 kpc.  Some positive outliers are seen at R$^{deproj}_{LMC}\sim$2 kpc with position angle 165$^{\circ}$$<$PA$<$180$^{\circ}$ and $\tau_{90}$$\sim$2 Gyr, and although this subset is spatially clustered near the N206 star forming region \citep{henize56}, their HI column densities \citep{kim03} are typical of the other fields at their R$^{deproj}_{LMC}$, and this location does not appear unusual in terms of stellar density or kinematics \citep{yumiring,choi22,niederhofer22}.  These fields constitute $<$10\% of the sample by projected area and $<$7\% by total mass so our conclusions are unaffected by their inclusion or exclusion.  More generally, the tendency for fields in the southern disk (darker colors in Fig.~\ref{csfh_rad_fig}; PA$\lesssim$$-$90$^{\circ}$ or PA$\gtrsim$90$^{\circ}$) to have assembled a given mass fraction by earlier lookback times than fields in the northern disk (lighter colors in Fig.~\ref{csfh_rad_fig}; $-$90$^{\circ}$$\lesssim$PA$\lesssim$90$^{\circ}$) at fixed R$^{deproj}_{LMC}$ is 
unsurprising since the southern part of the LMC disk, preferentially lacking recent enhancements in star formation triggered by interaction with the SMC, is older on average than the northern part \citep{ruizlara20,povick}.  In light of our field placement (see Fig.~\ref{map_fig}), this also explains minor offsets between our values of $\tau_{50}$ and the median age-radius relation from \citet[][see Appendix \ref{altmodelsect}]{povick}
\footnote{The \citet{povick} age-radius relation is fit versus an on-sky elliptical radius, whereas R$^{deproj}_{LMC}$ is a three-dimensional in-plane distance.  The conversion between the two is, in a strict sense, position dependent, but we use an average value of R$^{deproj}_{LMC}$$\approx$0.83R$_{\rm Povick}$, which is accurate to $\sim$5\% over our sightlines.}.

For the inner disk, the inside-out gradient is not well represented by a simple linear relation.  Specifically, for the more recent metrics ($\tau_{90}$, $\tau_{75}$), maximum likelihood fits of a linear relation versus R$^{deproj}_{LMC}$ yield $\chi^{2}$ per degree of freedom $>$5 when fitting to the full sample and $>$1.5 when fitting to northern fields alone. 
The large reduced $\chi^{2}$ values of the linear fits are not due to underestimated uncertainties, since we found that allowing for a fractional underestimate of the variance did not affect the fit parameters beyond their uncertainties, with an increase in the variance of $<$2\%.   Field-to-field variations at fixed R$^{deproj}_{LMC}$ are therefore real, stemming from the highly non-axisymmetric structure of the LMC \citep[e.g.][]{vdm02,miller22}, and are also predicted by models of more massive disks due to both internal and external processes \citep[e.g.][]{sanchezblazquez09,halle18,carr22}.  Meanwhile, for $\tau_{50}$, uncertainties do not allow constraints on the linearity of the inside-out gradient in the inner disk (see below).  In light of substantial field-to-field variations, we checked for correlations between the CSFH metrics ($\tau_{90}$, $\tau_{75}$, $\tau_{50}$) versus gas and dust properties as well as potential sources of field-to-field observational systematics.  These include HI column density \citep{kim03}, dust temperature and dust mass surface density \citep{dusttemp}, and observational properties including foreground and differential extinction (either recovered by \texttt{MATCH} or measured independently by \citealt{yumimap} and \citealt{skowron}), photometric depth in both filters (i.e., C80), and the line-of-sight distance residuals versus independently measured red clump distances (shown in the top row of Fig.~\ref{comp_dist_av_fig}), finding no visually apparent correlations.

To further quantify the significance of the radial trends in the inner disk in Fig.~\ref{csfh_rad_fig}, we calculated both the Pearson linear correlation coefficient $\rho$ and the Spearman rank correlation coefficient $r_{s}$ and their $p$-values\footnote{The $p$-value gives the probability of obtaining the resultant correlation coefficient given the null hypothesis of random data.} for fields with R$\leq$R$_{inv}$, with uncertainties based on Monte Carlo draws from the (asymmetric) Gaussian-distributed per-field errors (see Table \ref{csfhtab}).  For the more recent metrics ($\tau_{90}$, $\tau_{75}$), the inside-out trend has a moderate level of both anticorrelation ($-$0.53$\leq$$r_{s}$$\leq$$-$0.36) and linearity ($-$0.51$\leq$$\rho$$\leq$$-$0.33) at high confidence ($p$$\lesssim$0.01) when considering the northern fields ($\eta$$>$0) alone (across all three assumed evolutionary models).  Furthermore, in \textit{all} cases, the level of correlation worsened when including the southern fields ($\eta$$<$0) with R$<$R$_{inv}$, and the southern fields alone showed no significant correlations with R$^{deproj}_{LMC}$ ($p$$>$0.07 in all cases), although these constitute a minority (42\%) of the sample.  Meanwhile, for $\tau_{50}$ (and earlier lookback times not shown), uncertainties do not allow quantitative constraints on correlations, although the data are still qualitatively consistent with the inside-out trend seen at more recent lookback times, as well as ground-based results from \citealt{meschin} and \citealt{monteagudo}.  As an additional check for any systematic impact of differential extinction on our results, we divided our sample into two subsamples with independent measurements of differential extinction $(\sigma_{\rm 1}+\sigma_{\rm 2})$ from \citet{skowron} above and below the median across our sample ($\delta$A$_{V}$=0.29 mag).  The resulting values of $\rho$ and $r_{s}$ for each subsample are in agreement with each other and with the full-sample values in Table \ref{csfhtab} to within their 1-$\sigma$ uncertainties, 
and we found no residual correlation with $(\sigma_{\rm 1}+\sigma_{\rm 2})$ in the trend of the CSFH metrics versus R$^{deproj}_{LMC}$.  In summary, while radial age gradients in the inner disk are not formally well represented by linear fits due to field-to-field variations, Pearson and Spearman correlation coefficients reveal a statistically significant inside-out age gradient.

\begin{deluxetable}{ccccc}
\tablecaption{Correlation Coefficients of CSFH Metrics Versus R$^{deproj}_{LMC}$ For Inner Disk Fields (R$<$R$_{inv}$)\label{csfhtab}}
\tablehead{\colhead{} & \multicolumn{2}{c}{Pearson} & \multicolumn{2}{c}{Spearman} \\[-0.2cm] \colhead{Metric} & \colhead{$\rho$} & \colhead{Log$_{\rm 10}$ $p$-value} & \colhead{$r_{s}$} & \colhead{Log$_{\rm 10}$ $p$-value}} 
\startdata
\hline
\multicolumn{5}{c}{All Fields} \\
\hline
$\tau_{90}$ & -0.25$^{+0.05}_{-0.05}$ & -1.75$^{+0.57}_{-0.72}$ & -0.29$^{+0.06}_{-0.05}$ & -2.17$^{+0.68}_{-0.78}$  \\
$\tau_{75}$ & -0.33$^{+0.06}_{-0.06}$ & -2.77$^{+0.86}_{-1.07}$ & -0.40$^{+0.05}_{-0.04}$ & -3.84$^{+0.84}_{-0.92}$  \\
$\tau_{50}$ & -0.12$^{+0.08}_{-0.08}$ & -0.56$^{+0.38}_{-0.61}$ & -0.12$^{+0.08}_{-0.08}$ & -0.57$^{+0.40}_{-0.66}$  \\
\hline
\multicolumn{5}{c}{Northern Fields Only} \\
\hline
$\tau_{90}$  & -0.47$^{+0.09}_{-0.08}$ & -2.63$^{+0.84}_{-1.00}$ & -0.51$^{+0.08}_{-0.08}$ & -3.02$^{+0.91}_{-1.16}$  \\
$\tau_{75}$ & -0.51$^{+0.08}_{-0.08}$ & -3.02$^{+0.91}_{-1.12}$ & -0.53$^{+0.09}_{-0.08}$ & -3.25$^{+1.02}_{-1.22}$  \\
$\tau_{50}$  & -0.19$^{+0.12}_{-0.12}$ & -0.60$^{+0.42}_{-0.68}$ & -0.18$^{+0.12}_{-0.12}$ & -0.57$^{+0.40}_{-0.64}$  \\
\enddata
\tablecomments{These fits assume PARSEC evolutionary models, and analogous results assuming alternate stellar evolutionary models and LMC center locations are given in Appendix \ref{altmodelsect}.}
\end{deluxetable}

\subsection{Inversion in Radial Age Trends}

Beyond the inner disk (R$_{LMC}^{deproj}$$>$R$_{inv}$), our knowledge of CSFH trends hangs largely on the three ground-based outer disk fields in the northern LMC analyzed by \citet{gallart08} and \citet{meschin}, shown as green stars in Fig.~\ref{csfh_rad_fig}.  Our results, viewed in this context, reveal a striking reversal in lifetime CSFH metrics, to an outside-in gradient in the outer disk.  
The radial location of the inversion remains fixed over all of the lookback times we sample (shown in Fig.~\ref{csfh_rad_fig}), implying at least one driver that is neither spatially nor temporally localized.  In addition, the radial location of the inversion is also approximately coincident with the dropoff in HI column density (see Fig.~\ref{csfh_hi_fig}), predicted by N-body models of more massive disks \citep{roskar08}.

The inversion in radial age trends is unlikely to be an artifact of differences in observing setup, SFH fitting code or assumed stellar evolutionary models between our results and \citet{meschin} since the excellent agreement between our results and theirs in the range of radial overlap (R$_{LMC}^{deproj}$$\approx$3$-$4 kpc) persists across all choices of center location, evolutionary model and lookback time (see Appendix \ref{altmodelsect}).  Also, the inside-out trend in the inner disk is consistent with SFH results from ground-based fields analyzed by \citet{monteagudo}, shown as green diamonds in Fig.~\ref{csfh_rad_fig}, and the inversion in radial age trends is evident solely from the combination of the \citet{meschin} and \citet{monteagudo} results, at least for more recent CSFH metrics where the radial dependence is relatively steeper (i.e., $\tau_{75}$ and $\tau_{90}$).

While we cannot rule out the possibility that the inverted radial age trend in the LMC is impacted on some level by inhomogenous spatial sampling of the LMC disk, there are four arguments against a bias in our field placement affecting our results: First, preferential targeting of sightlines close to young, massive stars (i.e., our Scylla and METAL WFC3/UVIS fields) is counterbalanced by both our \textit{exclusion} of fields with severe differential extinction (correlating with $\Sigma_{\rm dust}$, see Sect.~\ref{distcompsect}) and our \textit{inclusion} of the WFPC2 LGSPA fields, which were culled from a variety of archival programs \citep{lgspa} and constitute half the sample (47\% by projected area), yielding a combined LMC CSFH identical to the WFC3/UVIS fields (Fig.~\ref{comp_uvis_wfpc2_fig}).  Second, the radial age trends we find are fully consistent with results from ground-based fields placed to sample spatial trends representative of the LMC field population \citep[Fig.~\ref{csfh_rad_fig};][]{meschin, monteagudo}.  Third, a global LMC CSFH produced by statistically combining SFH results from all of the individual fields is in good agreement with previous surveys mapping the LMC contiguously (Appendix \ref{amrsect}).  Fourth, the combined LMC CSFH is also consistent with theoretical predictions for LMC-mass galaxies infalling into a Milky Way-mass host \citep{engler22}.

\section{Discussion \label{discusssect}}

The age metrics illustrated in Fig.~\ref{csfh_rad_fig} span the entirety of the LMC-SMC interaction, likely extending even earlier.  The persistence of the inversion in radial age trends and its fixed value of R$_{LMC}^{deproj}$ across all the lookback times we sample argue for a driver of the radial age trend that is neither spatially localized nor short-lived.  Radial migration seems a likely contender, and models of more massive disks predict that an inverted radial age trend is a natural consequence of radial migration, selectively contributing larger fractions of migrators to more external present-day radii \citep[e.g.,][]{roskar08,minchev12,bird13}.  At LMC-like masses, one avenue for assessing the importance of radial migration is by constraining the frequency of inverted radial age gradients at fixed stellar mass, so in Sect.~\ref{lvsect} we examine radial age inversions in other Local Volume LMC-mass galaxies measured directly from spatially resolved SFH fits as we have done, and in Sect.~\ref{ifusect} we examine the frequency of radial age inversions based on larger samples gleaned primarily from integral field unit (IFU) surveys.  We find that radial age inversions in late-type LMC mass galaxies are common but not universal, 
so in Sect.~\ref{interactsect} we discuss the indirect role of galaxy-galaxy interactions in facilitating and boosting radial migration, particularly relevant to the LMC.  

\subsection{Radial Age Inversions in LMC-Mass Galaxies: Resolved SFH Studies \label{lvsect}}

There are three late-type Local Volume galaxies that have similar stellar masses as the LMC, and likely owing to their similarly moderate inclinations, also have radially resolved lifetime SFH information available from SFH fits to resolved CMDs.  We compare each of these to the LMC, and then move on to a broader comparison of radial age trends in late-type LMC-mass galaxies based on spectral fitting techniques.  
 
\subsubsection{Comparison to M33 \label{m33sect}}
    
Like the LMC, the inner disk of M33 has an inside-out gradient and a reversal to an outside-in gradient at larger radii \citep{barker07,williams09,barker11,phatter}, 
corroborated by an SFH analysis of long period variables \citep{javadi17}.  In the right-hand column of Fig.~\ref{csfh_rad_fig}, we compare radial trends in LMC CSFH metrics to measurements of M33 \citep{barker07,williams09,barker11}, normalizing deprojected radii to the present-day disk scalelength of each galaxy 
$h_{r}$ assuming 
$h_{r,LMC}$ = 1.67 kpc \citep{yumiring} and 
$h_{r,M33}$ = 2.59 kpc \citep{smercina23}\footnote{Although the measured disk scalelength can depend on the age of the tracer population used \citep[e.g.,][]{radburnsmith12,streich16}, this is unlikely to affect the comparison here, since the LMC value we employ was measured by \citet{yumiring} using red clump stars, which they verified to yield results qualitatively consistent with those obtained using red giants, used by \citet{smercina23} to fit the M33 density profile.}.  The radial age trends in M33 show a striking qualitative similarity to the LMC, with two quantitative differences: First, while inner disk radial trends are compatible to within their uncertainties for metrics probing recent lookback times ($\tau_{75}$, $\tau_{90}$), the innermost field of M33 had already assembled half of its mass over 10 Gyr ago.  Such rapid early assembly is not seen anywhere in the inner LMC, most likely offset by vigorous star formation over the last few Gyr \citep{weisz13,monteagudo,lazygiants,ruizlara20,mazzi21}.  Second, the inversion in radial age gradients in M33 may be less dramatic and more difficult to pinpoint, occurring as far out in the disk as R$_{inv}$$\sim$3.5$h_{r}$.  
Nevertheless, the general similarity in the trends seen for the LMC and M33 suggests that even the prolonged LMC-SMC interaction and recent collision has had a relatively minor impact on radial age gradients in the LMC.  Support for this theory comes from the fact that the LMC and M33 otherwise have numerous global properties in common (compiled in Table \ref{m33comptab}), effectively isolating recent interaction history as a variable.  Accordingly, the salient differences between M33 and the LMC are seen primarily in their gas morphology and kinematics as a consequence of their different recent interaction histories.  

However, the interaction history of M33 does remain somewhat controversial.  On one hand, orbit modeling based on recent mass estimates and proper motions argues strongly against any previous pericentric passages of M31 by M33 \citep{patel17,vdm19}.  On the other hand, a gaseous stream-like feature between M31 and M33 and a warp in the outer HI disk \citep[e.g.,][]{putman09,lewis13} and outer stellar disk \citep[e.g.,][]{mcconnachie10} of M33 may result from past interactions.  A possible resolution to this mystery is provided by \citet{smercina23}, who point out that these features, in addition to the asymmetry they detect in the spiral arms of M33, could result from a past interaction with the progenitor of the Giant Stellar Stream of M31.  This progenitor could have been accreted $\sim$2 Gyr ago, and may have been as massive as 1/3 the mass of M31 itself \citep{dsouzabell18}, impacting the gravitational potential of M31 and resulting orbital calculations.  Even in this case, M33 has not undergone such intense recent interaction as the LMC-SMC system, so the similarity of their radial age profiles suggests that the \textit{presence} of inversions in radial age profiles is not sensitive to the details of interaction history, although the exact shape of the inverted radial age profile may be affected on a case-by-case basis.

One alternative explanation put forth for the inversion in radial age trends in M33 was the presence of accreted stars that should be preferentially old and kinematically hot \citep{mostoghiu18}.  Recent spectroscopic observations are not consistent with this scenario, since the fraction of kinematically hot stars actually decreases at larger radii, with a smooth trend in stellar disk kinematics across the location of the age and surface brightness profile break \citep{gilbert22}.  In the LMC, accreted stars are also unlikely to contribute significantly to the inversion in radial age gradients, since spectroscopic abundances for stars out to $\sim$3 times farther from the LMC center than we probe are mostly inconsistent with an SMC origin \citep{munoz23,massana23}.  Another potential explanation for radial age inversions was based on simulations of a Milky Way-mass disk by \citet{sanchezblazquez09}, but also inconsistent with observations of M33.  They proposed that the outside-in gradient in the outer disk was mainly a result of decreased star formation at large galactocentric radii due to a lower gas density there.  However, the HI disk in M33 is quite extended relative to the location of R$_{inv}$ both presently and likely in the past, arguing against a lack of gas driving the outside-in gradient in the outer disk: 18\% of M33's HI mass is located outside $\sim$3$h_{r}$, and such an extended HI disk is unlikely to be due to a recent interaction \citep{putman09}.  
  
\begin{deluxetable}{lcc}
\tablecaption{Comparison of Global Properties for the LMC and M33\label{m33comptab}}
\tablehead{\colhead{Property} & \colhead{LMC} & \colhead{M33}} 
\startdata
First infall into massive host & Yes (1,2,3) & Yes (4,5) \\
M$_{\star}$/M$_{\odot}$ & $\sim$3 $\times$ 10$^{9}$ (6) & $\sim$4.8 $\times$ 10$^{9}$ (7) \\
Warped stellar disk & Yes (8,9,10) & Yes (11) \\
Mild negative radial metallicity gradient & Yes (12,13,14) & Yes (15,16,17) \\
Age-dependent spiral arm structure & Yes, dominant arm (18,19,20) & Yes, two asymmetric arms (21) \\
Bar & Tilted, offset (22,23,24) & Short and weak (21) \\
Stellar rotation curve & Flattens/declines beyond 4$-$5 kpc (25,26,27,28,29,30) & Flattens/declines beyond $\sim$9 kpc (7) \\
HI disk morphology & Truncated (31) & Warped, extended (32) \\
M$_{\rm HI}$/M$_{\odot}$ & $\sim$0.5 $\times$ 10$^{9}$ (6,33) &  $\sim$1.4 $\times$ 10$^{9}$ M$_{\odot}$ (32) \\
HI rotation curve & Flattens/declines beyond 4$-$5 kpc (6) & Rises out to $\sim$20 kpc (7) \\
\enddata
\tablecomments{References: (1) \citet{besla07}; (2) \citet{besla12}; (3) \citet{nitya13}; (4) \citet{patel17}; (5) \citet{vdm19}; (6) \citet{kim98}; (7) \citet{corbelli14}; (8) \citet{olsen}; (9) \citet{yumimap}; (10) \citet{saroon22}; (11) \citet{mcconnachie10}; (12) \citet{cioni09}; (13) \citet{choudhury21}; (14) \citet{grady21}; (15) \citet{cioni08}; (16) \citet{magrini10}; (17) \citet{beasley15}; (18) \citet{dev55}; (19) \citet{ruizlara20}; (20) \citet{mazzi21}; (21) \citet{smercina23}; (22) \citet{zhao00}; (23) \citet{zaritsky04}; (24) \citet{yumimap}; (25) \citet{vdm02}; (26) \citet{vdmk14}; (27) \citet{wan20}; (28) \citet{gaia21}; (29) \citet{choi22}; (30) \citet{schmidt22}; (31) \citet{salem15}; (32) \citet{putman09}; (33) \citet{bruns05};}
\end{deluxetable}

\subsubsection{Comparison to NGC 7793}

SFH fitting directly to resolved CMDs also reveals evidence for an inversion in radial age gradients in the 
Sculptor Group spiral NGC 7793 \citep{sacchi19}.  NGC 7793 has a similar mass as the LMC and M33 (3.2 $\times$ 10$^{9}$ M$_{\odot}$; \citealt{dale09}), and the inversion in age trends occurs at $R_{inv}$$\sim$2$-$3$h_{r}$ assuming a distance of 3.1 Mpc \citep{radburnsmith11} and $h_{r}$ = 1.2 kpc \citep{vlajic7793}, similar to the LMC and M33.  While the detection and characterization of the inversion in radial CSFH trends hangs entirely on the SFH results for the poorly sampled outermost radial bin from \citet{sacchi19}, the inside-out gradient they find in the inner disk of $\Delta$$\tau_{50}$/$\Delta$$h_{r}$ $\sim$ 1 is compatible with our LMC values.  In fact, NGC 7793 provides a complementary point of comparison with the LMC since they have several features in common that are not shared by M33: Like the LMC, 
NGC 7793 has an old stellar population that extends radially well beyond its HI gas disk \citep{radburnsmith12}, a rotation curve that decreases (or is warped and flattens) at $\sim$4 kpc \citep{bacchini19}, and may have undergone recent interactions based on kinematics and morphology of its gas \citep{carignan,dicaire} and young stars \citep{mondal}.

In the case of M33, the \citet{roskar08} model of a more massive (Milky Way-mass) disk was the only available model for \citet{williams09} to advocate for radial migration as the cause of its inverted radial age gradient (right-hand panel of Fig.~\ref{csfh_rad_fig}).  However, \citet{radburnsmith12} subsequently tailored such a model to the mass and disk scalelength of NGC 7793 and found substantial radial migration, strongly supporting radial migration as the cause of the inversion in the radial age gradient in NGC 7793.  Furthermore, although not run in a fully cosmological context, the \citet{radburnsmith12} NGC 7793 model also predicts a global CSFH in good agreement with our results for the LMC.  In particular, the NGC 7793 model predicts a slightly slower early assembly history for LMC-mass disks compared to the constant star formation rate predicted for Milky Way-mass disks, with a shift in $\tau_{50}$ to more recent times by $\sim$1 Gyr, in excellent agreement with our global LMC $\tau_{50}$$\approx$5$-$6 Gyr.  

\subsubsection{Comparison to NGC 300 \label{ngc300sect}}

Not all LMC-mass spirals host inversions in their age gradients.  NGC 300 
has an inside-out age gradient extending to at least $\sim$4$h_{r}$ measured directly from SFH fits to resolved CMDs  (\citealt{gogarten10}; also see \citealt{kang16}). 
While NGC 300 is not substantially less massive than the LMC, M33 or NGC 7793 (M$_{\star}$$\approx$ 2$\times$10$^{9}$M$_{\odot}$; \citealt{munozmateos15}), it is less luminous, and has a smaller disk scalelength, lower surface brightness and more flocculent spiral structure.  Unlike the case of NGC 7793, a model tailored to the parameters (mass, disk scalelength, and rotational velocity) of NGC 300 yielded little evidence for radial migration, supporting the idea that even at fixed stellar mass, the density of NGC 300 is too low to support the transient structures that contribute to radial migration, evidenced by its more flocculent spiral structure \citep{gogarten10,radburnsmith12,williams2403}.  
However, present data cannot rule out the possibility that R$_{inv}$ is simply located beyond 4$h_{r}$ and/or its radial age inversion is more subtle than in M33 or the LMC and remains yet to be characterized.  

Three of the four late-type LMC-mass galaxies discussed thus far show evidence for radial age inversions, suggesting that such inversions are relatively common.  All three of these galaxies show evidence for past interactions (unlike NGC 300), hinting that interaction history or environment plays a role in catalyzing sufficient radial migration to generate an inverted radial age gradient.  However, these conclusions are severely limited by sample size, so we now turn to evidence from larger, more distant samples, primarily from IFU surveys, to gain a statistical perspective on radial age inversions in late-type LMC-mass galaxies.  

\subsection{Radial Age Inversions in LMC-Mass Galaxies: Larger Samples \label{ifusect}}
 
Radial age gradient measurements of late-type LMC-mass galaxies from spectral fitting techniques support the idea that inversions in radial age profiles 
are common but not universal.  A flattening or inversion is apparent in both mass- and light-weighted radial age profiles from stacked SDSS-MaNGA spectra (of 143 galaxies each in stellar mass bins of Log$_{\rm 10}$ M$_{\star}$/M$_{\odot}$=8.6$-$9.2 and 9.2$-$9.6; \citealt{parikh21}, see their fig.~13).  In addition, among 17 star-forming 
disk galaxies with PHANGS-MUSE IFU data \citep{pessa23}, a few of the lowest-mass galaxies (9.7$\lesssim$Log$_{\rm 10}$ M$_{\star}$/M$_{\odot}$$\lesssim$10) show inversions in their radial age profiles, appearing in both light- and mass-weighted age measurements in at least two cases.  
However, while these results from IFU surveys support the high \textit{frequency} of radial age inversions seen in resolved SFH studies (Sect.~\ref{lvsect}), they find a \textit{location} of the radial age inversion that is somewhat more internal, at R$_{inv}$/$h_{r}$$\approx$1.5$-$2.\footnote{For a purely exponential disk, $R_{e} = 1.68h_{r}$, while \citet{leroy21} find that star-forming galaxies are slightly more concentrated, with $R_{e} = 1.41h_{r}$, so we use the approximation $R_{e}$$\approx$1.5$h_{r}$.}    
One possible resolution to any mild tension between 2$\lesssim$R$_{inv}/h_{r}$$\lesssim$3.5 from direct CMD-based SFH fits for the LMC, M33 and NGC 7793 versus 1.5$\lesssim$R$_{inv}/h_{r}$$\lesssim$2 from IFU surveys (at similar stellar mass) could simply relate to 
differences in methodology.  
For example, while SFHs from spectral fitting techniques generally agree well with those measured by fitting to resolved CMDs \citep[e.g.,][]{ruizlara18}, age and metallicity are particularly degenerate for low galaxy masses and low metallicities \citep{walcher11}, and inside-out radial age gradients recovered from IFU data can be biased towards artificially shallow slopes \citep{ibarramedel19}.   In this context it is noteworthy that simulations of more massive disks find values of R$_{inv}/h_{r}$ that are more external than those found by IFU surveys, and compatible with observed values for the LMC, M33 and NGC 7793 (R$_{inv}/h_{r}$$\sim$2.5$-$3; \citealt{sanchezblazquez09,ruizlara16a}). 

\subsection{The Indirect Role of Galaxy-Galaxy Interactions \label{interactsect}}

\begin{figure}
\gridline{\fig{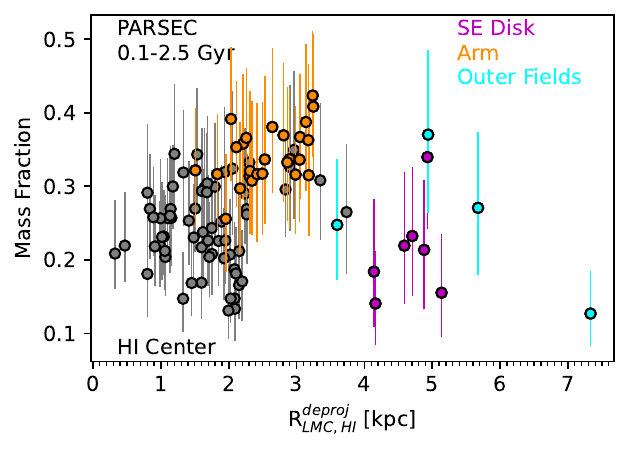}{0.6\textwidth}{}}
\caption{Fraction of total mass formed from 0.1-2.5 Gyr ago in each field, as a function of deprojected in-plane radius from the kinematic center of the LMC.  Fields located in the northern spiral arm are shown in orange (and enclosed by the orange dashed line in Fig.~\ref{map_fig}), supporting the stability of the arm on a $\sim$2.5 Gyr timescale observed by \citet{ruizlara20}.}
\label{armstability_fig}
\end{figure}

Both the ubiquity of radial age gradient inversions (Sects.~\ref{lvsect}-\ref{ifusect}) and the lack of time evolution of R$_{inv}$ (Fig.~\ref{csfh_rad_fig}) suggest that the inverted radial age gradient seen in the LMC is not specific to its interaction history with the SMC, despite a recent collision.  Such a negligible impact of the SMC on the radial age profile of the LMC may ultimately be unsurprising, since the inner LMC disk where most of our fields lie (R$^{deproj}_{LMC}$$\lesssim$6 kpc) is still well fit by straightforward inclined disk models both photometrically \citep[e.g.,][]{vdmcioni,yumiring} and kinematically \citep[e.g.,][]{vdm02,choi22,niederhofer22}.  However, interaction history can play an indirect role in setting age gradients.  Models predict that interactions can induce the spiral structure that drives migration, even in the case of a low-mass (mass ratio $\lesssim$0.1) companion, and such spiral structure can persist for Gyrs \citep{besla16,pettitt16}.  This prediction is consistent with the longevity of the LMC's spiral arm for $>$2 Gyr \citep{ruizlara20}, also supported by our results.  Fig.~\ref{armstability_fig} illustrates that fields located in the spiral arm (shown in orange, corresponding to the orange dashed lines in Fig.~\ref{map_fig}) preferentially formed more of their stellar mass over the last 2.5 Gyr than fields at either larger radii or fields closer to the LMC center where \citet{ruizlara20} were unable to probe due to crowding.    

More generally, since radial migration is driven by resonant scattering off of spiral arms \citep{sellwoodbinney}, the strength of radial migration and the presence of the resulting radial age inversion may correlate primarily with the ability of a galaxy to support spiral structure based on its density.  For example, despite nearly identical stellar masses, the lack of a radial age inversion in NGC 300 can be explained by its inability to drive substantial radial migration due to a lower density \citep{gogarten10}.  Conversely, the LMC and M33 show clear radial inversions accompanied by well-defined spiral arms.  For NGC 7793, despite its more flocculent structure, it has a higher density than NGC 300, so a model accounting for this difference in density invokes strong radial migration \citep{radburnsmith12}, resulting in an inverted radial age gradient.   

Over shorter timescales (a few hundred Myr), 
interactions with a satellite can also impact radial migration. 
In collisionless N-body simulations of the Milky Way-Sagittarius interaction, 
stars in the outer disk are preferentially heated during pericentric passages of the satellite, inducing radial excursions of nearly 50\% in galactocentric radius, although such disk heating is predicted to dissipate after a few hundred Myr \citep{carr22}.  In the LMC, this is almost certainly the case, since a comparison of the observed disk heating in the LMC (quantified via residual proper motion dispersion) with N-body SPH simulations of the LMC-SMC interaction \citep{besla12} places an upper limit of $\sim$250 Myr on the time since the most recent LMC-SMC collision \citep{choi22}.

\subsection{Disk Size Evolution \label{scalelengthsect}}

\begin{figure}
\gridline{\fig{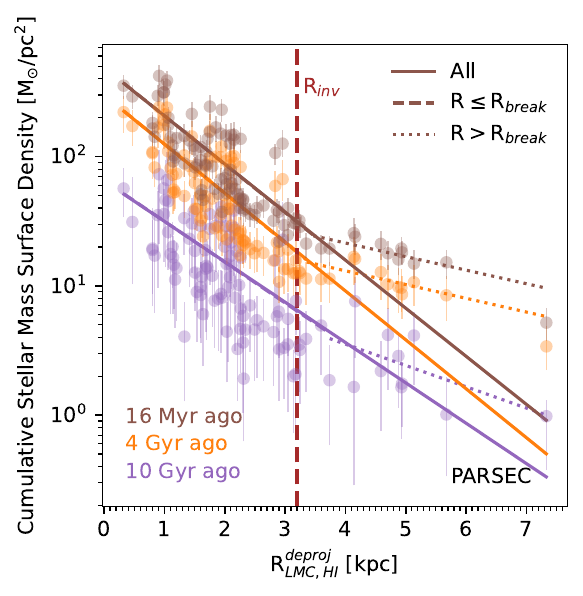}{0.34\textwidth}{}
          \fig{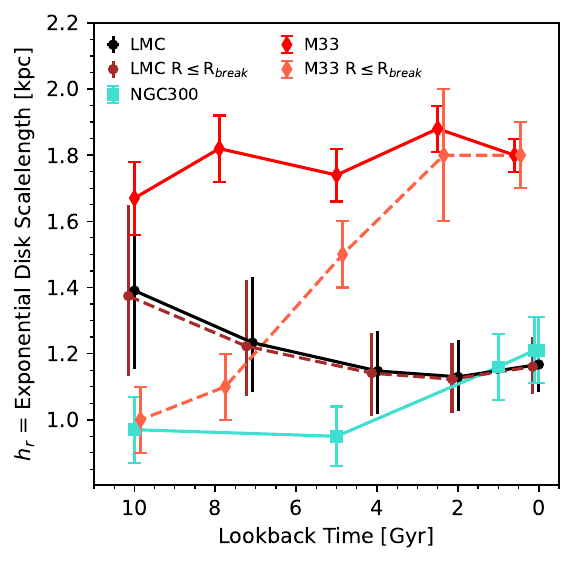}{0.33\textwidth}{}
          \fig{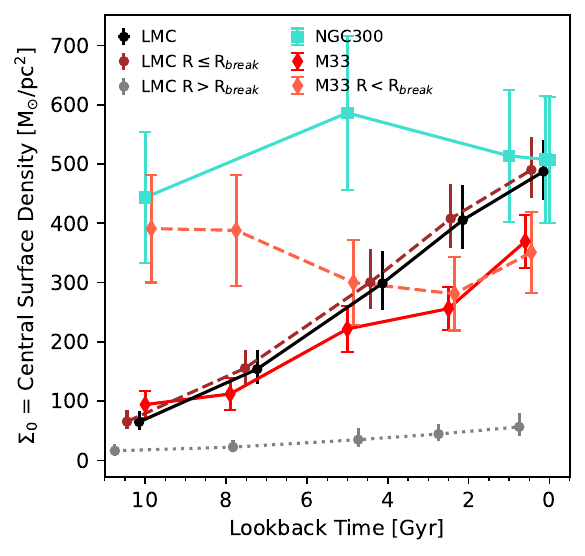}{0.33\textwidth}{}}
\caption{\textbf{Left:} Radial profiles of cumulative stellar mass surface density color-coded by three example lookback times, including the present day.  Exponential fits to fields inside R$_{break}$ (dashed lines) are essentially identical to fits to the full sample (solid lines), but fits to the fields at R$>$R$_{break}$ (dotted lines) are flatter.  \textbf{Center:} The evolution of the best-fit exponential disk scalelength $h_{r}$ with lookback time, shown for fits to the full sample (solid line) as well as fields inside R$_{break}$ (dashed line), offset by intervals of 0.15 Gyr to avoid overlapping errorbars.  Fits to LMC fields with R$>$R$_{break}$ are not shown since they all yield $h_{r}$$>$2.6 kpc due to the flattening of the profile seen in the left panel, with large ($>$25\%) uncertainties (see Table \ref{scalelentab}). Values for M33 \citep{williams09} and NGC 300 \citep{gogarten10} are overplotted.  When viewed globally, the LMC, M33 and NGC300 all show approximately constant $h_{r}$ over time, but unlike the LMC, the inner disk of M33 has grown over time, possibly due to its different interaction history.  \textbf{Right:} The evolution of the best-fitting central surface density, using the same symbols as in the middle panel.  Trends for the LMC are similar to M33 when fit as a single sample across R$_{break}$, but the \textit{outer} region of the LMC disk and the \textit{inner} region of M33's disk deviate from these trends, also possibly reflecting different interaction histories.}
\label{scalelength_fig}
\end{figure}

The evolution of the size of the LMC disk over time can provide additional insight into its assembly history and any potential impact of the interaction with the SMC.  Specifically, a comparison with M33 can reveal whether similar radial age gradient inversions are accompanied by similar trends in disk size evolution despite differences in recent interaction history.  We examine the time evolution of the disk size by integrating the SFHs for each of our LMC fields up to different lookback times, and fit exponential profiles as a function of R$^{deproj}_{LMC}$ at each lookback time, parameterized by disk scalelength $h_{r}$ and central density $\Sigma_{0}$.  Although caution is required when interpreting the results at face value in the presence of substantial radial migration \citep[e.g.,][]{frankel18}, the trends we find provide observables that can be compared across galaxies as well as to simulations.
In the left panel of Fig.~\ref{scalelength_fig}, we plot cumulative stellar mass surface density profiles as a function of R$_{LMC}^{deproj}$ for three example lookback times.  For each color-coded lookback time, a single exponential fit to the full sample is shown as a solid line.  Additionally, we also perform a two-part broken exponential fit, with the break radius R$_{break}$ allowed to float as a free parameter.  These piecewise fits, shown as dashed (R$\leq$R$_{break}$) and dotted (R$>$R$_{break}$) lines, are motivated by the apparent flattening seen beyond R$_{inv}$, and allow direct comparison to the broken exponential profile seen for 
M33 \citep{williams09}.  

In the LMC, the inner portion of the two-part fits (dashed lines in Fig.~\ref{scalelength_fig}) are essentially identical to the single full-sample fits (solid lines in Fig.~\ref{scalelength_fig}), due in part to the small fraction (15\%) of fields located beyond R$_{inv}$.  They show no significant evolution in the disk scalelength $h_{r}$ (middle panel of Fig.~\ref{scalelength_fig}), although uncertainties allow for a decreasing $h_{r}$ at younger ages (reported by \citealt{frankel24}).  Both the lack of evolution of $h_{r}$ over time as well as the steadily increasing central density (right-hand panel of Fig.~\ref{scalelength_fig}) are similar to trends seen for M33, although there is a hint of an inflection in the LMC $h_{r}$ values around the time the LMC and SMC likely began interacting several Gyr ago (more dense sampling of lookback times did not further constrain this feature).  Our best-fit near-present-day ($\sim$16 Myr, the most recent lookback time available across all three of our assumed evolutionary models) $h_{r}=$1.17$^{+0.07}_{-0.09}$ (or 1.45$^{+0.08}_{-0.07}$ when assuming the isopleth-based LMC center location; see Appendix \ref{altmodelsect}), is somewhat lower than $h_{r} = 1.667^{+0.002}_{-0.002}$ obtained from fitting a disk+bar model to the spatial distribution of LMC red clump stars (\citealt{yumiring}; also see \citealt{nidever19}).  However, this discrepancy is probably due to some combination of the apparent sensitivity of the results to the chosen center location, our non-uniform radial sampling, the residual features deviating from the rotating disk model and the use of a different tracer population (present-day red clump stars) by \citealt{yumiring}).

The outer portion of the two-part fits to the LMC disk's cumulative mass surface density (dotted lines in the left panel of Fig.~\ref{scalelength_fig}) reveals a flattening in the radial profiles (albeit at $\sim$2$-$3$\sigma$ given small number statistics).  A similar flattening is also seen in the radial stellar density profiles of \citet{vdm02}, although our fields do not probe the preferentially old ring-like stellar overdensity associated with this feature and characterized by \citet{yumiring}, with the possible exception of the four outer fields, outside the spatial coverage of their imaging.  In M33, a break is also seen in the cumulative mass surface density, but in M33 it is the fields \textit{inside} R$_{break}$ that show trends different from the full sample, and consistent with analytical models of disk growth, showing an increase in $h_{r}$ and decrease in $\Sigma_{0}$ of the inner disk over time \citep[e.g.,][]{mo98}.  Conversely, in the LMC, it is the fields \textit{outside} R$_{break}$ that show a different behavior than the full sample, with a central density that increases much more slowly than the inner disk (although inconsistent with the flattening predicted by the \citealt{mo98} analytical disk growth models at $\sim$2$\sigma$).  Given that the LMC and M33 have similar radial age profiles (Fig.~\ref{csfh_rad_fig}), the marked difference in the size evolution of their disks inside versus outside R$_{inv}$$\approx$R$_{break}$ suggests that the size evolution of the disk \textit{within a galaxy} may be a particularly sensitive probe of detailed interaction history.

For both the LMC and M33, R$_{break}$ appears to be approximately colocated with the age gradient inversion R$_{inv}$ (this was assumed to be the case for M33 in the \citealt{williams09} analysis, while we float R$_{break}$ as a free parameter in our fits).  For massive disks, some models predict the colocation of R$_{break}$ and R$_{inv}$ as a consequence of radial migration \citep[e.g.,][]{roskar08,roskar12}.  However, even the miniscule comparison sample of LMC-mass galaxies with age gradients measured directly from SFH fits (Sects.~\ref{m33sect}$-$\ref{ngc300sect}) reveals that radial density (and/or surface brightness) profile breaks are uncorrelated with radial age profile breaks: NGC 7793 has a density profile that breaks to a steeper slope like M33, but with R$_{break}$$>$R$_{inv}$ \citep{radburnsmith12}, while NGC 300 has a double-broken density profile with slope changes in the opposite sense as the LMC \citep{jang20} and no age gradient inversion \citep{gogarten10}.  In fact, both observations \citep{ruizlara16b,tang20} and simulations \citep{radburnsmith12,ruizlara16a,ruizlara17} demonstrate that the relationship between breaks in age gradients versus breaks in surface brightness profiles (or lack thereof) is complex and multifaceted.

\begin{deluxetable}{cccccccc}
\tablecaption{Time Evolution of the Disk Scalelength and Central Density\label{scalelentab}}
\tablehead{\colhead{} & \multicolumn{2}{c}{All Fields} & \multicolumn{2}{c}{R$\leq$R$_{break}$} & \multicolumn{2}{c}{R$>$R$_{break}$} & \colhead{} \\ \colhead{Lookback Time} & \colhead{$h_{r}$} & \colhead{$\Sigma_{0}$} & \colhead{$h_{r}$} & \colhead{$\Sigma_{0}$} & \colhead{$h_{r}$} & \colhead{$\Sigma_{0}$} & \colhead{R$_{break}$} \\ \colhead{Gyr} & \colhead{kpc} & \colhead{M$_{\odot}$/pc$^{2}$} & \colhead{kpc} & \colhead{M$_{\odot}$/pc$^{2}$} & \colhead{kpc} & \colhead{M$_{\odot}$/pc$^{2}$} & \colhead{kpc}}   
\startdata
10.0 & 1.39$^{+0.28}_{-0.23}$ & 64$^{+19}_{-12}$ & 1.37$^{+0.27}_{-0.24}$ & 65$^{+19}_{-12}$ & 2.63$^{+1.70}_{-0.63}$ & 16$^{+11}_{-9}$ & 3.73$^{+3.35}_{-0.49}$ \\
7.1 & 1.23$^{+0.19}_{-0.14}$ & 153$^{+31}_{-25}$ & 1.22$^{+0.19}_{-0.14}$ & 155$^{+31}_{-26}$ & 3.99$^{+2.51}_{-1.15}$ & 22$^{+12}_{-8}$ & 3.35$^{+0.41}_{-0.17}$ \\
4.0 & 1.14$^{+0.12}_{-0.13}$ & 298$^{+53}_{-45}$ & 1.14$^{+0.12}_{-0.12}$ & 300$^{+55}_{-45}$ & 4.06$^{+2.36}_{-1.16}$ & 34$^{+19}_{-12}$ & 3.35$^{+0.35}_{-0.13}$ \\
2.0 & 1.12$^{+0.11}_{-0.10}$ & 405$^{+58}_{-48}$ & 1.12$^{+0.11}_{-0.10}$ & 407$^{+58}_{-49}$ & 4.29$^{+1.87}_{-1.05}$ & 44$^{+17}_{-12}$ & 3.35$^{+0.25}_{-0.11}$ \\
0.016 & 1.16$^{+0.09}_{-0.08}$ & 487$^{+53}_{-48}$ & 1.16$^{+0.09}_{-0.08}$ & 489$^{+54}_{-46}$ & 4.12$^{+1.48}_{-0.97}$ & 56$^{+23}_{-15}$ & 3.46$^{+0.24}_{-0.15}$ \\
\enddata
\tablecomments{These fits assume PARSEC evolutionary models, and analogous results assuming alternate choices of stellar evolutionary models and LMC center location are given in Appendix \ref{altmodelsect}.}
\end{deluxetable}

\section{Summary and Future Prospects \label{futuresect}}

We have fit SFHs to two-filter imaging of 111 individual \textit{HST} pointings towards the LMC disk, and self-consistently measured lifetime CSFH metrics ($\tau_{50}$, $\tau_{75}$, $\tau_{90}$; Table \ref{tautab}).  Despite covering a small total projected area on the sky (0.2 deg$^{2}$$\approx$0.157 kpc$^{2}$), leveraging all of our individual fields together 
provides global star formation histories with a temporal precision comparable to contiguous, ground-based imaging surveys \citep{ruizlara20,mazzi21}, and an AMR precision comparable to results incorporating wide-field multi-object spectroscopic surveys \citep{povick}.  Our results can be summarized as follows:
\begin{enumerate}
    \item The radial age profile of the LMC (i.e., CSFH metrics versus deprojected in-plane radius) reveals an inversion, with a present-day inside-out (outside-in) age gradient in the inner (outer) disk.  The location of the inversion remains fixed at at R$_{inv}/h_{r}$$\sim$2 over lookback times sampling the full duration of the LMC-SMC interaction and perhaps earlier (Fig.~\ref{csfh_rad_fig}).  
    
    \item In the inner disk, radial age gradients are more statistically significant for the northern part of the LMC disk, but field-to-field variations are substantial, restricting the magnitude of linear and rank correlation coefficients to $<$0.6 over the lookback times we explore (Sect.~\ref{innerdisksect}, Table \ref{csfhtab}).  Such azimuthal scatter in age metrics at fixed galactocentric radius, predicted by models of more massive disks \citep{sanchezblazquez09,halle18,carr22}, is a natural consequence of non-axisymmetric structure, and likely contributes to the diversity of radial age profiles seen in the outer disks ($\gtrsim$$R_{e}$) of LMC-mass galaxies \citep[e.g.,][]{parikh21,pessa23}.   
    
    \item Inversions in radial age profiles are common but not universal in late-type LMC-mass galaxies, based on both SFH fits to CMDs of Local Volume galaxies and results from IFU surveys (Sect.~\ref{lvsect}-\ref{ifusect}).  We argue that inversions in radial age gradients are driven by stellar radial migration, supported by simulations of both LMC-mass disks \citep{radburnsmith12} and Milky Way-mass disks \citep[e.g.,][]{roskar08,roskar12}.  
    
    \item Galaxy-galaxy interactions play an indirect role, since they can both drive the structures that support migration over timescales of Gyrs as well as boost migration over shorter timescales (hundreds of Myrs) by dynamically heating the outer disk.  In the LMC, both of these mechanisms are likely currently operating, and minor galaxy-to-galaxy differences in the shape of inverted radial age profiles may reflect differences in the details of their interaction histories (Sect.~\ref{interactsect}).  
    
\end{enumerate}

A set of direct (i.e., CMD-based) homogenous radial age profile measurements in dwarf galaxies across different environments and morphologies at fixed stellar mass 
would further quantify the role of interactions in setting age gradients, and in turn, the spatially resolved mass assembly histories of dwarfs.  Currently, our ability to compare R$_{inv}$/$h_{r}$ values across all three galaxies where an age inversion is detected via SFH fitting (LMC, M33 and NGC 7793) is limited by the quality of available data (i.e., depth and/or field placement), and we cannot strongly exclude the possibility that all three galaxies have nearly identical locations of R$_{inv}/h_{r} \sim$2.5.  Meanwhile, comparisons to more distant targets may be hampered by the non-negligible impact of viewing angle on radial stellar population trends in dwarf galaxies \citep[e.g.,][]{mostoghiu18,graus}.  Deep imaging, beyond providing improved measurements of R$_{inv}$ (in terms of both quantity and quality) and any spatial dependence of SFH trends, are needed to probe the \textit{time evolution} of radial age gradients, which serves as a powerful diagnostic in the case of the LMC where independent orbital constraints are available.  Fortunately, improved results for M33 are forthcoming from the PHATTER survey \citep{phatter}, while new and forthcoming observational facilities bring spatially resolved lifetime SFHs beyond a few Mpc within reach.  Such an expanded dataset would allow tests of predictions from N-body simulations of massive disks that have yet to be validated at LMC-like (or lower) masses.  These predictions include correlations between migration strength and the radial location of arm/bar resonances and corotation \citep{roskar12,halle15,martinezbautista21} as well as quadropole azimuthal metallicity variations induced by an orbiting satellite \citep{carr22}.

An ensemble of direct, homogenous radial age profile measurements would also provide a new set of empirical constraints for the latest zoom-in cosmological simulations.  For example, \citet{yuxi23} present a method to independently constrain radial migration strength in external galaxies (including the LMC) using resolved stellar photometry and abundance information.  The simulations they analyze predict that migration strength is independent of galaxy mass when normalized to disk scalelength, and a larger dataset could test this prediction.  Furthermore, simulations have thus far parameterized age gradients in dwarf galaxies using linear relations \citep{graus}.  In light of the results discussed here, useful higher-order comparisons can be made against detailed simulations of galaxies across a broad swath of mass-morphology-environment parameter space.  Such model-data comparisons are particularly timely as the latest simulations are pushing to ever-increasing mass resolution, allowing comparisons to low-mass galaxies that are simulated in a cosmological context and successfully reproduce observed global scaling relations \citep[e.g.][]{applebaum}.     

\begin{acknowledgements}

It is a pleasure to thank Pol Massana, Tom{\'a}s Ruiz-Lara, Lara Monteagudo, Dan Weisz and Carme Gallart for sharing SFH results.  

Based on observations with the NASA/ESA Hubble Space Telescope obtained at the Space Telescope Science Institute, which is operated by the Association of Universities for Research in Astronomy, Incorporated, under NASA contract NAS 5-26555.  
Support for programs HST GO-15891, GO-16235 and GO-16786 were provided by NASA through a grant from the Space Telescope Science Institute, which is operated by the Association of Universities for Research in Astronomy, Inc., under NASA contract NAS 5-26555.  R.~E.~C. acknowledges support from Rutgers the State University of New Jersey.  

\end{acknowledgements}

\facilities{HST (ACS, WFC3, WFPC2)}
\software{Astropy \citep{astropy}, matplotlib \citep{matplotlib}, numpy \citep{numpy}, emcee \citep{emcee}}

\clearpage

\appendix

\section{Clusters Removed From Target Fields \label{clusterapp}}

In Table \ref{clustab}, we list all stellar clusters from the \citet{bica08} catalog (with their Object Type containing a C) overlapping with any of our target fields.  The name of each cluster in the SIMBAD database is given in the first column, followed by any alternate names.  We conservatively assumed a cluster radius corresponding to the catalog semi-major axis (given in the 5th column) and removed all stars within this distance from the cluster center (given in the 3rd and 4th columns).  For each cluster, the affected field(s) are given in the last column.  

\begin{deluxetable}{lcccccc}[h!]
\tablecaption{Known Clusters Excluded from Photometric Catalogs \label{clustab}}
\tablehead{
\colhead{Cluster} & \colhead{Other Names} & \colhead{RA (J2000)} & \colhead{Dec (J2000)} & \colhead{Radius} & \colhead{Field} \\ \colhead{} & \colhead{} & \colhead{$^{\circ}$} & \colhead{$^{\circ}$} & \colhead{$\arcmin$} & \colhead{}} 
\startdata
KMHK 988 & SL522 & 82.654167 & -67.186111 & 1.0 & LMC\_2 \\ 
GKK2003 O239 & &  82.925000  &  -67.316667 & 1.5 & LMC\_3 \\ 
BSDL 2161 & & 82.954167  & -67.296389 & 0.6 & LMC\_3 \\ 
KMHK 900 & & 81.620833 & -67.699444 & 0.5 & LMC\_15 \\ 
BSDL 1705 & & 81.625000 & -67.695556 & 0.3 & LMC\_15 \\ 
KMHK 534 & SL224 & 76.425000 & -70.325000 & 0.9 & LMC\_29 \\ 
NGC 1791 & SL155, KMHK391 & 74.775000 & -70.168056 & 1.2 & LMC\_32 \\ 
KMHK 294& SL110 & 73.933333  &  -69.156667 & 0.8 & LMC\_38 \\ 
H88 47 & & 74.250000 & -66.335556 & 0.6 & LMC\_M6 \\ 
NGC 1787 & ESO85-SC31, KMHK445, SL178 & 75.437500 & -65.823889 & 1.1 & LMC\_M9 \\
BSDL 2583 & & 84.204167 & -69.430556 & 0.6 & LMC\_M13 \\ 
GKK2003 O100 & & 82.750000 & -70.916667 & 1.5 & LMC\_M14 \\ 
BSDL 2144 & & 82.779167 & -71.130556  & 0.5 & LMC\_M18 \\ 
NGC 2102 & ESO57-SC29, KMHK1264, SL665  & -69.488056 & 85.583333 & 0.8 & LMC\_M23 \\ 
BSDL 2351 & & 83.533333 & -67.426389 & 0.5 & LMC\_M29 \\ 
KMHK 448 & SL176 ,LOGLE15 & 75.370833 & -68.712222 & 1.0 & LMC\_W39 \\ 
KMHK 467  &  HS92 & 75.575000 & -68.545000 & 0.6 & LMC\_W36 \\ 
NGC 1810 & ESO85-SC35,KMHK475, SL194 & 75.845833 & -66.381667 & 1.2 & LMC\_W61, W63 \\ 
H88 137 & LOGLE117, HS119 & 76.745833 & -69.320000 & 1.0 & LMC\_W41 \\ 
GKK2003 O69 & & 77.575000 & -71.250000 & 1.5 & LMC\_W27 \\ 
KMHK634 & H88-195 & 77.862500 & -65.449444 & 0.7 &  LMC\_W48 \\ 
KMHK 707 & & 79.054167 & -70.502778 & 0.8 & LMC\_W17 \\ 
BSDL 1099 &  & 79.120833 & -70.506389 & 0.5 & LMC\_W17 \\ 
H88 279 & LOGLE361 & 80.008333 & -69.261111 & 0.6 & LMC\_W19 \\ 
OGLE-CL LMC 363 & & 80.016667 & -69.265278 & 0.3 & LMC\_W19 \\ 
H88 281 & & 80.087500 & -69.246667 & 0.7 & LMC\_W19 \\ 
OGLE-CL LMC 413 & HS255 &  80.658333 & -69.744444 & 0.9 & LMC\_W26 \\ 
OGLE-CL LMC 419 & HS266  & 80.854167 & -69.835278 & 1.4 & LMC\_W24 \\ 
OGLE-CL LMC 425 & & 80.904167 & -69.821944 & 0.5 & LMC\_W24 \\ 
OGLE-CL LMC 431 & HS275 & 81.083333 & -69.773889 & 0.7 & LMC\_W2 \\ 
OGLE-CL LMC 434 & & 81.100000 & -69.780000 & 0.3 & LMC\_W2 \\ 
OGLE-CL LMC 442 & HS280 & 81.220833 & -69.829722 & 0.9 & LMC\_W20 \\ 
KMHK 906 & SL472 & 81.558333 & -70.381111 & 1.0 & LMC\_W21 \\ 
BSDL 2487 & & 83.858333 & -69.268056 & 0.5 & LMC\_W9, W10, W12, W14 \\ 
SN1987A & & 83.866750 & -69.269742 & 0.5 & LMC\_W9, W10, W12, W14 \\ 
KMHK 1566 & H88-377, BM189 & 89.583333 & -68.355000 & 0.8 & LMC\_W8 \\ 
KMHK 1704 & SL869, LW441 & 93.670833 & -69.801944 & 1.6 & LMC\_W53 \\ 
NGC 2041 & ESO86-SC15, KMHK1132, SL605  &    84.116667  &  -66.990556 & 2.6 & LMC\_40  \\ 
KMHK 931 & HS309 &   82.070833 &   -66.925556 &  1.1 & LMC\_45  \\ 
BSDL 2953   &  &   86.412500  &  -67.071667 &   0.6 & LMC\_49  \\ 
BSDL 137 & & 73.154167  &  -68.023889 & 0.5 & LMC\_59 \\ 
\enddata
\end{deluxetable}

\section{Tests of Consistency \label{amrsect}}

We performed several internal and external consistency checks to validate our SFH fitting results:

\begin{enumerate}
\item We performed an internal consistency check exploiting serendipitous imaging of a pair of fields with near-total ($\gtrsim$90\%) spatial overlap (LMC\_20 and LMC\_24).  Bearing in mind that photometry, artificial star tests and SFH fitting were performed independently for each field, their CSFH and AMR show good agreement to within their uncertainties (see Fig.~\ref{comp_overlap_fig}).

\begin{figure}
\gridline{\fig{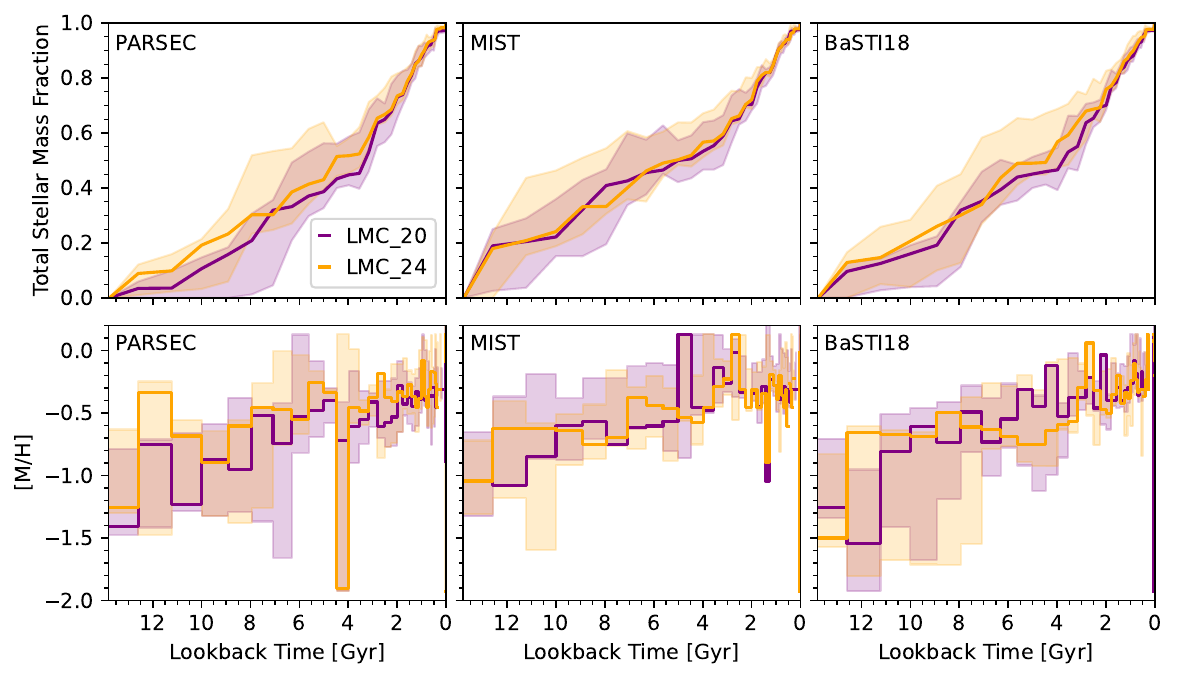}{0.8\textwidth}{}}
\caption{Comparison of CSFH (top row) and AMR (bottom row) for two fields (LMC\_20 and LMC\_24)  which were observed, reduced and analyzed separately but have $>$90\% spatial overlap.  Both the CSFHs and AMRs agree to within their 1$\sigma$ uncertainties, represented by the shaded regions, for all three assumed isochrone libraries (shown from left to right).}
\label{comp_overlap_fig}
\end{figure}

\item We also perform a second internal consistency check to test for any dependence on observing setup, comparing the combined CSFH from the archival WFPC2 fields (based on WFPC2 F555W+F814W imaging) against the combined CSFH from the Scylla and METAL fields (based on WFC3/UVIS F475W+F814W imaging).  The two CSFHs are compared for each assumed evolutionary model set in Fig.~\ref{comp_uvis_wfpc2_fig}, demonstrating that they are essentially identical.  

\begin{figure}
\gridline{\fig{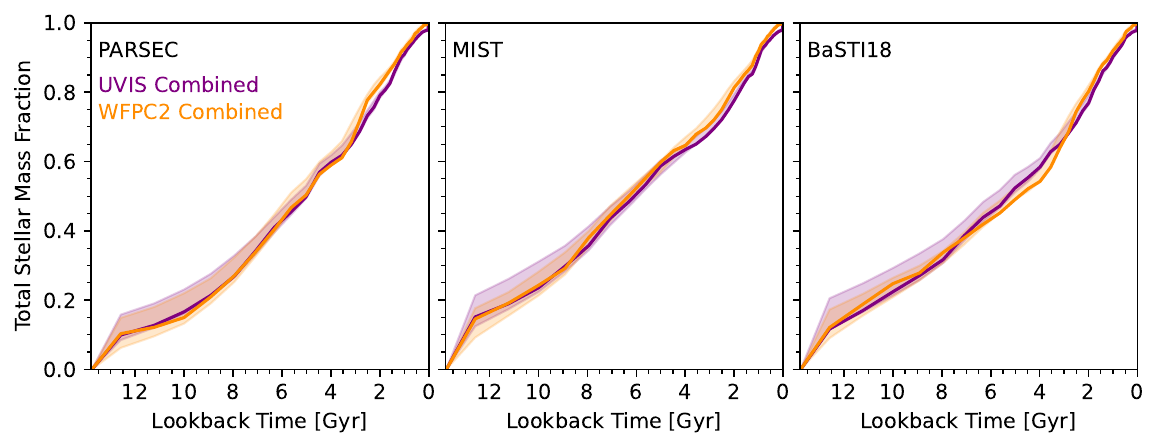}{0.8\textwidth}{}}
\caption{Comparison of combined CSFH for all WFC3/UVIS (i.e., Scylla and METAL) fields (purple) and all archival WFPC2 LGSPA fields (orange), showing no apparent dependence of recovered SFHs on observing setup for any of the three assumed evolutionary libraries (shown from left to right).}
\label{comp_uvis_wfpc2_fig}
\end{figure}

\item We compare our combined global LMC AMR to independent literature measurements in Fig.~\ref{combined_AMR_fig}.  
Bearing in mind that we placed no \textit{a priori} constraints whatsoever on the allowed AMR, our results are in good agreement with independent spectroscopic values from the bar and disk fields analyzed by \citet{carrera08}, 
overplotted in red and orange in Fig.~\ref{combined_AMR_fig}.  Our results are also in good agreement with 
the LMC field star AMR from \citet{pg13} based on Washington photometry,  
and the LMC cluster AMR from a homogenous cluster photometric analysis by \citet{perren17}.  This is especially true in light of the (albeit shallow) negative radial metallicity gradient in the LMC \citep[e.g.,][]{cioni09,choudhury21,grady21,skowron}, which could conspire with differences in field placement and radial coverage, particularly for the \citet{pg13} and the \citet{carrera08} disk fields, which extend to $\sim$8$^{\circ}$ from the LMC center.  
While there are several other cluster compilations available in the recent literature \citep[e.g.,][]{bica08,glatt10,nayak,sitek16}, including some with both age and metallicity information \citep{palma16,narloch22},  
a more detailed comparison to results from our lifetime SFH fits is hampered by the 
age gap at $\sim$4$-$10 Gyr seen in the age distribution of LMC clusters (e.g., \citealt{jensen88,dacosta91,ol91,geisler97}, but see \citealt{gatto22} and \citealt{piatti22}).    
Our AMR is also in excellent agreement with \citet{povick}, combining spectroscopically-derived stellar parameters with multi-band photometry and evolutionary models to calculate star-by-star ages for a sample comprised mostly of red giants.

\begin{figure}
\gridline{\fig{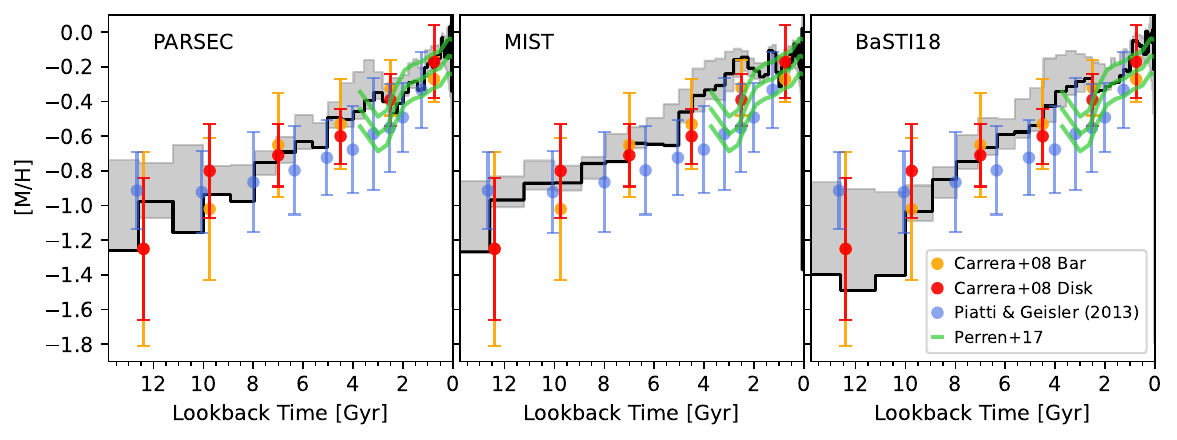}{0.85\textwidth}{}}
\caption{Best-fit age-metallicity relation from all of our LMC fields combined.  Each panel compares our best-fit age-metallicity relation assuming one of the stellar evolutionary libraries (black lines, with uncertainties shown with grey shading) with literature values.  These include LMC field disk and bar AMRs from \citet{carrera08}, shown in red and orange respectively, and the LMC field AMR based on Washington photometry from \citet{pg13} shown in blue.   
The LMC cluster AMR from \citet{perren17} is overplotted with its uncertainty interval using green lines.  Our recovered global LMC AMR is in good agreement with previous measurements, especially given differences in spatial coverage of the LMC.}
\label{combined_AMR_fig}
\end{figure}

\item In Fig.~\ref{compsfhlitfig}, we compare our combined SFH to ground-based results from the SMASH \citep{ruizlara20} and VMC \citep{mazzi21} surveys, as well as \citet{weisz13}, who analyzed a subset of eight of the LGSPA WFPC2 fields.  

\textbf{SMASH:} Star formation histories from SMASH were presented by \citet{ruizlara20} for two spatial regions of the LMC, denoted LMC North (placed to cover the northern spiral arm) and LMC South (placed to cover a comparable area on the opposite side of the LMC).  Although our field placement does not allow a useful comparison to their LMC South region, we compare our SFHs to their results from the LMC North region by normalizing our star formation rates to projected physical area, and combining our results for only the fields that lie in their LMC North footprint, indicated by the dashed orange line in Fig.~\ref{map_fig}.\footnote{For comparison to the SMASH results from their LMC North region \citep{ruizlara20}, we combined results from the following fields: LMC\_2, LMC\_3, LMC\_9, LMC\_16, LMC\_17, LMC\_20, LMC\_34, LMC\_35, LMC\_37, LMC\_39, LMC\_40, LMC\_45, LMC\_49, LMC\_50, LMC\_57, LMC\_59, LMC\_M1, LMC\_M3, LMC\_M6, LMC\_M7, LMC\_M9, LMC\_M28, LMC\_M29, LMC\_W48, LMC\_W49, LMC\_W61, LMC\_W63.}    
We compare our SFH to theirs in both cumulative and histogram form in the top row of Fig.~\ref{compsfhlitfig}.  Despite the fact that we sample a much smaller area (by more than two orders of magnitude), the depth of our observations reveals all of the features they detect in the lifetime SFH of the LMC's northern spiral arm.  These features include enhancements in star formation at very early ($>$12 Gyr) lookback times followed by a decreased but statistically significant star formation rate, enhancements at $\sim$5 and $\sim$8 Gyr mentioned by \citet{ruizlara20}, and an overall increase in star formation activity beginning $\sim$3$-$3.5 Gyr ago \citep[also seen by e.g.,][]{hz09,weisz13,monteagudo,mazzi21}, including multiple bursts discussed by \citet{massana22}.  Here we mention only briefly in a general context that the exact timing and amplitude of enhancements in star formation rate appear to be mildly model-dependent, and an analysis of the timing and spatial distribution of star formation enhancements in the LMC in the context of the LMC-SMC interaction will be presented elsewhere (C.~Burhenne et al., in prep.).  Furthermore, slight scale factor differences between our star formation rate and the results from \citet{ruizlara20} in the upper right-hand panel of Fig.~\ref{compsfhlitfig} are expected from non-identical radial sampling of the LMC disk given the exponential radial profile of cumulative stellar mass surface density (see Sect.~\ref{scalelengthsect}).

\textbf{VMC:} Our global SFH from combining all of our LMC fields is compared to SFH results from VMC survey $JK_{S}$ imaging \citep{mazzi21} in the middle row of Fig.~\ref{compsfhlitfig}, again in both cumulative and histogram form for all three evolutionary models we assume.  To perform as direct a comparison as possible, we calculate the VMC-based results including only the \citet{mazzi21} tile subregions containing one or more Scylla fields, weighting by the number of Scylla fields present in each subregion, normalizing the \citet{mazzi21} star formation rates to the projected area listed for each VMC tile subregion.  Again, we find decent agreement in general trends, and the slightly larger discrepancy 
compared to deeper SMASH imaging suggests some influence from either  shallower imaging, a different wavelength regime, and/or slightly different best-fit distances.  Meanwhile, \citet{mazzi21} underscore that the VMC results are, in turn, an improvement on SFHs from shallower data by \citet{hz09}.

\textbf{Weisz et al.~(2013):} In the bottom row of Fig.~\ref{compsfhlitfig}, we compare our combined global LMC CSFH to the CSFH from \citet{weisz13}, calculated by statistically combining SFH results for eight of the WFPC2 LGSPA fields (see Fig.~\ref{map_fig}).  Again, we find good agreement, especially in light of their use of a fixed line-of-sight distance across the LMC.  
  
\begin{figure}
\gridline{\fig{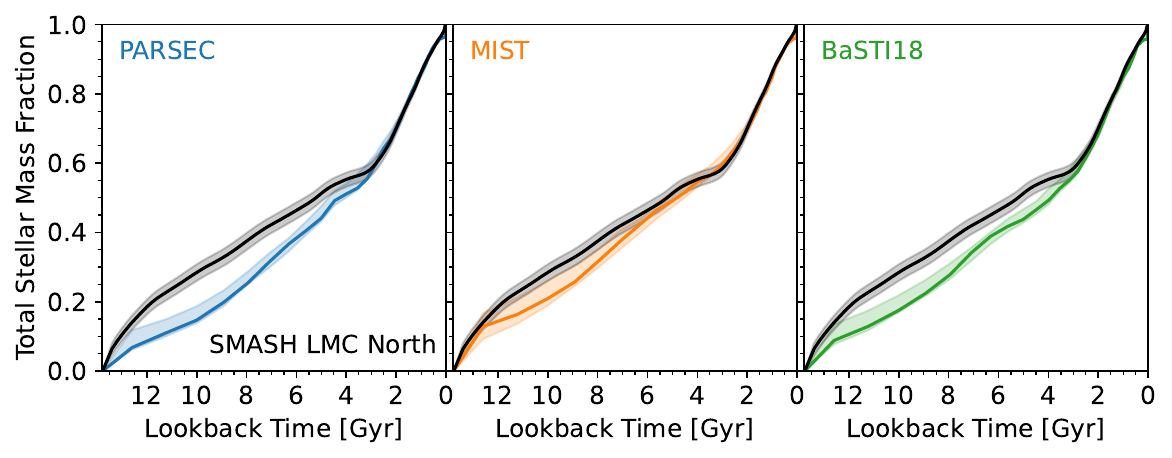}{0.495\textwidth}{}
          \fig{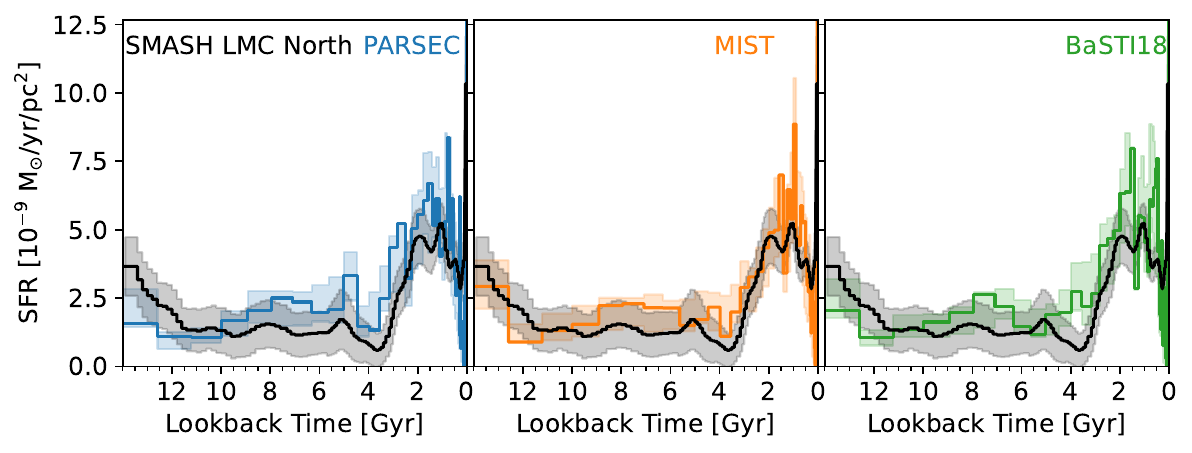}{0.495\textwidth}{}}
\vspace{-0.1cm}
\gridline{\fig{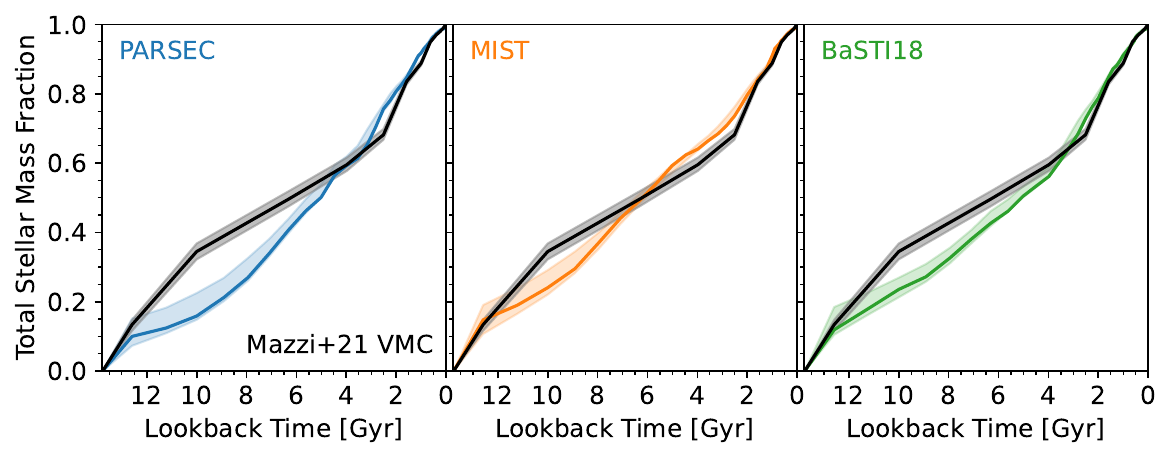}{0.495\textwidth}{}
          \fig{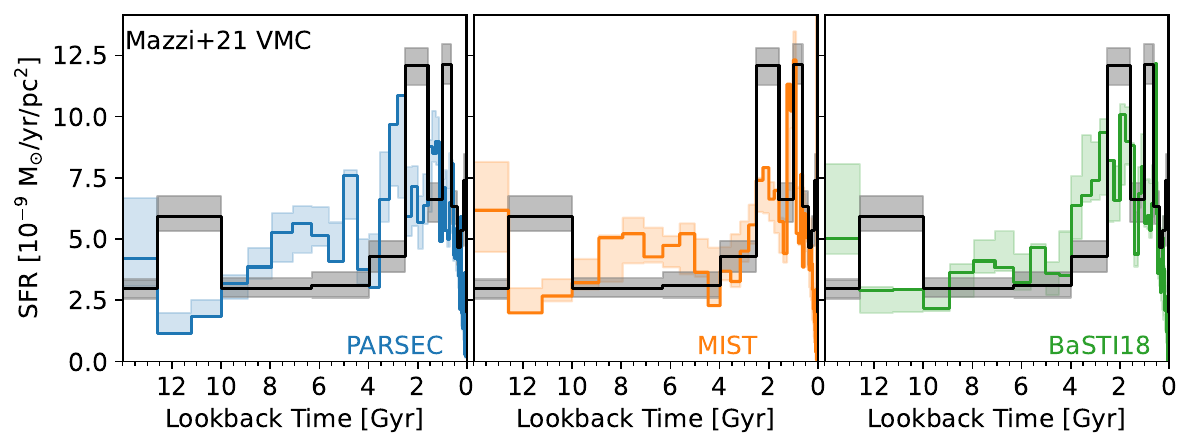}{0.495\textwidth}{}}
\vspace{-0.1cm}
\gridline{\fig{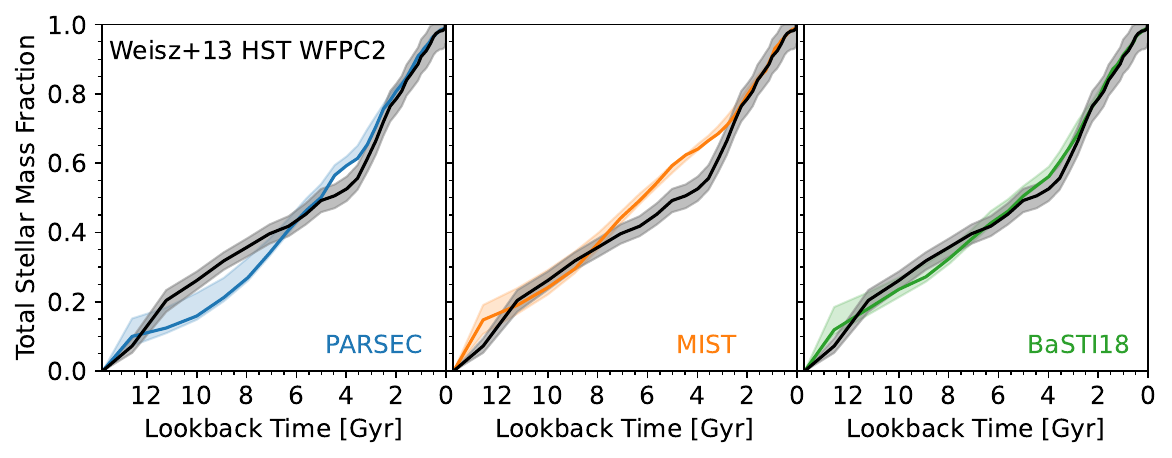}{0.495\textwidth}{}}
\caption{Comparison of our global SFHs to recent literature results.  For the top two rows, the left-hand column compares CSFHs and the right-hand column compares star formation rate versus lookback time.  In each plot, the three panels represent the three different evolutionary models we assume for our fits.  \textbf{Top Row:} Our results are compared with SMASH results from their LMC North area \citep{ruizlara20}, including only our fields in the same footprint (see text for details).  \textbf{Middle Row:} Comparison with VMC results from \citet{mazzi21}, normalized to our field distribution over the VMC tiles.  \textbf{Bottom Row:}  Comparison of our global combined CSFH with the LMC CSFH from \citet{weisz13}.}  
\label{compsfhlitfig}
\end{figure}

\end{enumerate}

\section{Sensitivity of Radial Gradients to Assumed Center Location and Stellar Evolutionary Model \label{altmodelsect}}
\begin{figure}
\gridline{\fig{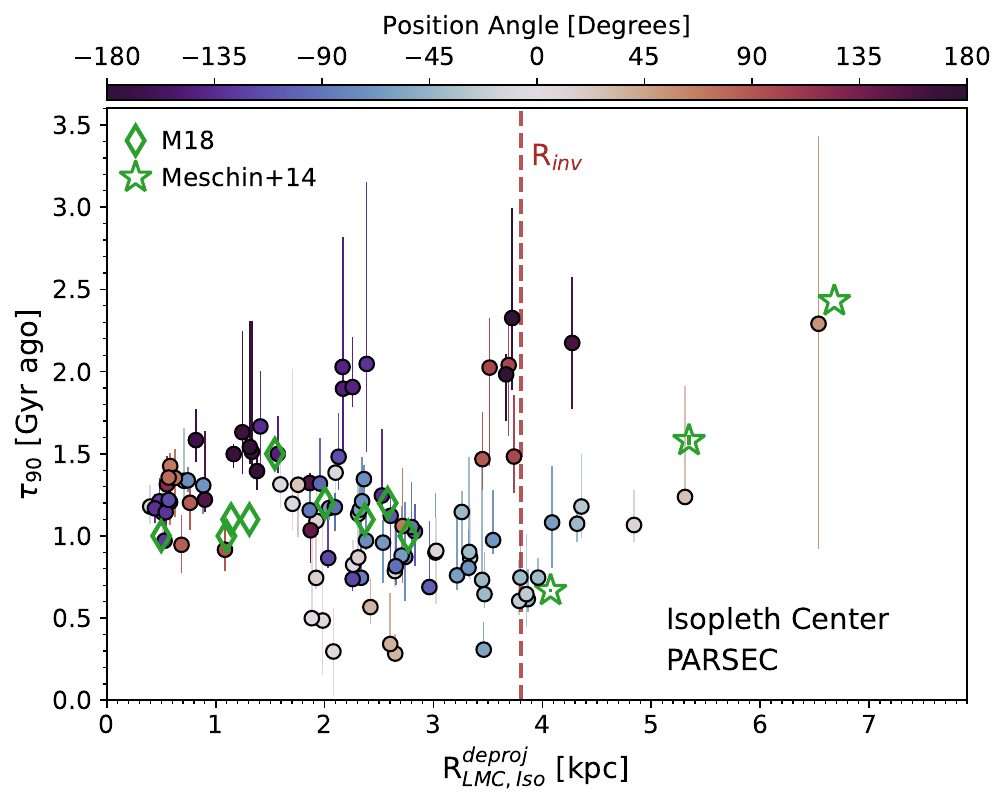}{0.5\textwidth}{}}
\vspace{-1.0cm}
\gridline{\fig{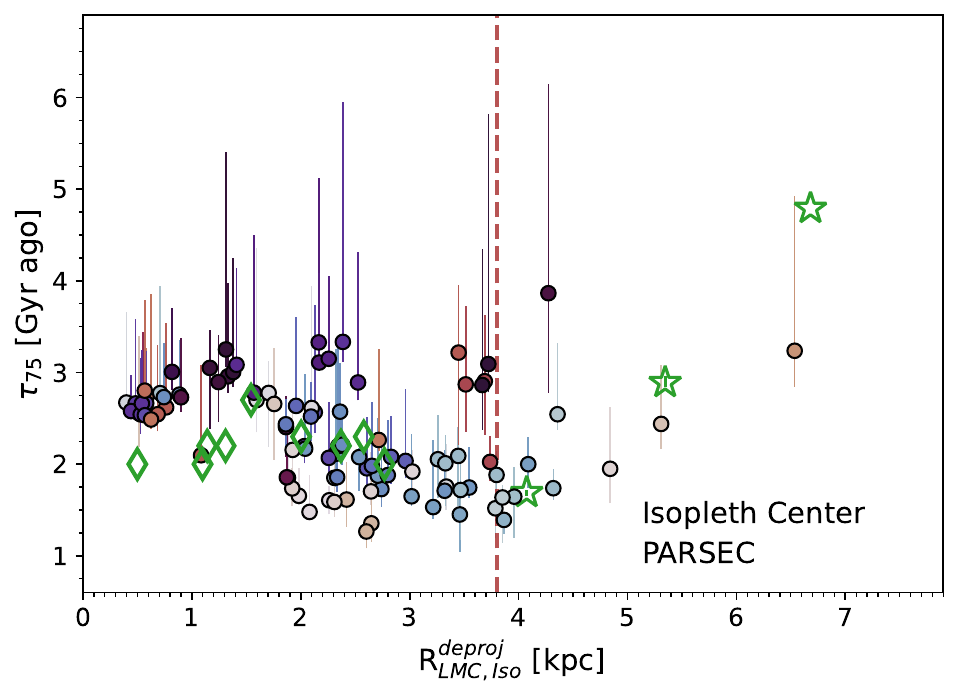}{0.49\textwidth}{}}
\vspace{-1.0cm}
\gridline{\fig{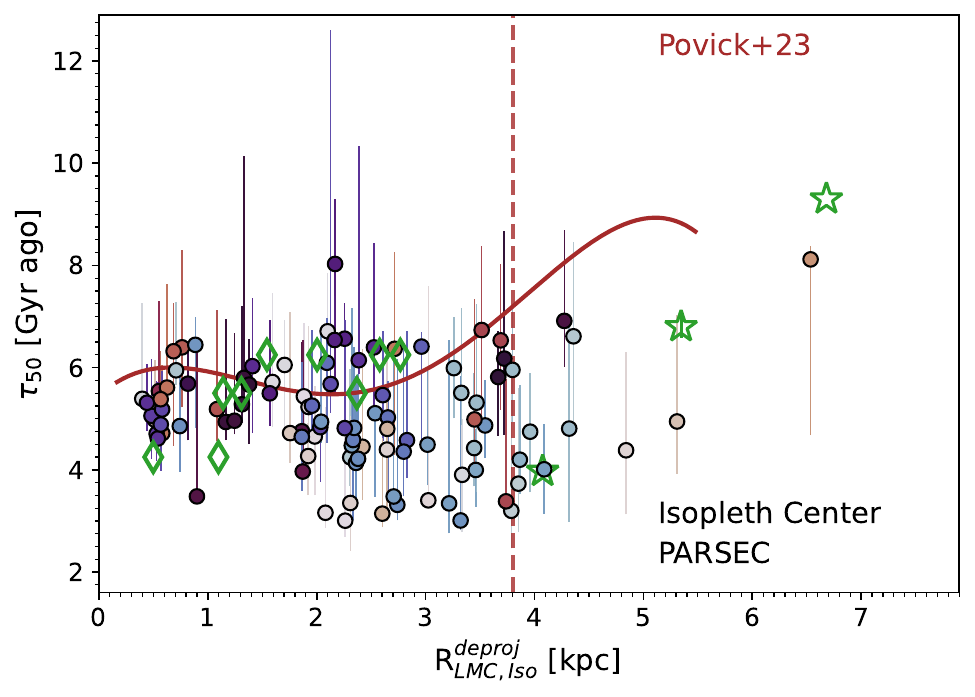}{0.49\textwidth}{}}
\caption{As in the left column of Fig.~\ref{csfh_rad_fig}, but assuming the isopleth-based LMC center location from \citet{vdmcioni}.  The median age-radius relation from \citet{povick} is overplotted as a brown curve in the bottom panel.} 
\label{csfh_rad_fig_alt1}
\end{figure}

\begin{figure}
\gridline{\fig{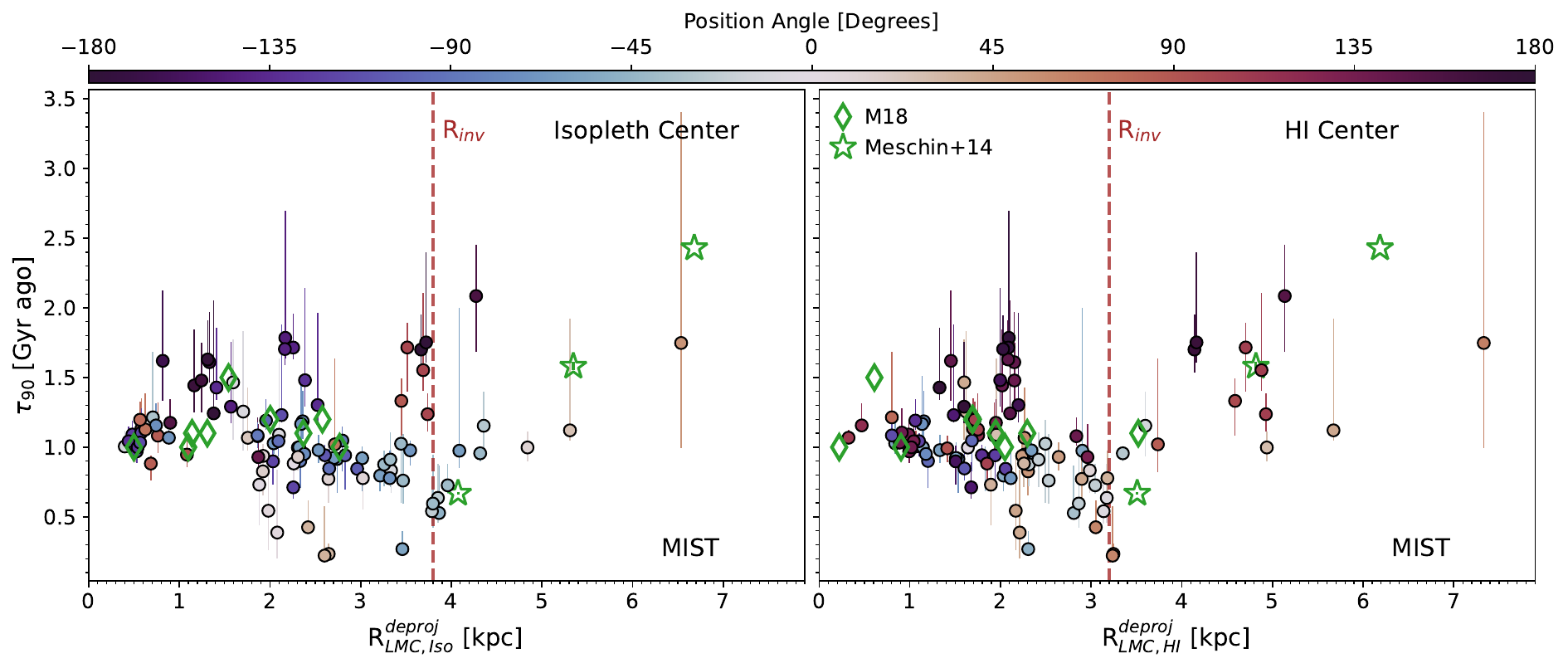}{0.99\textwidth}{}}
\vspace{-1.0cm}
\gridline{\fig{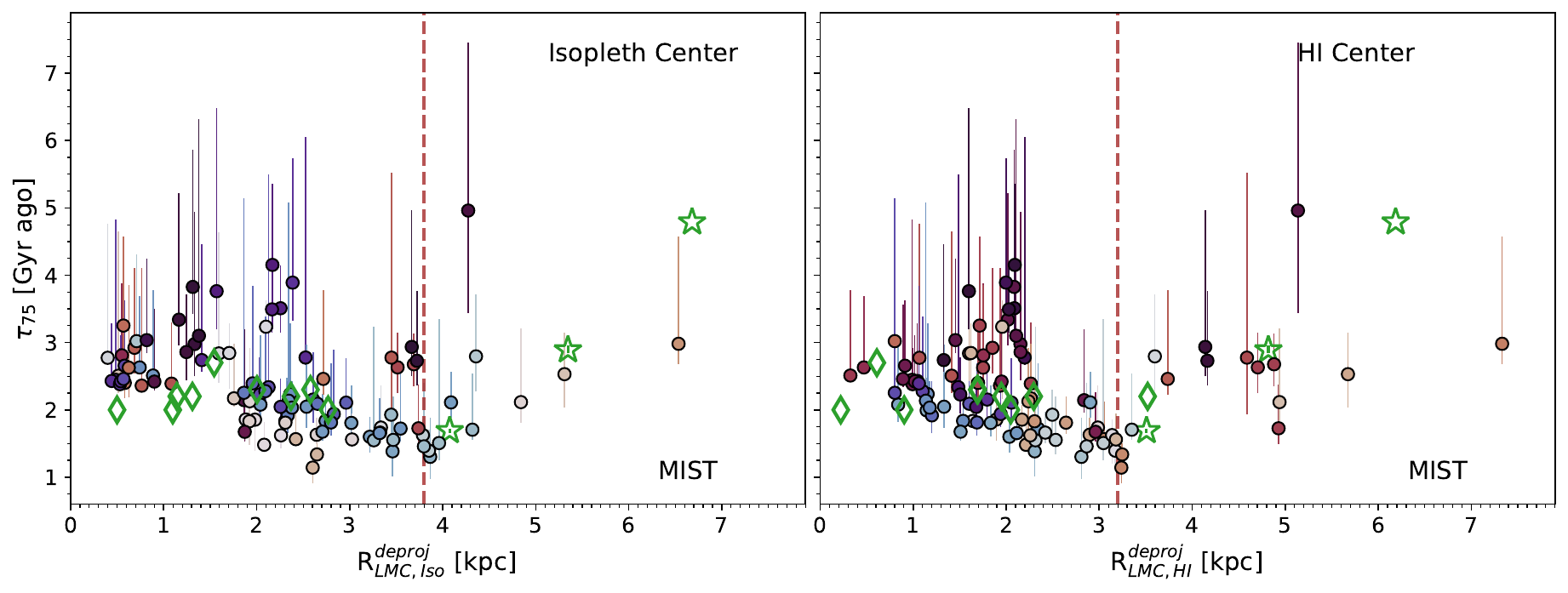}{0.99\textwidth}{}}
\vspace{-1.0cm}
\gridline{\fig{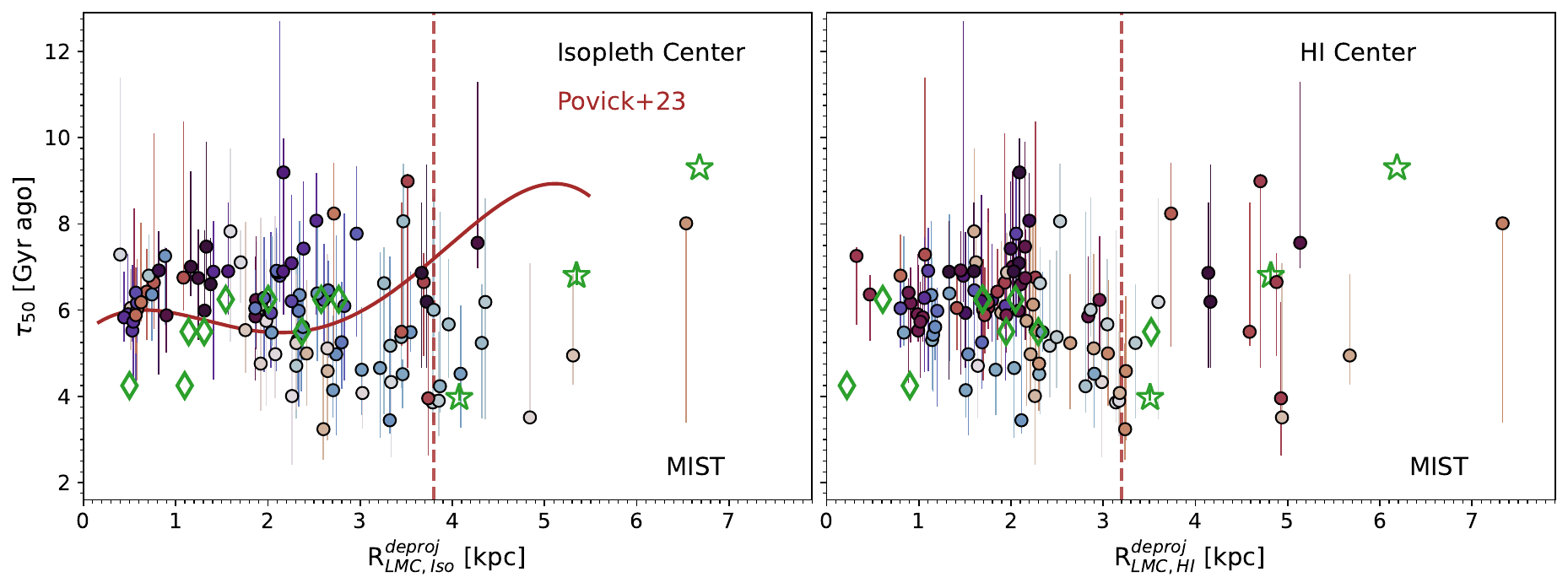}{0.99\textwidth}{}}
\caption{As in Fig.~\ref{csfh_rad_fig_alt1}, but assuming the MIST evolutionary models and adopting the isopleth-based center (left column) or the HI kinematic center (right column).}
\label{csfh_rad_fig_mist}
\end{figure}

\begin{figure}
\gridline{\fig{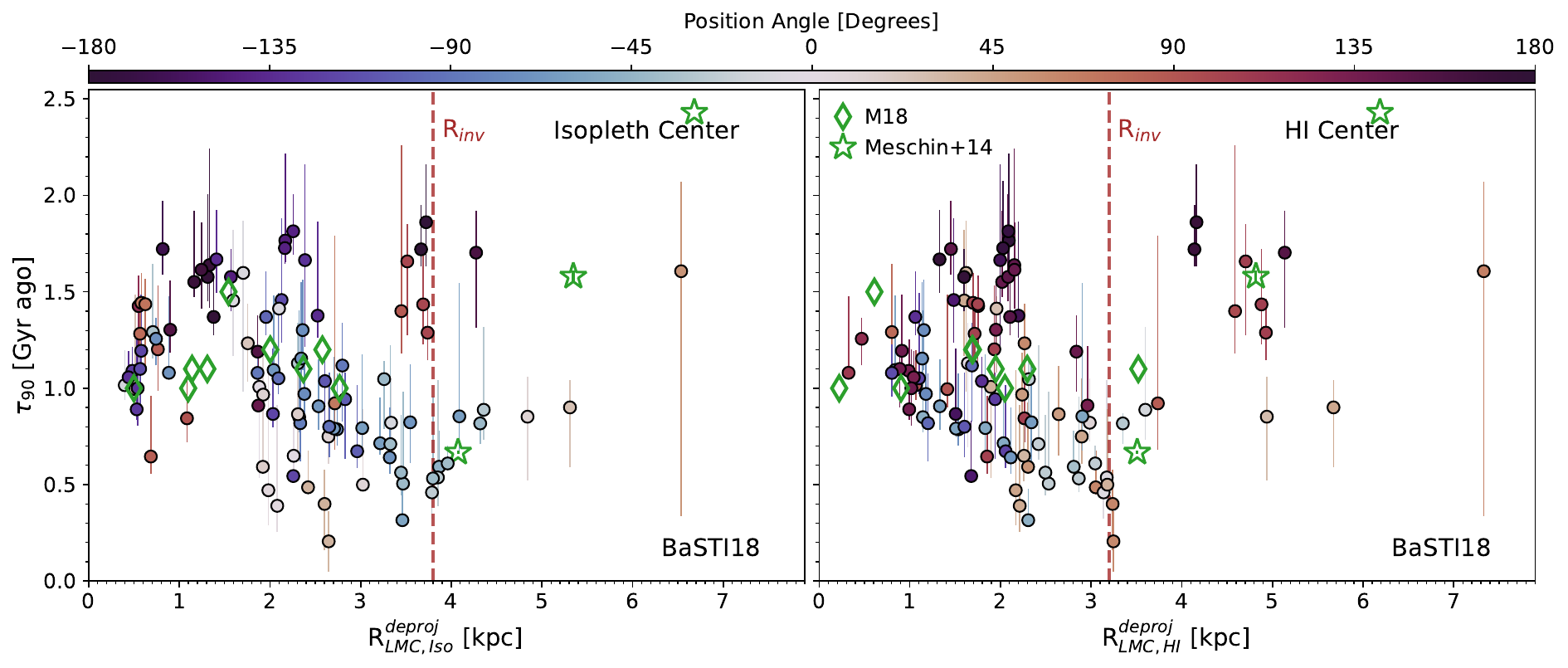}{0.99\textwidth}{}}
\vspace{-1.0cm}
\gridline{\fig{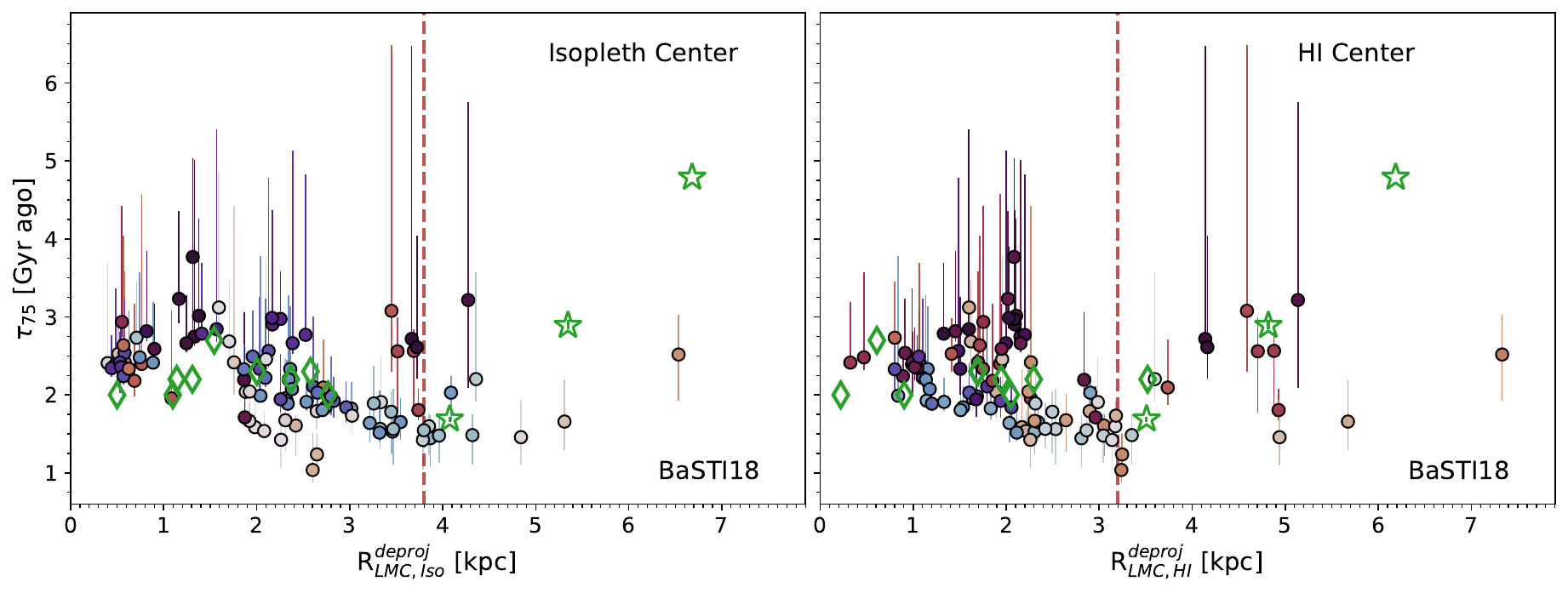}{0.99\textwidth}{}}
\vspace{-1.0cm}
\gridline{\fig{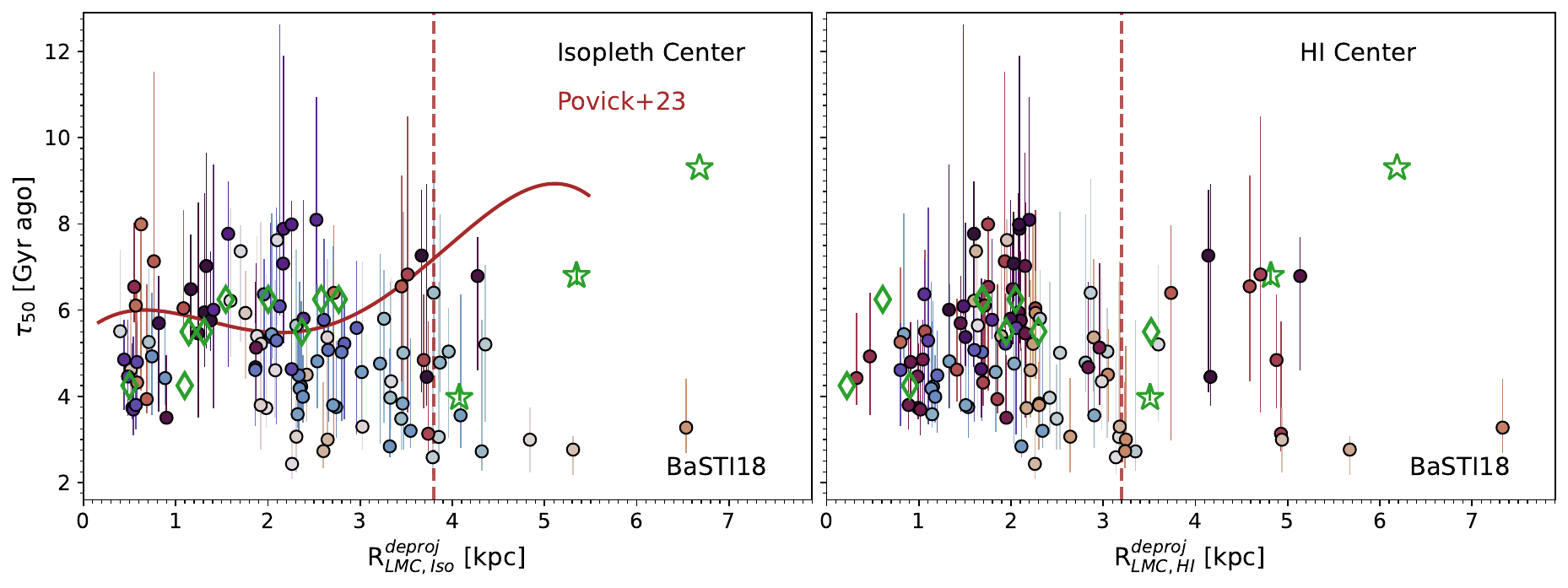}{0.99\textwidth}{}}
\caption{As in Fig.~\ref{csfh_rad_fig_mist}, but assuming BaSTI18 evolutionary models.}
\label{csfh_rad_fig_basti}
\end{figure}

In Figs.~\ref{csfh_rad_fig_alt1}-\ref{csfh_rad_fig_basti}, we present radial trends of lifetime SFH metrics as in the left column of Fig.~\ref{csfh_rad_fig}, adopting both the HI kinematic center and the \citet{vdmcioni} isopleth-based center and assuming MIST or BaSTI18 evolutionary models.  In Table \ref{csfhtab_alt} we present the Pearson linear correlation coefficient $\rho$ and Spearman rank correlation coefficient $r_{s}$ along with their $p$-values from SFH fits to inner disk (R$\leq$R$_{inv}$) fields as in Table \ref{csfhtab}, also assuming alternate stellar evolutionary models and center locations.  In Table \ref{scalelentab_alt}, we present the results of exponential fits to the cumulative stellar mass surface density profiles as in Table~\ref{scalelentab}, also given for both center locations and alternate evolutionary models.

\clearpage

\begin{deluxetable}{ccccc}
\tablecaption{Correlation Coefficients of Inner Disk CSFH Metrics Versus Deprojected Radius Assuming Alternate Stellar Evolutionary Models and LMC Centers \label{csfhtab_alt}}
\tablehead{\colhead{} & \multicolumn{2}{c}{Pearson} & \multicolumn{2}{c}{Spearman} \\[-0.2cm] \colhead{Metric} & \colhead{$\rho$} & \colhead{Log$_{\rm 10}$ $p$-value} & \colhead{$r_{s}$} & \colhead{Log$_{\rm 10}$ $p$-value}} 
\startdata
\hline
\multicolumn{5}{c}{PARSEC Models, Isopleth Center, All Fields} \\
\hline
$\tau_{90}$ & -0.20$^{+0.05}_{-0.05}$ & -1.22$^{+0.46}_{-0.59}$ & -0.31$^{+0.06}_{-0.05}$ & -2.52$^{+0.77}_{-0.90}$  \\
$\tau_{75}$ & -0.41$^{+0.06}_{-0.06}$ & -4.14$^{+1.18}_{-1.44}$ & -0.52$^{+0.05}_{-0.05}$ & -6.78$^{+1.33}_{-1.60}$  \\
$\tau_{50}$ & -0.13$^{+0.08}_{-0.08}$ & -0.70$^{+0.46}_{-0.72}$ & -0.19$^{+0.08}_{-0.08}$ & -1.16$^{+0.67}_{-0.91}$  \\
\hline
\multicolumn{5}{c}{PARSEC Models, Isopleth Center, Northern Fields Only} \\
\hline
$\tau_{90}$ & -0.33$^{+0.08}_{-0.08}$ & -1.71$^{+0.67}_{-0.82}$ & -0.36$^{+0.09}_{-0.08}$ & -1.99$^{+0.78}_{-0.96}$  \\
$\tau_{75}$ & -0.51$^{+0.08}_{-0.08}$ & -3.95$^{+1.21}_{-1.44}$ & -0.52$^{+0.07}_{-0.06}$ & -4.08$^{+1.06}_{-1.22}$  \\
$\tau_{50}$ & -0.23$^{+0.11}_{-0.11}$ & -0.98$^{+0.61}_{-0.89}$ & -0.25$^{+0.10}_{-0.11}$ & -1.13$^{+0.62}_{-0.91}$  \\
\hline
\multicolumn{5}{c}{MIST Models, Isopleth Center, All Fields} \\
\hline
$\tau_{90}$ & -0.23$^{+0.04}_{-0.04}$ & -1.52$^{+0.46}_{-0.55}$ & -0.34$^{+0.05}_{-0.04}$ & -2.98$^{+0.74}_{-0.82}$  \\
$\tau_{75}$  & -0.38$^{+0.07}_{-0.07}$ & -3.61$^{+1.19}_{-1.44}$ & -0.53$^{+0.05}_{-0.05}$ & -6.92$^{+1.47}_{-1.75}$  \\
$\tau_{50}$  & -0.20$^{+0.08}_{-0.08}$ & -1.22$^{+0.67}_{-0.90}$ & -0.23$^{+0.08}_{-0.08}$ & -1.48$^{+0.77}_{-1.04}$  \\
\hline
\multicolumn{5}{c}{MIST Models, Isopleth Center, Northern Fields Only} \\
\hline
$\tau_{90}$ & -0.33$^{+0.07}_{-0.06}$ & -1.77$^{+0.56}_{-0.67}$ & -0.39$^{+0.08}_{-0.07}$ & -2.29$^{+0.78}_{-0.92}$  \\
$\tau_{75}$ & -0.47$^{+0.12}_{-0.10}$ & -3.30$^{+1.42}_{-1.76}$ & -0.53$^{+0.08}_{-0.07}$ & -4.25$^{+1.24}_{-1.46}$  \\
$\tau_{50}$ & -0.25$^{+0.11}_{-0.11}$ & -1.14$^{+0.65}_{-0.94}$ & -0.28$^{+0.11}_{-0.11}$ & -1.32$^{+0.72}_{-1.03}$  \\
\hline
\multicolumn{5}{c}{MIST Models, HI Kinematic Center, All Fields} \\
\hline
$\tau_{90}$ & -0.24$^{+0.06}_{-0.05}$ & -1.61$^{+0.60}_{-0.68}$ & -0.30$^{+0.05}_{-0.05}$ & -2.23$^{+0.68}_{-0.77}$  \\
$\tau_{75}$ & -0.29$^{+0.06}_{-0.06}$ & -2.14$^{+0.77}_{-0.97}$ & -0.39$^{+0.06}_{-0.05}$ & -3.58$^{+0.94}_{-1.11}$  \\
$\tau_{50}$ & -0.13$^{+0.09}_{-0.08}$ & -0.66$^{+0.46}_{-0.69}$ & -0.13$^{+0.09}_{-0.08}$ & -0.63$^{+0.45}_{-0.71}$  \\
\hline
\multicolumn{5}{c}{MIST Models, HI Kinematic Center, Northern Fields Only} \\
\hline
$\tau_{90}$ & -0.50$^{+0.13}_{-0.09}$ & -2.87$^{+1.21}_{-1.21}$ & -0.55$^{+0.09}_{-0.08}$ & -3.46$^{+1.07}_{-1.32}$  \\
$\tau_{75}$ & -0.48$^{+0.09}_{-0.08}$ & -2.70$^{+0.88}_{-1.06}$ & -0.52$^{+0.10}_{-0.09}$ & -3.13$^{+1.04}_{-1.29}$  \\
$\tau_{50}$ & -0.18$^{+0.13}_{-0.13}$ & -0.56$^{+0.40}_{-0.70}$ & -0.20$^{+0.13}_{-0.13}$ & -0.64$^{+0.45}_{-0.73}$  \\
\hline
\multicolumn{5}{c}{BaSTI18 Models, Isopleth Center, All Fields} \\
\hline
$\tau_{90}$ & -0.29$^{+0.05}_{-0.05}$ & -2.22$^{+0.67}_{-0.81}$ & -0.36$^{+0.05}_{-0.05}$ & -3.20$^{+0.83}_{-0.98}$  \\
$\tau_{75}$ & -0.39$^{+0.09}_{-0.07}$ & -3.78$^{+1.41}_{-1.52}$ & -0.55$^{+0.05}_{-0.04}$ & -7.57$^{+1.46}_{-1.60}$  \\
$\tau_{50}$ & -0.12$^{+0.08}_{-0.08}$ & -0.59$^{+0.42}_{-0.68}$ & -0.17$^{+0.08}_{-0.08}$ & -0.98$^{+0.61}_{-0.87}$  \\
\hline
\multicolumn{5}{c}{BaSTI18 Models, Isopleth Center, Northern Fields Only} \\
\hline
$\tau_{90}$ & -0.36$^{+0.09}_{-0.08}$ & -2.04$^{+0.82}_{-0.99}$ & -0.41$^{+0.08}_{-0.07}$ & -2.54$^{+0.83}_{-0.98}$  \\
$\tau_{75}$ & -0.45$^{+0.15}_{-0.10}$ & -3.08$^{+1.57}_{-1.65}$ & -0.55$^{+0.06}_{-0.06}$ & -4.61$^{+1.09}_{-1.30}$  \\
$\tau_{50}$ & -0.20$^{+0.12}_{-0.12}$ & -0.81$^{+0.56}_{-0.88}$ & -0.24$^{+0.12}_{-0.11}$ & -1.05$^{+0.65}_{-0.92}$  \\
\hline
\multicolumn{5}{c}{BaSTI18 Models, HI Kinematic Center, All Fields} \\
\hline
$\tau_{90}$ & -0.29$^{+0.05}_{-0.05}$ & -2.23$^{+0.66}_{-0.75}$ & -0.30$^{+0.05}_{-0.05}$ & -2.35$^{+0.69}_{-0.78}$  \\
$\tau_{75}$ & -0.30$^{+0.06}_{-0.06}$ & -2.31$^{+0.76}_{-0.99}$ & -0.41$^{+0.05}_{-0.05}$ & -3.94$^{+0.87}_{-1.01}$  \\
$\tau_{50}$ & -0.02$^{+0.08}_{-0.08}$ & -0.22$^{+0.16}_{-0.35}$ & -0.03$^{+0.08}_{-0.09}$ & -0.25$^{+0.18}_{-0.39}$  \\
\hline
\multicolumn{5}{c}{BaSTI18 Models, HI Kinematic Center, Northern Fields Only} \\
\hline
$\tau_{90}$ & -0.53$^{+0.09}_{-0.08}$ & -3.32$^{+1.08}_{-1.23}$ & -0.56$^{+0.08}_{-0.08}$ & -3.60$^{+1.05}_{-1.27}$  \\
$\tau_{75}$ & -0.51$^{+0.10}_{-0.09}$ & -3.06$^{+1.06}_{-1.26}$ & -0.58$^{+0.08}_{-0.07}$ & -4.01$^{+1.05}_{-1.23}$  \\
$\tau_{50}$ & -0.13$^{+0.13}_{-0.13}$ & -0.39$^{+0.30}_{-0.60}$ & -0.15$^{+0.13}_{-0.12}$ & -0.46$^{+0.34}_{-0.60}$  \\
\enddata
\end{deluxetable}

\begin{deluxetable}{cccccccc}
\tablecaption{Time Evolution of the Disk Scalelength and Central Density Assuming Alternate Stellar Evolutionary Models and LMC Center Locations \label{scalelentab_alt}}
\tablehead{\colhead{} & \multicolumn{2}{c}{All Fields} & \multicolumn{2}{c}{R$\leq$R$_{break}$} & \multicolumn{2}{c}{R$>$R$_{break}$} & \colhead{} \\ \colhead{Lookback Time} & \colhead{$h_{r}$} & \colhead{$\Sigma_{0}$} & \colhead{$h_{r}$} & \colhead{$\Sigma_{0}$} & \colhead{$h_{r}$} & \colhead{$\Sigma_{0}$} & \colhead{R$_{break}$} \\ \colhead{Gyr} & \colhead{kpc} & \colhead{M$_{\odot}$/pc$^{2}$} & \colhead{kpc} & \colhead{M$_{\odot}$/pc$^{2}$} & \colhead{kpc} & \colhead{M$_{\odot}$/pc$^{2}$} & \colhead{kpc}}   
\startdata
\multicolumn{7}{c}{PARSEC models, Isopleth Center} \\
\hline
10.0 & 1.49$^{+0.26}_{-0.20}$ & 58.$^{+12.}_{-10.}$ & 1.48$^{+0.26}_{-0.24}$ & 59.$^{+13.}_{-11.}$ & 2.14$^{+0.71}_{-0.40}$ & 20.$^{+14.}_{-9.3}$ & 5.76$^{+4.78}_{-3.04}$ \\
7.1 & 1.40$^{+0.19}_{-0.13}$ & 127$^{+18.}_{-17.}$ & 1.40$^{+0.19}_{-0.15}$ & 129$^{+19.}_{-18.}$ & 2.90$^{+1.48}_{-0.82}$ & 29.$^{+22.}_{-11.}$ & 3.96$^{+1.34}_{-0.90}$ \\
4.0 & 1.38$^{+0.11}_{-0.10}$ & 230$^{+22.}_{-23.}$ & 1.38$^{+0.11}_{-0.11}$ & 231$^{+23.}_{-24.}$ & 2.81$^{+1.16}_{-0.69}$ & 47.$^{+30.}_{-21.}$ & 4.08$^{+1.22}_{-0.53}$ \\
2.0 & 1.37$^{+0.08}_{-0.09}$ & 307$^{+27.}_{-25.}$ & 1.37$^{+0.09}_{-0.09}$ & 308$^{+30.}_{-25.}$ & 3.24$^{+1.93}_{-0.98}$ & 48.$^{+33.}_{-28.}$ & 4.31$^{+2.14}_{-0.72}$ \\
0.0 & 1.45$^{+0.08}_{-0.07}$ & 365$^{+27.}_{-26.}$ & 1.45$^{+0.08}_{-0.07}$ & 365$^{+28.}_{-26.}$ & 10.7$^{+4.95}_{-3.15}$ & 9.4$^{+1.1}_{-0.9}$ & 6.14$^{+0.56}_{-0.52}$ \\
\hline
\multicolumn{7}{c}{MIST models, Isopleth Center} \\
\hline
10.0 & 1.55$^{+0.22}_{-0.19}$ & 79.$^{+15.}_{-11.}$ & 1.54$^{+0.20}_{-0.20}$ & 79.$^{+16.}_{-11.}$ & 2.02$^{+0.47}_{-0.30}$ & 22.$^{+14.}_{-7.8}$ & 8.84$^{+8.45}_{-4.98}$ \\
7.1 & 1.38$^{+0.13}_{-0.13}$ & 161$^{+21.}_{-19.}$ & 1.38$^{+0.14}_{-0.15}$ & 161$^{+23.}_{-19.}$ & 2.50$^{+0.92}_{-0.63}$ & 40.$^{+25.}_{-19.}$ & 4.27$^{+2.93}_{-0.95}$ \\
4.0 & 1.34$^{+0.10}_{-0.10}$ & 253$^{+24.}_{-25.}$ & 1.34$^{+0.10}_{-0.11}$ & 253$^{+28.}_{-26.}$ & 2.68$^{+1.06}_{-0.73}$ & 49.$^{+43.}_{-24.}$ & 4.27$^{+1.84}_{-0.93}$ \\
2.0 & 1.37$^{+0.08}_{-0.09}$ & 303$^{+30.}_{-25.}$ & 1.37$^{+0.08}_{-0.10}$ & 305$^{+34.}_{-27.}$ & 2.98$^{+1.37}_{-0.93}$ & 55.$^{+38.}_{-29.}$ & 4.27$^{+2.29}_{-0.80}$ \\
0.0 & 1.43$^{+0.08}_{-0.07}$ & 366$^{+25.}_{-26.}$ & 1.43$^{+0.08}_{-0.07}$ & 366$^{+27.}_{-26.}$ & 10.4$^{+4.78}_{-3.30}$ & 9.4$^{+1.2}_{-0.9}$ & 6.04$^{+0.58}_{-0.49}$ \\
\hline
\multicolumn{7}{c}{MIST models, HI Kinematic Center} \\
\hline
10.0 & 1.25$^{+0.22}_{-0.17}$ & 102$^{+27.}_{-20.}$ & 1.24$^{+0.21}_{-0.17}$ & 102$^{+28.}_{-20.}$ & 2.72$^{+1.37}_{-0.79}$ & 21.$^{+12.}_{-10.}$ & 3.56$^{+3.48}_{-0.32}$ \\
7.1 & 1.11$^{+0.18}_{-0.12}$ & 215$^{+41.}_{-40.}$ & 1.10$^{+0.18}_{-0.13}$ & 217$^{+42.}_{-40.}$ & 3.58$^{+2.17}_{-1.07}$ & 28.$^{+19.}_{-11.}$ & 3.24$^{+0.35}_{-0.24}$ \\
4.0 & 1.08$^{+0.11}_{-0.10}$ & 340$^{+49.}_{-48.}$ & 1.08$^{+0.11}_{-0.09}$ & 341$^{+51.}_{-48.}$ & 4.20$^{+1.77}_{-1.14}$ & 34.$^{+17.}_{-9.4}$ & 3.34$^{+0.22}_{-0.21}$ \\
2.0 & 1.10$^{+0.10}_{-0.10}$ & 408$^{+60.}_{-52.}$ & 1.10$^{+0.10}_{-0.09}$ & 411$^{+61.}_{-52.}$ & 4.36$^{+1.75}_{-1.06}$ & 43.$^{+17.}_{-11.}$ & 3.32$^{+0.18}_{-0.19}$ \\
0.0 & 1.14$^{+0.09}_{-0.08}$ & 495$^{+56.}_{-52.}$ & 1.13$^{+0.09}_{-0.08}$ & 499$^{+58.}_{-53.}$ & 4.03$^{+1.26}_{-0.92}$ & 60.$^{+22.}_{-16.}$ & 3.35$^{+0.24}_{-0.21}$ \\
\hline
\multicolumn{7}{c}{BaSTI18 models, Isopleth Center} \\
\hline
10.0 & 1.79$^{+0.24}_{-0.20}$ & 66.$^{+10.}_{-9.4}$ & 1.74$^{+0.23}_{-0.20}$ & 68.$^{+11.}_{-9.8}$ & 2.17$^{+0.41}_{-0.45}$ & 15.$^{+11.}_{-6.8}$ & 15.8$^{+2.84}_{-6.09}$ \\
7.1 & 1.64$^{+0.16}_{-0.15}$ & 119$^{+16.}_{-13.}$ & 1.64$^{+0.16}_{-0.15}$ & 119$^{+16.}_{-13.}$ & 2.79$^{+0.26}_{-0.50}$ & 15.$^{+5.4}_{-2.5}$ & 8.40$^{+2.04}_{-1.21}$ \\
4.0 & 1.48$^{+0.11}_{-0.10}$ & 203$^{+20.}_{-19.}$ & 1.48$^{+0.11}_{-0.10}$ & 203$^{+20.}_{-19.}$ & 2.65$^{+0.32}_{-0.63}$ & 24.$^{+26.}_{-3.7}$ & 7.13$^{+1.10}_{-0.66}$ \\
2.0 & 1.45$^{+0.10}_{-0.09}$ & 283$^{+28.}_{-23.}$ & 1.45$^{+0.10}_{-0.09}$ & 284$^{+28.}_{-23.}$ & 4.73$^{+2.11}_{-2.06}$ & 14.$^{+34.}_{-4.3}$ & 6.38$^{+0.55}_{-1.07}$ \\
0.0 & 1.52$^{+0.09}_{-0.08}$ & 350$^{+25.}_{-22.}$ & 1.52$^{+0.09}_{-0.08}$ & 350$^{+25.}_{-23.}$ & 10.0$^{+2.48}_{-1.68}$ & 9.0$^{+0.8}_{-0.6}$ & 6.61$^{+0.27}_{-0.44}$ \\
\hline
\multicolumn{7}{c}{BaSTI18 models, HI Kinematic Center} \\
\hline
10.0 & 1.42$^{+0.26}_{-0.23}$ & 84.$^{+22.}_{-17.}$ & 1.41$^{+0.27}_{-0.23}$ & 84.$^{+23.}_{-17.}$ & 2.16$^{+0.58}_{-0.40}$ & 21.$^{+12.}_{-10.}$ & 8.39$^{+5.58}_{-4.95}$ \\
7.1 & 1.40$^{+0.20}_{-0.17}$ & 143$^{+25.}_{-22.}$ & 1.40$^{+0.21}_{-0.17}$ & 144$^{+25.}_{-23.}$ & 2.33$^{+0.65}_{-0.46}$ & 36.$^{+14.}_{-13.}$ & 3.73$^{+6.46}_{-0.38}$ \\
4.0 & 1.21$^{+0.15}_{-0.13}$ & 260$^{+51.}_{-37.}$ & 1.20$^{+0.16}_{-0.13}$ & 262$^{+51.}_{-38.}$ & 2.85$^{+1.07}_{-0.73}$ & 48.$^{+20.}_{-16.}$ & 3.45$^{+0.67}_{-0.21}$ \\
2.0 & 1.16$^{+0.10}_{-0.11}$ & 380$^{+58.}_{-43.}$ & 1.15$^{+0.10}_{-0.10}$ & 381$^{+59.}_{-42.}$ & 3.59$^{+0.96}_{-0.70}$ & 51.$^{+18.}_{-12.}$ & 3.35$^{+0.24}_{-0.10}$ \\
0.0 & 1.18$^{+0.10}_{-0.08}$ & 479$^{+50.}_{-49.}$ & 1.17$^{+0.10}_{-0.08}$ & 482$^{+51.}_{-51.}$ & 3.82$^{+1.73}_{-0.81}$ & 61.$^{+23.}_{-21.}$ & 3.47$^{+0.26}_{-0.17}$ \\
\enddata
\end{deluxetable}

\end{document}